\documentclass[twocolumn]{aastex7}
\usepackage{graphicx}
\usepackage{float}
\usepackage{amsmath}
\usepackage{amssymb}
\usepackage{cases}
\usepackage{mathrsfs}
\usepackage[flushleft]{threeparttable}
\usepackage{pifont}
\usepackage{epstopdf}
\usepackage{array}
\usepackage{xcolor}
\usepackage[all]{hypcap}
\usepackage{mwe,tikz}
\usepackage{bm}
\usepackage{tabularx}
\usepackage{multirow}
\usepackage{makecell}
\usepackage{longtable}
\usepackage{xspace}
\usepackage{rotating}
\usepackage{upgreek}
\usepackage{booktabs}
\usepackage{textcomp}
\usepackage{enumitem}
\usepackage[utf8]{inputenc}
\usepackage{csquotes}
\usepackage{soul}
\usepackage{bbding}
\usepackage[T1]{fontenc}
\usepackage{float}
\usepackage{tikz}
\usepackage{appendix}
\usepackage{cleveref}
\usepackage[percent]{overpic} 
\usepackage{ulem} 
\usepackage{CJKutf8}
\usepackage{placeins} 
\usepackage{subcaption}
\usepackage{xcolor}

\newcommand{\chinesename}{\textnormal{(}张钊\textnormal{)}}

\newcommand{\ud}{\mathrm{d}}

\newcommand{\e}{\text{e}}

\newcommand{\p}{\partial}

\hypersetup{
    colorlinks=true,
    linkcolor=blue,
    citecolor=blue,
    urlcolor=blue
}

\shorttitle{FRB Local DM from 1D SNR Models}
\shortauthors{Zhang et al.}

\begin{document}
\begin{CJK}{UTF8}{gbsn}
\title{Probing the Dispersion and Rotation Measure Contributions from Supernova Remnants in Fast Radio Burst Source Environments with 1D SNR Simulation}

\correspondingauthor{Zhao Joseph Zhang}
\email{zzhang@astro-osaka.jp}

\author[0000-0001-6869-2996]{Zhao Joseph Zhang \chinesename}
\affiliation{Theoretical Astrophysics, Department of Earth and Space Science, The University of Osaka, 1-1 Machikaneyama, Toyonaka, Osaka 560-0043, Japan}
\email{zzhang@astro-osaka.jp}
\author[orcid=0009-0006-1396-8397,sname='Kawashima']{Gaku Kawashima}
\affiliation{Department of Astronomy, Kyoto University, Sakyo-ku, Kyoto 606-8502, Japan}
\email[show]{kawashima@kusastro.kyoto-u.ac.jp}  

\author[orcid=0000-0002-2899-4241,sname='Lee']{Shiu-Hang Lee}
\affiliation{Department of Astronomy, Kyoto University, Sakyo-ku, Kyoto 606-8502, Japan}
\affiliation{Kavli Institute for the Physics and Mathematics of the Universe (WPI), The University of Tokyo, Kashiwa 277-8583, Japan}
\email{herman@kusastro.kyoto-u.ac.jp}

\author[0000-0001-7457-8487]{Kentaro Nagamine}
\affiliation{Theoretical Astrophysics, Department of Earth and Space Science, The University of Osaka, 1-1 Machikaneyama, Toyonaka, Osaka 560-0043, Japan}
\affiliation{Theoretical Joint Research, Forefront Research Center, Graduate School of Science, The University of Osaka, Toyonaka, Osaka 560-0043, Japan}
\affiliation{Kavli IPMU (WPI), UTIAS, The University of Tokyo, Kashiwa, Chiba 277-8583, Japan}
\affiliation{Department of Physics \& Astronomy, University of Nevada, Las Vegas, 4505 S. Maryland Pkwy, Las Vegas, NV 89154-4002, USA}
\affiliation{Nevada Center for Astrophysics, University of Nevada, Las Vegas, 4505 S. Maryland Pkwy, Las Vegas, NV 89154-4002, USA}
\email{kn@astro-osaka.jp}

\author[0000-0002-9725-2524]{Bing Zhang}
\affiliation{The Hong Kong Institute for Astronomy and Astrophysics, The University of Hong Kong, Pokfulam Rd, Hong Kong SAR, P. R. China}
\affiliation{Department of physics, The University of Hong Kong, Pokfulam Rd, Hong Kong SAR, P. R. China}
\affiliation{Nevada Center for Astrophysics, University of Nevada, Las Vegas, 4505 S. Maryland Pkwy, Las Vegas, NV 89154-4002, USA}
\affiliation{Department of Physics \& Astronomy, University of Nevada, Las Vegas, 4505 S. Maryland Pkwy, Las Vegas, NV 89154-4002, USA}
\email{bzhang1@hku.hk}

\author[0009-0007-8699-8973]{Yusei Fujimaru}
\affiliation{Department of Astronomy, Kyoto University, Sakyo-ku, Kyoto 606-8502, Japan}
\email{fujimaru@kusastro.kyoto-u.ac.jp}
\begin{abstract}
Fast radio bursts (FRBs) provide a sensitive probe of ionized baryons through their dispersion measure (DM). In addition to slowly evolving cosmological terms, {at least three repeaters now show clear secular DM-decrease episodes: FRB~20190520B, FRB~20220529A, and FRB~20121102} (the latter with a two-stage trend: early mild rise and late decline), supporting a dense, dynamically evolving local environment.
We adopt a \emph{forward-modeling} approach and use time-dependent 1D SNR simulations for a young magnetar embedded in SN ejecta, combining single-star and binary-stripped progenitors with {hydrodynamic (HD) and non-equilibrium ionization (NEI)} calculations to follow shock structure, ionization, and electron density.
The shocked region contributes only limited DM ($\lesssim10\,{\rm pc\,cm^{-3}}$), while the dominant time-varying component is the unshocked ejecta, whose early behavior follows ${\rm DM}\propto t^{-\alpha}$ with $\alpha\simeq1.8$--$1.9$. Although shocked-region DM is small, shock-amplified magnetic fields can still generate substantial RM; in our shock-only RM framework, only the $11\,M_\odot$ SS model reproduces the FRB~20121102 RM evolution.
Binary-stripped progenitors generally yield smaller DM than single-star models at fixed $M_{\rm ZAMS}$, with composition-dependent mean molecular weights introducing non-monotonic mass trends. {Matching the observed ${\rm dDM}/{\rm d}t$ of FRB~20190520B, FRB~20220529A, and the late-stage slope of FRB~20121102, we infer local SNR DM contributions of $\sim 10$ to a few $10^2$ ${\rm pc\,cm^{-3}}$.}
We also find GHz escape is allowed in most models, with $\tau_{\rm ff}=1$ typically reached by $t_{\rm esc}\lesssim70$ yr; for weakly ionized ejecta, the source can be nearly transparent from very early times.
These results support a young CCSN/SNR origin for a substantial fraction of ${\rm DM}_{\rm source}$ and highlight that physically consistent local-environment modeling is essential for robust FRB cosmological DM inferences.
\end{abstract}

\keywords{fast radio bursts --- supernova remnants --- magnetars --- dispersion measure}

\section{Introduction}\label{sec:intro}

FRBs are short ($\sim$ms), bright radio transients with large dispersion measures (DMs), implying extragalactic or even cosmological distances.
Since the first Lorimer burst \citep{Lorimer2007}, the DM budget of FRBs has been recognized as a key observable, connecting FRBs to the baryon content of the Universe \citep{McQuinn2014, DZ2014, Zheng2014,Li2019...876...146, Macquart2020Nature, Simha2020, Takahashi2021MNRAS, Medlock2021, Zhu2021, Lee2022, KHRYKIN2024, Medlock2024, Connor2024, ZZ2025, Konietzka2025arXiv} as well as to the physical conditions in their local environments (see also the review by \citealt{Zhan2023RvMP}).
Repeating FRBs with well-sampled monitoring campaigns now reveal that the DM is not constant over time; it can vary secularly on timescales of years, suggesting that at least part of the DM originates in a dynamical, evolving medium near the source.

A natural framework is that at least some FRBs are produced by young magnetars {\citep{Margalit2018MNRAS}} embedded in SNRs and their surrounding circumstellar medium (CSM).
In this picture, the total observed FRB DM can be decomposed into several contributions:
\begin{equation}
\label{eq:DM_total}
    {\rm DM}_{\rm FRB}
    =
    {\rm DM}_{\rm MW}
    +
    {\rm DM}_{\rm Halos}
    +
    {\rm DM}_{\rm IGM}
    +
    \frac{{\rm DM}_{\rm HG} + {\rm DM}_{\rm source}}{1+z},
\end{equation}
Where ${\rm DM}_{\rm MW}$ is the contribution from the Milky Way, including both the interstellar medium (${\rm DM}_{\rm MW,ISM}$) and the Milky Way's circumgalactic halo (${\rm DM}_{\rm MW,halo}$).
${\rm DM}_{\rm Halos}$ represents the cumulative contribution from all intervening galactic halos along the line of sight, 
i.e., 
\begin{equation*}
{\rm DM}_{\rm Halos} = \sum_i \frac{{\rm DM}_{\rm halo,i}}{1+z_i},
\end{equation*}
where each term corresponds to the circumgalactic medium (CGM) of a foreground galaxy intersected by the FRB sight line, ${\rm DM}_{\rm IGM}$ collects the contributions from the intergalactic medium, ${\rm DM}_{\rm HG}$ is the contribution from the host galaxy, and ${\rm DM}_{\rm source}$\footnote{We note that in \cite{ZZ2025}, the host-galaxy term ${\rm DM}_{\rm HG}$ was defined 
to include the ${\rm DM}_{\rm source}$.  
Here we intentionally adopt a different convention: 
${\rm DM}_{\rm HG}$ is treated as part of the cosmological contribution, 
while ${\rm DM}_{\rm source}$ is kept separate.  
This is done to avoid confusion between the two papers.} is the local contribution in the immediate vicinity of the FRB engine (typically on parsec or sub-parsec scales).
On timescales of $\sim$years, {temporal variations in} all components except ${\rm DM}_{\rm source}$ are expected to be negligible \citep{Yang&Zhang2017}, so any observed secular change in DM is naturally attributed to the evolution of the source environment. However, the ${\rm DM}_{\rm source}$ inferred from the observed DM-decay slope is itself model dependent; therefore, an apparent local DM excess can remain partially degenerate with ${\rm DM}_{\rm Halos}$ from foreground structures, leading to ongoing debate in some sources. FRB~20190520B provides a representative example, with a source-dominated interpretation \citep{ZYZhao2021b} versus a substantial foreground-halo/cluster contribution \citep{Lee2023}.

Previous theoretical work \citep{Yang2017ApJ...839L..25Y, Murase2016, Kashiyama2017, Yang&Zhang2017} has explored a variety of possible contributors to ${\rm DM}_{\rm source}$.
These include (i) the SNR ejecta and shocked shell 
(e.g., \citealt{Piro2016, Piro&Gaensler2018, Niu2025}), (ii) a nebula powered by the relativistic pulsar/magnetar wind (e.g., \citealt{Yang2016, Margalit2018, Metzger2019, ZYZhao2021b, Mahlmann2022}),  (iii) dense star-forming or H\,II regions in the host galaxy (e.g., \citealt{Tendulkar2017, Michilli2018Natur, Bassa2017}), and (iv) small-scale plasma lensing structures (e.g., \citealt{Cordes2017, Main2018Natur}).
Among these, the SNR scenario is particularly attractive for sources with large host DM and significant negative $\mathrm{dDM}/\mathrm{d}t$, in which an expanding ionized shell naturally leads to a declining DM over time.
Analytic models based on the McKee \& Truelove treatment of SNR evolution have been used to estimate the time dependence of the SNR DM in different expansion phases---ejecta-dominated (ED), Sedov--Taylor (ST), and radiative snowplow---\citep{McKee&Truelove1995PhR,McKee&Truelove1999,Piro2016, Piro&Gaensler2018}, and to place constraints on the age and environment of individual FRB sources.

On the observational side, several FRBs show a substantial excess or variations in their DM or Rotation Measure (RM) \citep{Masui2015Natur, Hilmarsson2021,Niu2022,Wang2022NatCo, Mckinven2023}, indicating the presence of a dense and magnetized local environment. FRB~20190520B shows a clear long-term secular decrease in DM, with a nearly linear decline over $\sim 4$ years at ${\rm dDM}/{\rm d}t \simeq -12\,{\rm pc\,cm^{-3}\,yr^{-1}}$ \citep{Niu2025}. {Recent monitoring of FRB~20121102 reveals secular DM evolution with a two-stage behavior: an early segment with slight DM increase and a later phase showing DM decline \citep{Wang2025arXiv250715790W}. FRB~20220529A now provides a third well-monitored case of secular DM decline: a $3.2$-yr campaign finds ${\rm dDM}/{\rm d}t=-0.881\pm0.001\,{\rm pc\,cm^{-3}\,yr^{-1}}$, together with a short-lived DM excursion associated with enhanced RM activity \citep{Pandhi2026}.} In this work, when performing SNR slope matching for FRB~20121102, we use the late (declining) segment. The slope turnover between the early and late stages may indicate additional local-plasma components (e.g., MWN-related contributions), consistent with hybrid interpretations discussed in \citet{Dai2017}, \citet{Beloborodov2017}, and \citet{Yang2019}. We also note recent attempts to place active repeaters within a unified framework, including FRB~20190417A, an active repeater with periodic activity that shows oscillatory RM with sign reversals and is generally interpreted in a magnetar--massive-star binary environment, as well as persistent radio source (PRS)-associated systems \citep{Wang2025arXiv251207140W}.

Beyond these analytical estimates of the SNR contribution to ${\rm DM}_{\rm source}$, numerical simulations provide a natural and complementary approach for studying the dynamical evolution of SNRs and its imprint on FRB DM. Modern stellar evolution calculations, such as those performed with MESA \citep[version 12115;][]{Paxton2011ModulesMESA,Paxton2013ModulesStars,Paxton2015ModulesExplosions,Paxton2018ModulesExplosions,Paxton2019ModulesConservation,Jermyn2023ModulesInfrastructure}, self-consistently follow massive stars from the zero-age main sequence (ZAMS) to core collapse (CC), thereby providing realistic progenitor structures and ejecta properties as inputs for
subsequent CCSNR simulations.

At the same time, numerical simulations of supernova remnants have rapidly advanced, ranging from one-dimensional (1D) models that incorporate detailed shock physics, radiative cooling, non-equilibrium ionization (NEI), and cosmic ray (CR) acceleration
(e.g., \citealt{Patnaude2010, Lee+2012, Lee+2013, Lee+2014, Diesing2019, Diesing2024, Diesing2025}) to multi-dimensional (2D/3D) simulations that reveal the importance of pulsar wind nebulae (PWN), hydrodynamic instabilities, ejecta clumping, asymmetric ejecta--CSM interactions (e.g., \citealt{Blondin2001, Ferrand2010, Orlando2012, Ferrand2012, Ferrand2019}).

While previous analytical studies have provided valuable insights into the global shock structure and approximate DM scalings, they often adopt simplified progenitor and CSM configurations.
In particular, massive single stars and binary-stripped (BS) progenitors are expected to produce markedly different CSM density profiles due to their distinct mass-loss histories and wind properties \citep{Patnaude2017,Jacovich+2021, Laplace+2021, Farmer+2023, Kawashima2026arXiv}. Such differences can lead to SNR density and ionization structures, and hence DM evolution, that deviate substantially from the idealized time-scaling relations commonly assumed in analytical models.

Motivated by these considerations, in this work we take a step beyond analytic scalings and investigate the contribution of SNR to ${\rm DM}_{\rm source}$ using detailed 1D HD simulations with NEI and radiative cooling. We aim to test whether, and under what conditions, self-consistent HD SNR models can reproduce the range of ${\rm DM}_{\rm source}$, $\mathrm{dDM}/\mathrm{d}t$, and ${\rm RM}_{\rm source}$ inferred from FRB observations. In particular, we investigate whether specific combinations of ZAMS mass, ejecta mass, and circumstellar density structure, as expected from single-star (SS) and BS progenitors, can satisfy the observational constraints on both the absolute DM and its time evolution.

This draft is organized as follows. In Section~\ref{sec:DM_components}, we briefly summarize different contributions to the FRB DM budget and motivate the focus on the local ${\rm DM}_{\rm source}$ term. Section~\ref{sec:methods} presents the numerical methodology adopted in this work, including the CSM modeling, SNR simulation setup, and the calculation methods for DM, optical depth, and RM. Section~\ref{sec:results} presents results for two cases $M_{\rm ZAMS}=11\,M_\odot$ and $M_{\rm ZAMS}=30\,M_\odot$, including the time evolution of DM, its derivatives, free-free optical depth, RM evolution and ${\rm RM}_{\rm source}$ constraints, comparisons with analytic/semi-analytic models, and shocked-region electron-source decomposition. We discuss model limitations and future directions in Section~\ref{sec:discussion}, and provide a summary in Section~\ref{sec:conclusions}.

\section{FRB DM Components and Local Source Environment}
\label{sec:DM_components}

\subsection{Decomposition of the FRB DM budget}

For a given FRB, the total observed DM can be written as in Eq.~\eqref{eq:DM_total}.
For convenience, we define the non-local (large-scale, slowly varying) component,
\begin{equation}
    {\rm DM}_{\rm diff}
    \equiv
    {\rm DM}_{\rm IGM}
    +
    {\rm DM}_{\rm Halos},
\end{equation}
\begin{equation}
    {\rm DM}_{\rm nonlocal}
    \equiv
    {\rm DM}_{\rm MW}
    +
    {\rm DM}_{\rm diff}
    +
    \frac{{\rm DM}_{\rm HG}}{1+z},
\end{equation}
so that
\begin{equation}
    {\rm DM}_{\rm FRB}
    =
    {\rm DM}_{\rm nonlocal}
    +
    \frac{{\rm DM}_{\rm source}}{1+z}.
\end{equation}
Here ${\rm DM}_{\rm diff}$ traces large-scale structure along the line of sight and is the component that can be {reasonably well} estimated estimated from cosmological simulations or empirical DM--$z$ relations. These non-local terms are expected to vary negligibly over human timescales. By contrast, ${\rm DM}_{\rm source}$ is associated with the local environment of the FRB engine, on scales from $\lesssim 0.1$~pc (PWN and magnetar wind nebula) to a few pc (SNR shell and surrounding \ion{H}{2} region).
Because this material can evolve dynamically on timescales of years--decades, ${\rm DM}_{\rm source}$ can in principle be responsible for the secular DM evolution seen in some repeating FRBs.
Moreover, since ${\rm DM}_{\rm source}$ is sensitive to the progenitor history and explosion properties, it may carry information about whether the FRB source formed in a SS or BS channel.

In the following, we focus on modeling ${\rm DM}_{\rm source}$ originating in the SNR and its shocked surroundings.
Other local contributors (PWN, \ion{H}{2} regions, plasma lenses) will be briefly discussed in Section~\ref{sec:loc_env} as complementary or additional components.

\subsection{Candidate local environments: SNR, PWN, H\,II region, plasma lensing} \label{sec:loc_env}

Previous work has explored several possible local environments that could contribute significantly to the FRB DM and its time evolution:

\begin{itemize}
    \item \textit{SNR:}  
    An expanding SNR ejecta and its forward shock shell 
    provide a natural reservoir of dense ionized gas whose column density 
    decreases over time.  In previous FRB–SNR theoretical studies, the temporal 
    evolution of the SNR has often been modeled using the self-similar solutions 
    of \citet{McKee&Truelove1995PhR}, which describes the ED, 
    ST, and radiative snowplow phases.  Within this analytic 
    framework, the resulting DM evolution is commonly written as 
    ${\rm DM}_{\rm SNR} \propto t^{-\alpha}$ with $0 \lesssim \alpha \lesssim 2$\footnote{We emphasize, however, that these analytic trends need not hold for our numerical 1D SNR simulations, whose detailed ionization, shock structure, and density evolution may deviate from the idealized self-similar solutions.}, 
    and the corresponding $\mathrm{dDM}/\mathrm{d}t$ values can fall in the range 
    discussed by \citet{Piro2016, Piro&Gaensler2018, Yang&Zhang2017}.  
    
    \item \textit{PWN / magnetar wind nebula (MWN):}   
    A relativistic wind from the central neutron star can inflate a nebula whose internal plasma contributes to ${\rm DM}_{\rm source}$.  Two types of PWN models have been discussed.  In rotation-powered pulsar–wind bubbles 
    \citep[e.g.,][]{Murase2016, Kashiyama2017}, the nebula is dominated by $e^\pm$ pairs, and the lepton density is set by the Goldreich–Julian density, the pair multiplicity, and the spin-down power, generally yielding modest DM.  In contrast, flare-powered, baryon-loaded nebulae  
    \citep[e.g.,][]{Margalit2018} contains an ion–electron plasma whose density 
    depends on the baryon-loading factor and energy-injection history, and can 
    produce substantially larger DM (and RM). The PWN/MWN can not only enhance the $e^\pm$ pair-plasma density near the magnetosphere, but also inject baryon-loaded plasma into the surrounding SNR ejecta/CSM, further modifying the local DM and RM.

    \item \textit{\ion{H}{2} region and star-forming complex:}  
    If the FRB source is located in a compact \ion{H}{2} region or a dense star-forming clump, the associated ionized gas can contribute a substantial, but relatively slowly varying, DM.

    \item \textit{Plasma lensing structures:}  
    Small-scale overdense plasma sheets or filaments can both refract and disperse FRB radiation, potentially producing burst-to-burst DM variations that are not strictly secular.
    Although dense plasma anywhere along the line of sight can, in principle, cause plasma lensing, 
    "strong" lensing requires extremely compact, overdense plasma structures \citep{Cordes2017, Main2018Natur}, which are found 
    almost exclusively in the immediate local environment of the FRB (from sub-AU scales to a few--tens of parsecs); hence their DM contribution is naturally assigned to ${\rm DM}_{\rm source}$.

\end{itemize}

\citet{Yang&Zhang2017} presented a unified framework for these different contributors and emphasized that the cosmological and host-galaxy terms are nearly time-independent, while the SNR and PWN components can produce measurable $\mathrm{dDM}/\mathrm{d}t$.
In this study, we concentrate on the SNR\footnote{Central compact-object effects, such as PWN/MWN contributions, are discussed only as an ionization-enhancement upper-limit scenario in this work.} scenario and ask: given realistic SNR evolution, what range of ${\rm DM}_{\rm source}$ and $\mathrm{dDM}/\mathrm{d}t$ can be produced, and how does this depend on the progenitor channel? We further examine whether the time evolution of ${\rm DM}_{\rm SNR}$ alone can account for the observed secular decrease of FRB~20190520B.
\section{Analytic benchmarks for DM evolution in SNR environments}
\label{sec:analytic_models}

To establish a reference baseline for the subsequent comparison with our numerical SNR simulations, we summarize four commonly used (semi-)analytic DM prescriptions drawn from the literature. These models capture different physical regimes, including (i) Entire ionized ejecta in free expansion, (ii) a shocked ionized ejecta contribution, and (iii--iv) wind contributions in the presence of a forward shock, either excluding or including the unshocked wind outside the shock. The corresponding expressions are summarized in Eqs.~\eqref{eq:dm_full}–\eqref{eq:DM_Zhao2021_wind_unshocked_core}.

\paragraph{(1) Entire Ionized ejecta.}
In this scenario, the ejecta shell (including the unshocked component) is assumed to be ionized, and the DM is estimated as the electron column across the ejecta layer, $\mathrm{DM_{SNR}}\sim n_e \Delta R$, where $\Delta R=R_b-R_r$, with $R_b$ and $R_r$ denoting the blast-wave and reverse-shock radii, respectively. \citet{Piro2016} provided an analytic treatment of $\mathrm{DM_{SNR}}$ and its temporal evolution in this free-expansion limit. Here we adopt the thin-shell approximation described by \citet{Yang&Zhang2017}, which provides an explicit time scaling, given by
\begin{equation}
\mathrm{DM}_{\rm SNR}(t) \simeq 2.6\times10^4\,
\chi_e \left(\frac{\mu_m}{1.2}\right)^{-1} M_1^2\,E_{51}^{-1}\,
t_{\rm yr}^{-2}
\,{\rm pc\ cm^{-3}}. \label{eq:dm_full}
\end{equation}
where $\chi_e$ is the average ionization fraction of the medium in the SNR, $\mu_m$ is the mean molecular weight (for reference, $\mu_m\simeq 1.2$ corresponds to a solar-composition, approximately neutral gas), $M_1 \equiv M_{\rm ej}/(10\,M_\odot)$, $E_{51}\equiv E/(10^{51}\,{\rm erg})$ and $t_{\rm yr}\equiv t/(1\,{\rm yr})$.

\paragraph{(2) Shocked ionized ejecta.}
In this scenario, only the shocked ejecta layer between the reverse shock and the contact discontinuity is ionized, i.e., $\Delta R = R_c - R_r$. Assuming a uniform ambient ISM with constant number density $n_0$, this prescription leads to a more gradual temporal decay in the DM contribution from the ionized region \citep{Piro&Gaensler2018}  :
\begin{equation}
\mathrm{DM}_{\rm SNR}(t) = 52.6\left(\frac{\mu_m}{\mu_e}\right)\,
E_{51}^{-1/4}\,M_1^{3/4}\,n_0^{1/2}\,t_{\rm yr}^{-1/2}
\ {\rm pc\ cm^{-3}}, \label{eq:dm_shock}
\end{equation}
where $\mu_e$ is the mean molecular weight per electron, respectively.

\paragraph{(3) Wind case: shocked-ejecta contribution in a circumstellar wind.}
If a proto-stellar wind is present, the wind density profile follows $\rho_w = K r^{-2}$, where $K = \dot{M}/(4\pi v_w)$, $\dot{M}$ is the mass loss rate and $v_w$ is the wind velocity. In this case, the dominant DM contribution is also the ionized ejecta layer from the reverse shock. Compared to the constant-density case, the resulting DM declines more steeply with time \citep{Piro&Gaensler2018}:
\begin{equation}
\mathrm{DM}_{\rm SNR}(t) \simeq 1.3 \times 10^{4}\,
\mu_e^{-1}\,E_{51}^{-3/4}\,M_{1}^{5/4}\,K_{13}^{1/2}\,t_{\rm yr}^{-3/2}
\ \mathrm{pc\ cm^{-3}},
\label{eq:dm_wind}
\end{equation}
where $K_{13} \equiv K/(10^{13}\,\mathrm{g\,cm^{-1}})$. The contribution from the shocked circumstellar medium, $\mathrm{DM}_{\rm sh,CSM}$, arises from the forward-shock--heated wind material and exhibits a shallower temporal decay. In this case, $\mathrm{DM}_{\rm sh,CSM}$ becomes important only at sufficiently late times, while it remains subdominant over the time range of interest considered in this work.

\paragraph{(4) Wind case: shocked ejecta and shocked CSM with self-similar evolution.}
In this Scenario, the SNR DM receives contributions from both the reverse-shock--heated ejecta and the forward-shock--heated CSM.
Following the self-similar solution of ejecta--wind interaction
\citep{ZYZhao2021a,ZYZhao2021b,Tang&Chevalier2017}, with the relevant
characteristic scales and self-similar constants summarized in
Appendix \ref{app:wind_self_similar}, we write
\begin{equation}
\mathrm{DM}_{\rm SNR}(t)
=
\mathrm{DM}_{\rm sh,ej}(t)
+
\mathrm{DM}_{\rm sh,CSM}(t)
+
\mathrm{DM}_{\rm un,ej}(t)
\,,
\label{eq:DM_SNR_wind_decomposition}
\end{equation}
where $\mathrm{DM}_{\rm sh,ej}$ and $\mathrm{DM}_{\rm sh,CSM}$ denote the contributions from the shocked ejecta and shocked CSM, respectively, and $\mathrm{DM}_{\rm un,ej}$ is the DM contribution from the unshocked ejecta (cf. \citealt{ZYZhao2021a}).

The shocked-ejecta contribution is
\begin{equation}
\mathrm{DM}_{\rm sh,ej}(t)=
\frac{2K (n-3)(n-4)}{\mu_a m_p}
\left(\frac{1}{r_2}-1\right)
\frac{1}{\zeta_c R_{\rm ch}}
\left(\frac{t_{\rm ch}}{t}\right)^{\frac{n-3}{\,n-2\,}}
\,,
\label{eq:DM_Zhao2021_wind_shocked_ejecta}
\end{equation}
where $r_2 \equiv R_{\rm r}/R_{\rm c}$, $\mu_a$ is the mean atomic weight, $m_p$ is the proton mass, $R_{\rm ch}$ is the SNR characteristic radius (i.e., the characteristic length scale in the SSDW normalization), and $t_{\rm ch}$ is the corresponding characteristic timescale, as defined by Eqs.~\eqref{eq:Rch} and \eqref{eq:tch}, while the dimensionless constant $\zeta_c$ follows Eq.~\eqref{eq:zeta_c_def}.

The shocked-CSM (i.e., forward-shocked wind) contribution is
\begin{equation}
\mathrm{DM}_{\rm sh,CSM}(t)=
\frac{4K}{\mu_a m_p}
\left(1-\frac{1}{r_1}\right)
\frac{1}{\zeta_c R_{\rm ch}}
\left(\frac{t_{\rm ch}}{t}\right)^{\frac{n-3}{\,n-2\,}}
\,,
\label{eq:DM_Zhao2021_wind_shocked_CSM}
\end{equation}
where $r_1 \equiv R_1/R_{\rm c}$ is the ratio of the forward-shock radius to the contact-discontinuity radius for the wind case ($s=2$), and the numerical values of $r_1$ (and $r_2$) for a given $(n,s)$ can be taken from \citet{Chevalier1982}.

The unshocked-ejecta contribution can be decomposed into the DM of the unshocked core and the DM of the unshocked outer power-law envelope,
\begin{equation}
\mathrm{DM}_{\rm un,ej}(t)=\mathrm{DM}_{\rm core}(t)+\mathrm{DM}_{\rm pl}(t)
\,.
\label{eq:DM_Zhao2021_wind_unshocked_decomp}
\end{equation}
For the broken power-law ejecta profile in Eq.~\eqref{eq:f_w}, we may write
\begin{equation}
\begin{split}
\mathrm{DM}_{\rm un,ej}(t)
&=
\int_{0}^{R_{\rm core}}
\frac{\chi_e\,M_{\rm ej}}{\mu_a m_p\,R_{\rm ej}^{3}}\,f_0\,\ud r\\
&\quad+
\int_{R_{\rm core}}^{R_{\rm ej}}
\frac{\chi_e\,M_{\rm ej}}{\mu_a m_p\,R_{\rm ej}^{3}}\,f_0\left(\frac{r}{R_{\rm core}}\right)^{-n}\ud r\,.
\end{split}
\label{eq:DM_Zhao2021_wind_unshocked_ejecta}
\end{equation}
where $R_{\rm core}=w_{\rm core}R_{\rm ej}$ and we take $\chi_e$ to be the average electron ionization fraction of the unshocked ejecta. 
In the free-expansion (FE) solution, $R_{\rm ej}=R_{\rm c}$, and the unshocked-core contribution becomes
\begin{equation}
\mathrm{DM}_{\rm core}(t)
=
\frac{M_{\rm ej}}{\mu_a m_p}
\chi_e f_0 w_{\rm core}\,\lambda_c^{-2}\,v_{\rm ch}^{-2}\,t^{-2}
\,.
\label{eq:DM_Zhao2021_wind_unshocked_core}
\end{equation}
The reverse shock radius in the FE solution can be written as $R_{\rm r}=q_r R_{\rm c}$, where $q_r\equiv q_b/\ell_{\rm ED}$.
For an ``apples-to-apples'' comparison with the semi-analytic model of \citet{ZYZhao2021a}, hereafter Zhao+21, we evaluate the self-similar constants using their fiducial parameters $(n,s)=(10,2)$, even though our numerical simulations adopt $n=11$ (see Section~\ref{subsec:SNtoSNRtransiton}).

Although an additional DM contribution from the unshocked CSM is present, it is subdominant in this regime.
We therefore use Eqs.~\eqref{eq:DM_Zhao2021_wind_shocked_ejecta}--\eqref{eq:DM_Zhao2021_wind_unshocked_core} to provide a more complete analytic estimate for the wind scenario.

\section{Numerical Methods}
\label{sec:methods}

\subsection{Progenitor models: SS and BS channels}
\label{subsec:progenitors}

We adopt progenitor models computed by \citet{Kawashima2026arXiv}, which are primarily based on the detailed stellar evolution calculations of \citet{Farmer+2023} using MESA from the ZAMS to core collapse. These models self-consistently follow the mass-loss history, pre-supernova structure, and explosion properties of massive stars, and provide physically motivated initial conditions for the subsequent SNR evolution. 

We consider two main evolutionary channels for producing CCSNe that host young FRB central engines. For both channels, we adopt progenitors with initial masses $M_{\rm ZAMS}=11$ and $30\,M_\odot$, which serve as representative low- and high-mass cases:

\begin{enumerate}
    \item \textit{Single-star (SS) channel:}  
     Massive stars that evolve in isolation, losing part of their envelopes via stellar winds and exploding as hydrogen-rich Type II SNe. Their wind velocity, mass-loss history, and stellar radius vary significantly over the stellar lifetime and are taken directly from the \texttt{MESA} outputs, in contrast to the steady-wind approximations adopted in some previous studies (e.g.,\citealt{Jacovich+2021}). The final pre-SN structure and ejecta mass $M_{\rm ej}$ therefore reflect the integrated history of wind-driven mass loss.

    \item \textit{Binary-star (BS) channel:}
    Massive binary systems in which the primary star loses most of its hydrogen-rich envelope via Roche-lobe overflow (RLOF) and winds prior to core collapse, becoming a stripped helium star prior to core collapse (Type Ib/c). The binary mass ratio is set to $M_2/M_1 = 0.8$ followed by \citet{Farmer+2023} and \citet{Kawashima2026arXiv}. Binary evolution is treated in a simplified yet physically motivated manner: the companion is modeled as a point mass until the end of core helium burning, after which the primary continues its evolution without further interaction. Mass transfer during RLOF is parameterized by a mass-loss condition \citep[e.g.,][] {Paczyski1967, Heuvel1969} and because the exact fraction of mass retained by the companion is uncertain, we adopt the assumption that half of the transferred mass is accreted while the remaining half is ejected from the system and contributes to the CSM. This treatment captures the key effect of binary interaction---a substantially reduced $M_{\rm ej}$ a more compact progenitor at collapse---without introducing additional free parameters.
\end{enumerate}
Note that we include both single and binary progenitors because a substantial fraction of massive stars reside in interacting binaries \citep{Sana2012Sci}, and for each $M_{\rm ZAMS}$, the SS and BS channels can yield markedly different ejecta masses and composition structures, which directly affect the density normalization of the SNR and the resulting electron density ($n_e$) profile relevant for ${\rm DM}_{\rm source}$. Table~\ref{tab:models} summarizes the main properties of the progenitor models used in the present paper. 
\begin{table*}
    \centering
    \caption{Summary of SS and BS progenitor models.}
    \label{tab:models}
    \begin{threeparttable}
    \begin{tabular}{lccccccccc}
        \toprule
        Channel & $M_{\rm ZAMS}$ & $\Delta M_{\rm wind}$ & $\Delta M_{\rm RLOF}$ & $M_{\rm ej}$ & $E_{\rm SN}$ & $M_\mathrm{rem}$ & SN type (expected) & Fate\\
        & ($M_\odot$) & ($M_\odot$) & ($M_\odot$) & ($M_\odot$) & ($10^{51}$ erg) &($M_\odot$)& & \\
        \midrule
        Single & 11 & 1.64 & --   & 7.83 & 1.0 &1.53 & IIP-like & NS\\
        Binary & 11 & 0.70 & 7.12 & 1.81 & 1.0 &1.62 & Ib/c-like & NS\\
        Single & 30 & 15.74 & --   & 12.27 & 1.0 &1.99 & IIb-like & NS\\
        Binary & 30 & 7.50 & 10.94   & 9.85 & 1.0 &1.71 & Ib/c-like & NS\\
        \bottomrule
    \end{tabular}
    \begin{tablenotes}
    \item SN-type labels are indicative and based on pre-SN envelope composition (especially residual H/He), not on radiative-transfer light-curve/spectral modeling.
    \end{tablenotes}
    \end{threeparttable}
\end{table*}
\subsection{SNe to SNR transition}
\label{subsec:SNtoSNRtransiton}
Modeling the transition from the SN explosion to the SNR phase requires specifying the initial ejecta structure at the onset of the remnant evolution. Following common practice (e.g., \citealt{McKee&Truelove1999}), we do not attempt to model the explosion mechanism itself. Instead, we assume that the ejecta have already reached a homologous expansion (FE) phase. The ejecta density structure is characterized by a prescribed density profile defined, consisting of a flat inner core and a steep outer envelope, defined by Eq.~\eqref{eq:f_w} and detailed in Appendix~\ref{APP:Contact_constant}. The outer ejecta follow a power-law profile $\rho_{\rm ej} \propto r^{-n}$, and we adopt $n=11$ for both evolutionary channels. This choice is motivated by the compact pre-supernova progenitors considered in this work, particularly for the binary-stripped channel.

The SNR simulations are initialized at $t_{\rm init}=3~\mathrm{yr}$ after explosion, by which time the ejecta are well described by homologous expansion. The total ejecta mass $M_{\rm ej}$ and explosion energy $E_{\rm SN}$ are set by the progenitor models and are listed in Table~\ref{tab:models}.
\subsection{CSM}
\label{subsec:csm}
The CSM is constructed by explicitly following the interaction between the progenitor outflows and the surrounding ISM, rather than prescribing an analytic density profile (e.g., $\rho \propto r^{-2}$ for a wind-shaped medium), as commonly adopted in both analytical models and some numerical treatments (see Section~\ref{sec:analytic_models}). As described in \citet{Kawashima2026arXiv}, time-variable mass-loss rates and wind velocities during different evolutionary stages can generate complex CSM structures, including piled-up shells, which strongly influence the subsequent SNR evolution.

The CSM structure is computed using 1D HD simulations of stellar winds expanding into a uniform ISM, employing the 1D hydrodynamics code \texttt{VH-1} \citep{Blondin2001,Blondin&Ellison2001}. Radiative cooling is included, allowing dense, geometrically thin shells to form as wind material accumulates. The adopted ISM number density is $n_{\rm ISM}=1.0\,\mathrm{cm^{-3}}$, and both the wind and ISM temperatures are set to $T=10^4\,\mathrm{K}$. {This choice is intended to represent a warm photoionized ambient medium and to avoid introducing an artificially cold initial pressure floor.} The resulting CSM profiles provide the ambient density structure into which the SN ejecta expand in our subsequent SNR simulations. Further details of the CSM construction and parameter choices are described in \citet{Kawashima2026arXiv}.
\subsection{SNR simulation settings}
\label{subsec:hydro}
The time-dependent evolution of the SNR is computed using a 1D, spherically symmetric HD+NEI code.
The numerical framework is conceptually based on the \textcolor{gray}{ChN} (CR-hydro-NEI) code \citep{Ellison2007, Patnaude2010, Lee+2012,Lee+2013,Lee+2014, Lee+2015, Jacovich+2021, Court2024}, and is closely related to the implementations employed by \citet{Kawashima2026arXiv}.
The code follows the NEI of 30 elements and accounts for radioactive decay by converting 162 unstable isotopes from the MESA yields to their stable counterparts using the Python library \texttt{radioactivedecay} \citep{Amaku2010}, based on the ICRP decay tables \citep[][]{ICRP2008NuclearCalculations}.

The code follows the coupled evolution of the forward and reverse shocks as they propagate through the CSM and SN ejecta, and self-consistently tracks the energy partition, electron--ion temperature equilibration, and the time-dependent ionization states of multiple elements in each radial zone.
In this work, we disable the cosmic-ray (CR) evolution/acceleration module; consequently, there is no self-consistent magnetic-field ($B$) evolution.
This simplified setup is nevertheless sufficient for our purpose, since we focus on the thermal plasma structure--- $n_e$, $T$, and ionization fraction $\chi_e\equiv n_e/(\rho/m_p)$---that determines the FRB DM, rather than on detailed non-thermal emission or magnetic-field diagnostics.

The simulations output 1D radial profiles of the hydrodynamical and microphysical quantities in each zone, including the mass and radius coordinates, density, pressure, velocity, electron and proton temperatures, electron and ion number densities, and element-by-element ion fractions. These profiles are used in post-processing to compute the FRB DM and optical depths as functions of time (See section \ref{subsec:dm_tau}).

To adequately resolve the early-time evolution of the SNR contribution while maintaining computational efficiency, we adopt a non-uniform output cadence. Motivated by the fact that the most significant and rapid evolution of the SNR contribution to the FRB DM occurs at early times (e.g., \citealt{Piro&Gaensler2018}) we use a finer time spacing during the first $\sim300$~yr after explosion, followed by a coarser spacing at later times. Specifically, we start from the simulation initialization time $t_{\rm init}=3$~yr, snapshots are recorded at $\Delta t=1$~yr intervals from $t=3$ to $300$~yr, and at $\Delta t=4$~yr intervals from $300$ to $500$~yr, resulting in a total of $\sim350$ output time bins. This sampling captures the relevant DM evolution and its temporal variability while keeping the overall computational cost tractable.
\subsection{DM and optical depth calculations}
\label{subsec:dm_tau}
Given the radial profiles of electron density $n_e(r,t)$ produced by the numerical simulations, we compute the DM contributed by the SNR as
\begin{equation}
    {\rm DM}_{\rm SNR}(t)
    =
    \int_{R_{\rm min}(t)}^{R_{\rm max}(t)} n_e(r,t)\,{\rm d}\ell,
\end{equation}
where $R_{\rm min}$ and $R_{\rm max}$ denote, respectively, the inner and outer boundaries of the ionized region in the simulation. For the shocked region, we take $R_{\rm min}=R_r(t)$ and $R_{\rm max}=R_b(t)$, i.e., the reverse- and forward-shock radii. For the entire ionized region, we take $R_{\rm min}=R_{\rm un, min}(t)$ (the inner boundary of the unshocked ejecta) and $R_{\rm max}=R_{\rm CSM, max}$ (the fixed outer boundary of the unshocked CSM). In our setup, $R_{\rm CSM, max}=39.9\,${pc} is set by the initial conditions, {which is within a reasonable range for the surrounding ISM and SNR, before going beyond into the host galaxy regime rather than the local source}. Here ${\rm d}\ell$ is the line element along the FRB line of sight.
For now we assume a magnetar as the FRB's central engine located at the origin\footnote{In this work, we neglect the effect of magnetar natal kick, as our analysis is restricted to very early SNR evolution ($\lesssim500\, \text{yr}$), during which the kick-induced displacement remains much smaller than the radial extent of ionized shock region} of the SNR, such that the DM is independent of the line-of-sight direction and can be written as a radial integral of the $n_e$.

Given the extremely high energy budget of FRBs, viable progenitors are generally expected to possess ultra-strong magnetic fields and highly active crustal or magnetospheric environments, pointing to a young magnetar origin. As a result, FRB radio emission can escape only once the SNR has expanded to become sufficiently optically transparent, while remaining dynamically young. Therefore, by modeling the time evolution of the optical depth of the SNR, one can place constraints on the FRB escape timescale ($t_{\rm esc}$).

We consider two sources of opacity in the SNR: electron scattering and free–free absorption, such that the total optical depth is given by
\begin{equation}
\tau_{\rm tot} = \tau_{\rm es} + \tau_{\rm ff}. \label{eq:tau_tot}
\end{equation}
At early times, when the SNR is still young and the ionized region remains dense, electron scattering can contribute significantly to the optical depth. In this regime, FRB propagation is primarily affected by scattering-induced temporal broadening rather than true absorption. We approximate the electron scattering optical depth using Thomson scattering,
\begin{equation}
\tau_{\rm es} = \int_{R_{\rm min}(t)}^{R_{\rm max}(t)} n_e \sigma_T \, dl, \label{eq:tau_es}
\end{equation}
where $\sigma_T$ is the Thomson cross section.

As the SNR expands and the electron density decreases, the contribution from electron scattering rapidly diminishes. At sufficiently late times, free–free absorption becomes the dominant opacity source and ultimately determines whether FRB radio emission can escape. We therefore compute the free–free optical depth at an observing frequency $\nu$ as
\begin{equation}
    \tau_{\rm ff}(\nu,t)
    =
    \int_{R_{\rm min}(t)}^{R_{\rm max}(t)}
    \alpha_{\rm ff}(\nu, T_e, n_e, n_{ij}, Z_{ij})\,{\rm d}\ell,
\end{equation}
where $\alpha_{\rm ff}$ is the thermal free–free absorption coefficient.
For radio frequencies satisfying $h\nu \ll k_{\rm B}T_e$, we adopt the 
standard expression from \citet{RybickiLightman1979},
\begin{equation}
    \alpha_{\rm ff}(\nu) =
    1.9\times10^{-2}\,
    T_e^{-3/2}\,\nu^{-2}\,
    n_e\,\sum_{i,j} Z_{ij}^{2}\, n_{ij}\,
    g_{\rm ff}(\nu, T_e, Z_{ij}),
    \label{eq:alpha_ff}
\end{equation}

where $Z_{ij}$ denotes the ionic charge of element $i$ in its $j$-th 
ionization state, and $n_{ij}$ is the corresponding number density.  where $Z_{ij}$ is the ion charge, $n_i$ is the ion number density, and 
$g_{\rm ff}$ is the Gaunt factor, whose detailed behavior is described in Appendix~\ref{App:Gaunt}. Here, in practice, we evaluate the Gaunt factor using the mean ionic charge $\langle Z\rangle = \sum_{ij} Z_{ij}n_{ij}/\sum_{ij}n_{ij}$ at each radius without loss of accuracy. Equation~(\ref{eq:alpha_ff}) can then be rewritten as
\begin{equation}
    \alpha_{\rm ff}(\nu) =
    1.9\times10^{-2}\,
    T_e^{-3/2}\,\nu^{-2}\,
    n_e\,g_{\rm ff}(\nu, T_e, \langle Z \rangle)\, \sum_{i,j} Z_{ij}^{2}\, n_{ij},
    \label{eq:alpha_ff_approx}
\end{equation}

The FRB emission is able to escape once $\tau_{\rm tot}\lesssim 1$ at the observing frequency.
Therefore, the combination of ${\rm DM}_{\rm SNR}(t)$ and $\tau_{\rm tot}(t)$ provides both the instantaneous DM contribution and an estimate of the ``$t_{\rm esc}$'' for GHz emission (throughout this work, we adopt $\nu=1\,{\rm GHz}$ as a representative frequency for the GHz observing band for FRBs).

Finally, we compute the first and second time derivatives of the SNR DM numerically:
\begin{equation}
\begin{aligned}
\frac{{\rm dDM}}{{\rm d}t}(t_i)
&\approx \frac{{\rm DM}_{i+1}-{\rm DM}_{i-1}}{t_{i+1}-t_{i-1}},\\[4pt]
\frac{{\rm d}^2{\rm DM}}{{\rm d}t^2}(t_i)
&\approx \frac{{\rm DM}_{i+1}-2{\rm DM}_i+{\rm DM}_{i-1}}{(t_{i+1}-t_i)^2},
\end{aligned}
\label{eq:dm_derivatives}
\end{equation}

which will be compared with {observationally inferred} $\mathrm{dDM}/\mathrm{d}t$ and constraints on higher-order variations.

\subsection{RM Calculation and Observational Matching Method}
\label{subsec:rm_method}
The rotation measure (RM) is defined as
\begin{equation}
{\rm RM}=\frac{e^3}{2\pi m_e^2 c^4}\int n_e\,B_\parallel\,dl,
\end{equation}
which is commonly approximated as
\begin{equation}
{\rm RM}\simeq 0.81\int n_e\,B_\parallel\,dl\quad ({\rm rad\,m^{-2}}),
\end{equation}
{where $n_e$ is in cm$^{-3}$, $B_\parallel$ is the magnetic-field component parallel to the line of sight in $\mu$G, and dl is in pc. We note that we assume an effective line-of-sight magnetic field and do not explicitly model small-scale field tangling.}

In this work, we compute the RM evolution using only shocked-region profiles. The magnetic field is parameterized as a fixed fraction of the ram pressure,
\begin{equation}
\frac{B^2}{8\pi} = \epsilon_B \frac{\rho v^2}{2},
\end{equation}
so that $B=\sqrt{4\pi\,\epsilon_B\,\rho v^2}$, evaluated cell by cell in the shocked layer. We also track the shocked-region mean magnetic field as an electron-density-weighted, cell-by-cell average,
\begin{equation}
\langle B_{\rm sh}\rangle = \frac{\sum_i n_{e,i} B_i}{\sum_i n_{e,i}},
\end{equation}
where $i$ labels the $i$th cell in the shocked region. The RM is then obtained by integrating $n_e B_\parallel$ along the line of sight over $r_r<r<r_b$. We exclude the unshocked ejecta from the RM integral because, in this prescription, it is approximately in free expansion and therefore does not carry a ram-pressure-supported magnetic field. For ionization, we directly use the NEI-simulated ionization state in the shocked cells. We explore $\epsilon_B=0.01$--$0.3$, with a fiducial value of $\epsilon_B=0.1$, consistent with \citet{Piro&Gaensler2018}. The adopted range $\epsilon_B = 0.01\text{--}0.3$ spans values commonly inferred in young supernova remnants and shock-powered synchrotron sources, where magnetic amplification by turbulence and cosmic-ray streaming can bring the post-shock magnetic energy density to a few percent up to nearly equipartition with the ram pressure.

To identify the simulated RM interval that matches the observed evolution of FRB~20121102, we adopt a slope-matching approach based on the logarithmic RM decay index. We define the logarithmic slope
\begin{equation}
\beta(t) \equiv \frac{d \log {\rm RM}(t)}{d \log t}.
\end{equation}
For the simulated RM evolution, the instantaneous slope is computed numerically in log--log space using finite differences,
\begin{equation}
\beta_{\rm sim}(t_i) \approx
\frac{\log {\rm RM}_{i+1} - \log {\rm RM}_{i-1}}
{\log t_{i+1} - \log t_{i-1}},
\end{equation}
and interpolation is applied to obtain a smooth representation of $\beta_{\rm sim}(t)$. Using the RM measurements reported in \citet{Hilmarsson2021} and \citet{Wang2025arXiv250715790W}, we fit the post-peak observational data in log--log space and derive an observed decay slope of
\begin{equation}
\beta_{\rm obs} \approx -0.36.
\end{equation}
We then search for the epoch $t_0$ in each single-star (SS) model that satisfies
\begin{equation}
\beta_{\rm sim}(t_0) = \beta_{\rm obs},
\end{equation}
which yields
\begin{equation}
t_0 \approx 9.54~{\rm yr}
\end{equation}
for the $11\,M_\odot$ model and
\begin{equation}
t_0 \approx 9.99~{\rm yr}
\end{equation}
for the $30\,M_\odot$ model. To compare the temporal evolution directly, we shift the observational time axis such that the median observed RM is placed at $t_0$. The mapped time is
\begin{equation}
t_{\rm mapped} = t_0 + (t_{\rm obs} - t_{\rm ref}),
\end{equation}
where $t_{\rm ref}$ corresponds to the epoch of the median observed RM, while the observed RM amplitudes are kept unchanged.

\section{Results}
\label{sec:results}
Motivated by the analytical considerations in Section \ref{sec:analytic_models}---which suggest that ${\rm DM}_{\rm SNR}$ may be dominated either by the shocked ejecta/shell or by the entire ionized region (including unshocked material)---we carry out the analysis for both cases.
\subsection{Shocked region}
\label{subsec:shocked}
\subsubsection{DM evolution in the shocked region}
\label{subsubsec:DM}
\begin{figure}[htbp]
  \centering
  \includegraphics[width=1\linewidth]{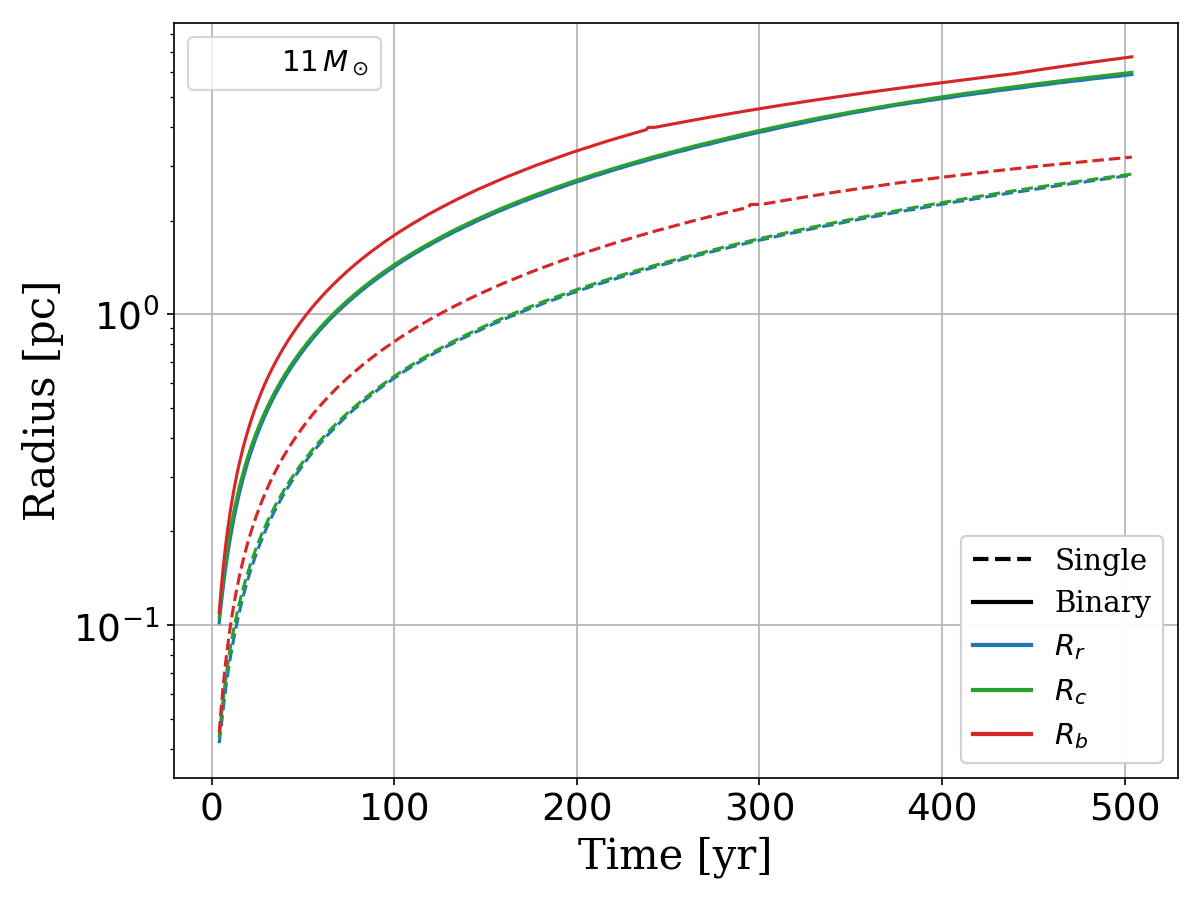}
  \includegraphics[width=1\linewidth]{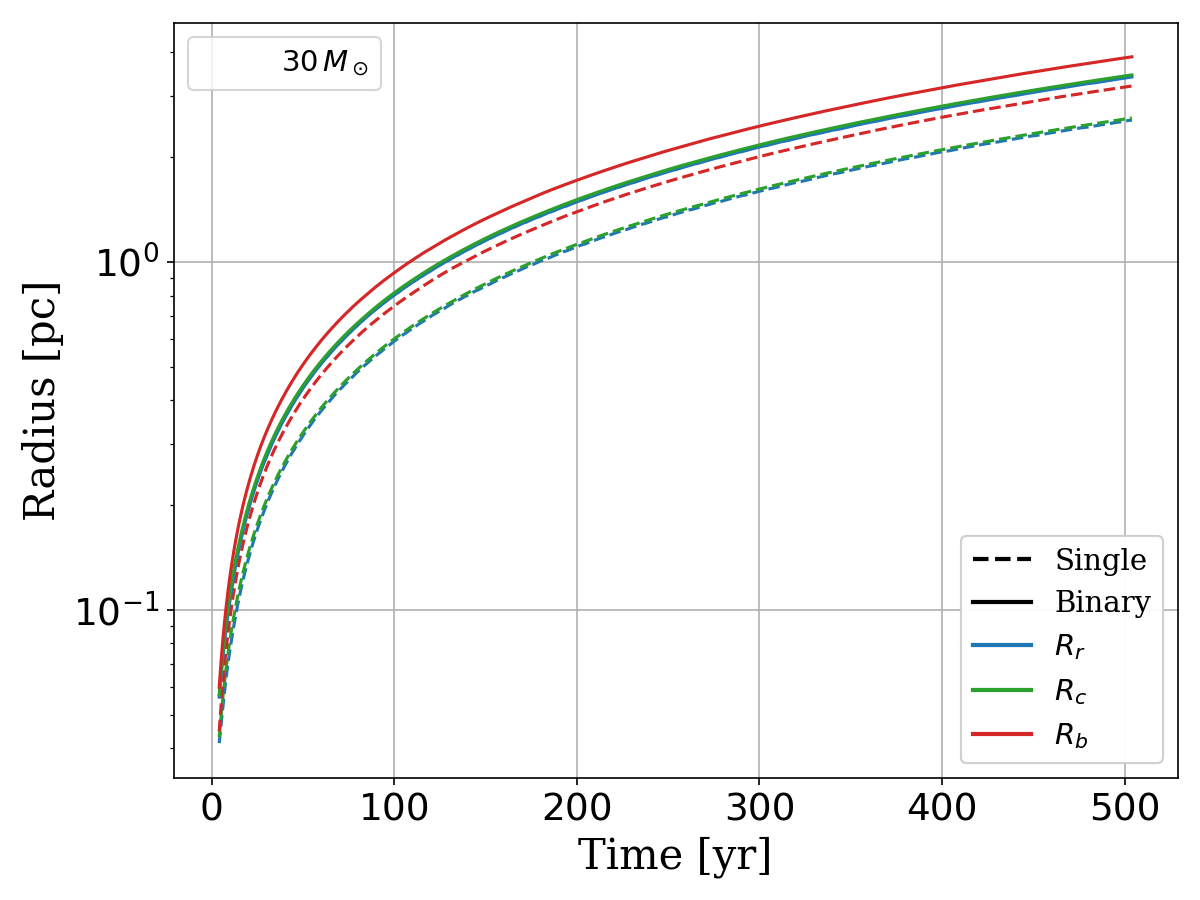}
  \caption{Radius evolution for the $11\,M_\odot$ (top) and $30\,M_\odot$ (bottom) models. Dashed: single-star; solid: binary-stripped. Blue, green, and red curves indicate the $R_r$, $R_c$, and $R_b$, respectively.}
  \label{fig:radius_comparison}
\end{figure}

\begin{figure}
    \centering
    \includegraphics[width=1\linewidth]{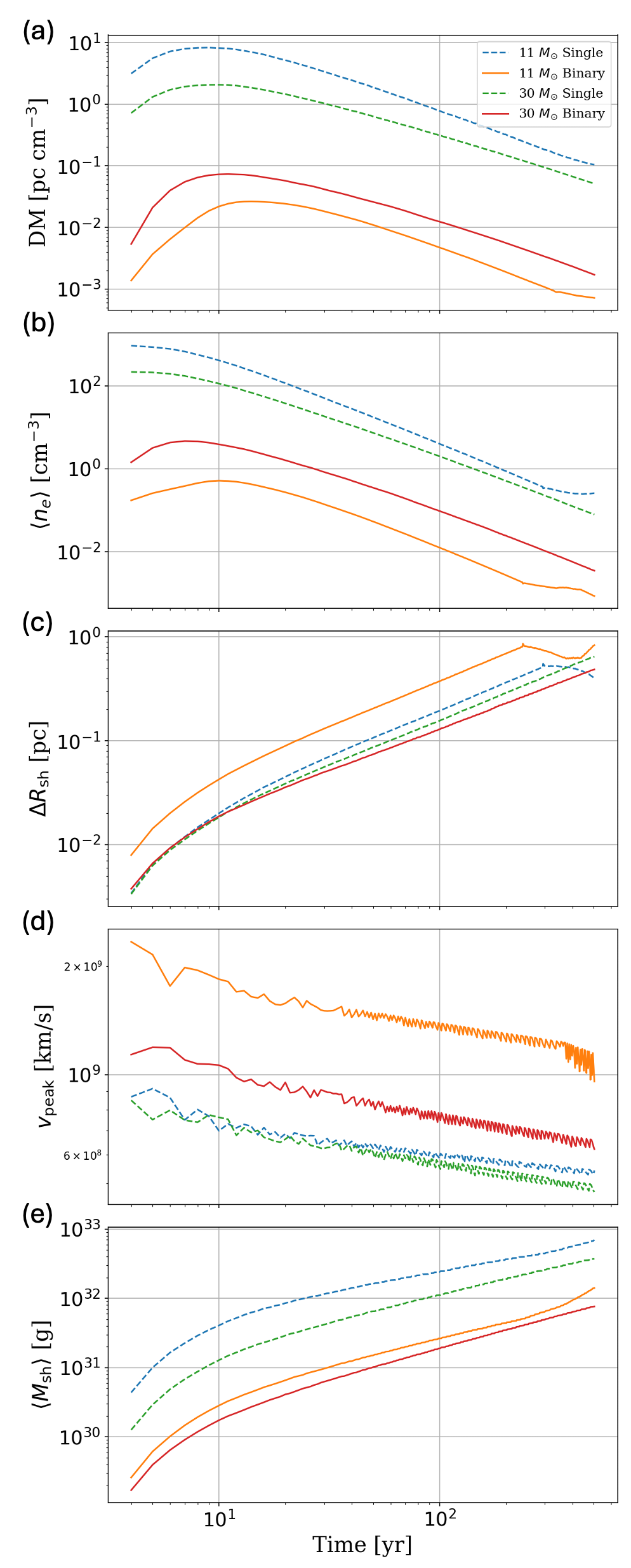}
    \caption{
        Comparison of the time evolution for $11\,M_\odot$ and $30\,M_\odot$ progenitor models in the SS and BS channels.
        (a) DM.
        (b) Volume-averaged electron density $\langle n_e \rangle$.
        (c) Effective thickness of the shocked ionized region, $\Delta R_{\rm sh}$.
        (d) Peak (maximum) SNR expansion velocity.
        (e) Cumulative swept-up (shocked) shell mass.
        Solid curves correspond to the fiducial unshocked-ionization case $\chi_{e,\mathrm{unsh}}=1$ (HH in Table~\ref{tab:toccur_tesc}).
    }
    \label{fig:dm_mass_model_comparison}
\end{figure}

Fig.~\ref{fig:radius_comparison} shows the temporal evolution of the characteristic radii in our 1D SNR simulations, including the reverse shock ($R_r$), contact discontinuity ($R_c$), and forward shock ($R_b$), for both the SS and BS progenitors. These radii define the extent of the shocked layer and its effective thickness, $\Delta R_{\rm sh} \equiv R_b - R_r$ (see Fig.~\ref{fig:dm_mass_model_comparison}c), and therefore set the integration boundaries used in our DM and optical-depth calculations. The radii increase rapidly at early times and then evolve more smoothly as the remnant expands.

Fig.~\ref{fig:dm_mass_model_comparison} compares the DM evolutions of each case. Aside from the initial peak due to the NEI during the earliest expansion stage, both models exhibit a monotonic decline in ${\rm DM}_{\rm SNR}(t)$ as the ejecta expand and their density decreases. For both $M_{\rm ZAMS}=11$ and $30\,M_\odot$, the SS channel produces a larger DM than the BS channel, consistent with its larger ejecta mass (Table~\ref{tab:models}) and hence a denser ionized ejecta layer.

Within the SS channel, the $11\,M_\odot$ model produces a larger DM than the $30\,M_\odot$ model because ${\rm DM}_{\rm SNR}\propto \int n_e\,\ud \ell$ depends on both the electron density ($n_e\propto \chi_e\,\rho$) and the effective path length through the shocked/ionized layer (characterized here by $\Delta R_{\rm sh}$). However, in the thin-shell approximation, these two factors are not independent. Using $\langle \rho\rangle_{\rm sh}\sim M_{\rm sh}/V_{\rm sh}$ with
$V_{\rm sh}\simeq (4\pi/3)(R_b^3-R_r^3)\approx 4\pi R^2\Delta R_{\rm sh}$,
we obtain $\langle n_e\rangle\propto \chi_e\,M_{\rm sh}/(4\pi R^2\Delta R_{\rm sh})$ and, therefore,
${\rm DM}_{\rm SNR}\sim \langle n_e\rangle\,\Delta R_{\rm sh}\propto \chi_e\,M_{\rm sh}/(4\pi R^2)$,
i.e., the explicit dependence on $\Delta R_{\rm sh}$ largely cancels.
Consequently, the DM difference between the $11\,M_\odot$ and $30\,M_\odot$ SS cases is primarily controlled by (i) the ionization degree $\chi_e$ and (ii) the ratio $M_{\rm sh}/R^2$.

The shocked volume is dominated by the geometric factor $R^2$ rather than by the shell thickness. At the same age the $11\,M_\odot$ SS remnant has a noticeably larger characteristic radius than the $30\,M_\odot$ case (see Fig.~\ref{fig:radius_comparison}); for a fixed shocked mass, this larger $R$ would reduce $\langle \rho\rangle_{\rm sh}$ and hence DM. In our models, however, the $11\,M_\odot$ remnant has a larger characteristic expansion speed (Fig.~\ref{fig:dm_mass_model_comparison}d), so at the same age it sweeps up more material and attains a higher shocked mass $M_{\rm sh}$ (Fig.~\ref{fig:dm_mass_model_comparison}e) and hence a higher $\rho_{\rm sh}$, despite the larger shocked volume.

In addition, the systematically larger $M_{\rm sh}$ in the SS channel than in the BS channel can be physically understood from binary stripping. In BS progenitors, Roche-lobe overflow removes much of the pre-SN H-rich envelope and suppresses dense pre-SN winds, yielding a more rarefied surrounding medium. The weaker external density profile leads to weaker SNR-shell deceleration and therefore a slower reverse-shock penetration into the ejecta, while the lower ambient density also reduces the shocked CSM component. Equivalently, writing $M_{\rm sh}=M_{\rm sh,ej}+M_{\rm sh,CSM}$, both terms are reduced in BS models, which naturally drives $M_{\rm sh,BS}<M_{\rm sh,SS}$ even when the total ejecta mass is not the sole controlling factor.

We also find that the $11\,M_\odot$ model reaches a higher ionization degree (e.g., $\chi_e\sim 0.9$) than the $30\,M_\odot$ model ($\chi_e\sim 0.7$; see Fig.~\ref{fig:dm_cum_single_binary}). This is consistent with NEI physics, since the approach to ionization equilibrium is governed by the ionization timescale parameter $n_e t$: the higher post-shock electron density in the $11\,M_\odot$ shell leads to a larger $n_e t$ at a given age, and thus a higher $\chi_e$.
Aside from the early-time NEI-driven rise, the long-term DM evolution resembles a power-law decay over hundreds of years, consistent with expectations from expanding SNR dynamics.

\subsubsection{Optical depth evolution in the shocked region}
\label{subsubsec:tau_shocked}
\begin{figure*}[!t]
    \centering
    \includegraphics[width=0.495\textwidth]{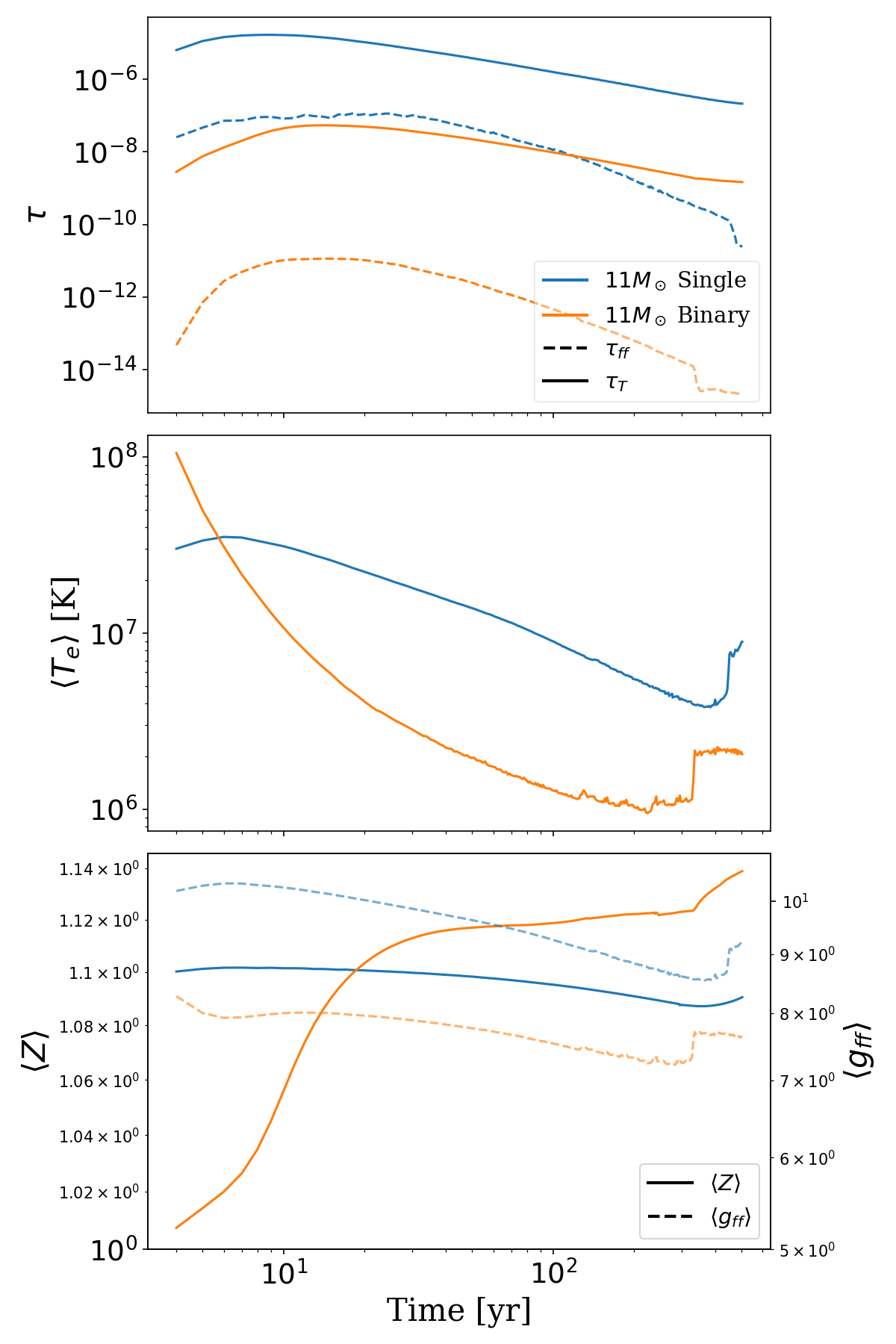}\hfill
    \includegraphics[width=0.495\textwidth]{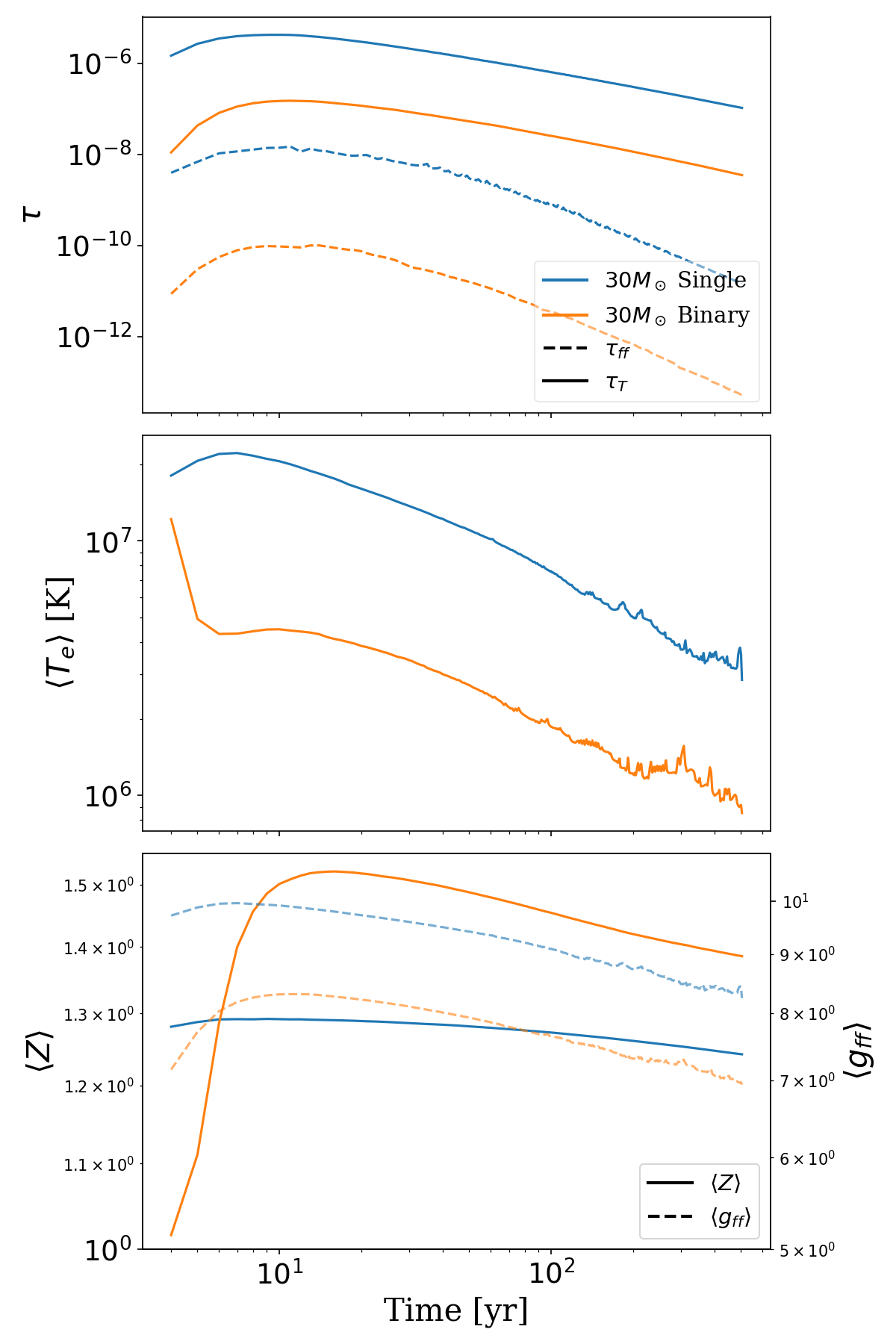}
    \caption{Time evolution of optical depths and mass-weighted averages in the shocked region for the $11\,M_\odot$ (left) and $30\,M_\odot$ (right) progenitor models. Each panel shows $\tau_{\rm es}$ and $\tau_{\rm ff}$, together with $\langle T_e\rangle$, $\langle Z\rangle$, and $\langle g_{\rm ff}\rangle$, comparing the SS and BS channels. Throughout this work, the Gaunt factor is evaluated at $\nu=10^{9}\,{\rm Hz}$, representative of the GHz band relevant for FRBs.}
    \label{fig:optical_depth_evolution}
\end{figure*}

Within the binary-stripped channel, on practical FRB timescales (tens to hundreds of years) the DM contribution from the SNR is negligible and exhibits essentially no measurable secular evolution. The $30\,M_\odot$ stripped model nevertheless yields a higher DM than the $11\,M_\odot$ stripped model (Fig.~\ref{fig:dm_mass_model_comparison}a). This difference is mainly driven by the much smaller ejecta mass in the $11\,M_\odot$ stripped case, which results in a particularly low post-shock electron density. The larger $\Delta R_{\rm sh}$ in that model further increases the shocked volume; under the thin-shell scaling, $V_{\rm sh}\propto R^2\Delta R_{\rm sh}$, this increase tends to dilute $n_e$.

Fig.~\ref{fig:optical_depth_evolution} shows the time evolution of the optical depths and several mass-weighted averages in the shocked region. The total optical depth remains $\tau_{\rm tot}\ll 1$ at GHz frequencies throughout the simulation, implying essentially no free--free ``escape time'' when considering only the shocked region, i.e., GHz FRB emission could escape almost immediately after the explosion. While the shocked region makes only a limited contribution to the total DM in our current setup, it provides a useful diagnostic of the plasma state: the mass-weighted electron temperature $\langle T_e\rangle$ declines with time in all cases, reflecting ongoing radiative cooling. We also find that $\langle Z\rangle$ is slightly above unity and differs modestly among the progenitor channels, while the Gaunt factor (evaluated at $\nu=10^{9}\,{\rm Hz}$) stays in the range $\langle g_{\rm ff}\rangle\sim 7$--$10$. We defer a more detailed discussion of the composition/ionization origin of these trends to Section~\ref{subsec:ionization}.

These results highlight the qualitative difference between the SS and BS channels: for the same $M_{\rm ZAMS}$, a stripped progenitor with a small ejecta mass produces both a much lower DM and a substantially more transparent shocked environment. Even at early times ($t\sim 10$~yr) when the shocked-region contribution reaches its maximum, the peak value remains modest: for the $11\,M_\odot$ SS model we find ${\rm DM}_{\rm peak}\simeq 8.3\,{\rm pc\,cm^{-3}}$, while the binary-stripped models exhibit significantly smaller peaks ($< 0.1\,{\rm pc\,cm^{-3}}$) that occur during the earliest evolution.
\subsection{Entire region}
\label{subsec:full}
\subsubsection{DM evolution in the entire ionized region}
\label{subsubsec:DM_full}
As demonstrated above, the shocked region alone cannot account for the large inferred ${\rm DM}_{\rm source}$ and its observed secular evolution in FRB~20190520B or FRB 20121102. This motivates us to extend the calculation to the entire ionized region of the SNR, incorporating contributions from the unshocked ejecta and the surrounding unshocked CSM/ISM.

Because our SNR simulations are primarily designed to resolve the detailed thermodynamic and ionization evolution in the shocked region, the unshocked ejecta and the surrounding unshocked CSM/ISM are not evolved with a physically self-consistent ionization history. In these regions, the simulation provides the density structure and elemental abundances, but does not self-consistently track the temperature evolution and ionization balance. We therefore treat the ionization fraction in the unshocked regions as a parameter and explore physically motivated bounds.

For the unshocked ejecta, \citet{Chevalier2017} estimated an ionization fraction of $\chi_{e,\mathrm{unej}} \sim 0.03$ for SN~1993J, in which the ionization is primarily driven by photoionization from radiation produced at the reverse shock. 

More recently, \citet{Laming2020} modelled the pre--reverse-shock (i.e., unshocked) inner ejecta of Cas~A under a photoionization--recombination (PR) equilibrium, with the photoionizing radiation field provided by UV-to-X-ray emission from both the forward- and reverse-shocked plasma. In their PR models, the inferred charge-state distributions that produce the observed IR fine-structure lines imply that the unshocked ejecta can maintain a non-negligible electron fraction, potentially at the level of $\chi_{e,\mathrm{unej}} \gtrsim 0.1$ depending on density/temperature conditions and the strength of the ionizing radiation field.

Furthermore, the ionization state of the unshocked ejecta in young core-collapse supernova remnants can depend sensitively on several factors, including the progenitor mass, mass-loss rate, wind velocity, and the resulting shock luminosity. In addition, the presence of a central compact object (CCO), such as a neutron star or magnetar, may provide additional high-energy radiation through spin-down-powered emission or a wind nebula (MWN/PWN), potentially enhancing the ionization level beyond that inferred from reverse-shock irradiation alone.

Given these uncertainties, and in the absence of a fully self-consistent radiative transfer calculation in our hydrodynamic models, we treat the ionization fraction of the unshocked ejecta parametrically and adopt a conservative yet physically motivated range:
\begin{equation}
0.01 \le \chi_{e,\mathrm{unej}} \le 1 ,
\end{equation}
where $\chi_{e,\mathrm{unej}} \sim 0.01$ is taken as the weak-ionization lower limit and $\chi_{e,\mathrm{unej}} \sim 1$ represents a fully ionized case, motivated by PR-equilibrium scenarios under strong photoionizing conditions 
(e.g., \citealt{Laming2020}) and/or potential additional ionizing contributions from a CCO.

For the unshocked CSM/ISM, we adopt
\begin{equation}
10^{-4} \le \chi_{e,\mathrm{ISM}} \le 1 ,
\end{equation}
spanning environments from cold neutral media to fully ionized hot plasma.
which spans plausible astrophysical environments. Specifically, $\chi_{e,\mathrm{ISM}} \sim 10^{-4}$ corresponds to a cold neutral medium (CNM), $\chi_{e,\mathrm{ISM}} \sim 10^{-2}$ to a warm neutral medium (WNM), $\chi_{e,\mathrm{ISM}} \sim 0.1$--$0.5$ to a partially ionized warm ionized medium (WIM), and $\chi_{e,\mathrm{ISM}} \sim 1$ to a fully ionized hot ionized medium (HIM). These bounds allow us to bracket the resulting DM contribution from the unshocked components and to quantify the systematic uncertainty associated with the poorly constrained ionization state in these regions. Table~\ref{tab:toccur_tesc} summarizes a subset of the ionization configurations explored in this work.
\begin{figure}[ht]
    \centering
    \includegraphics[width=0.495\textwidth]{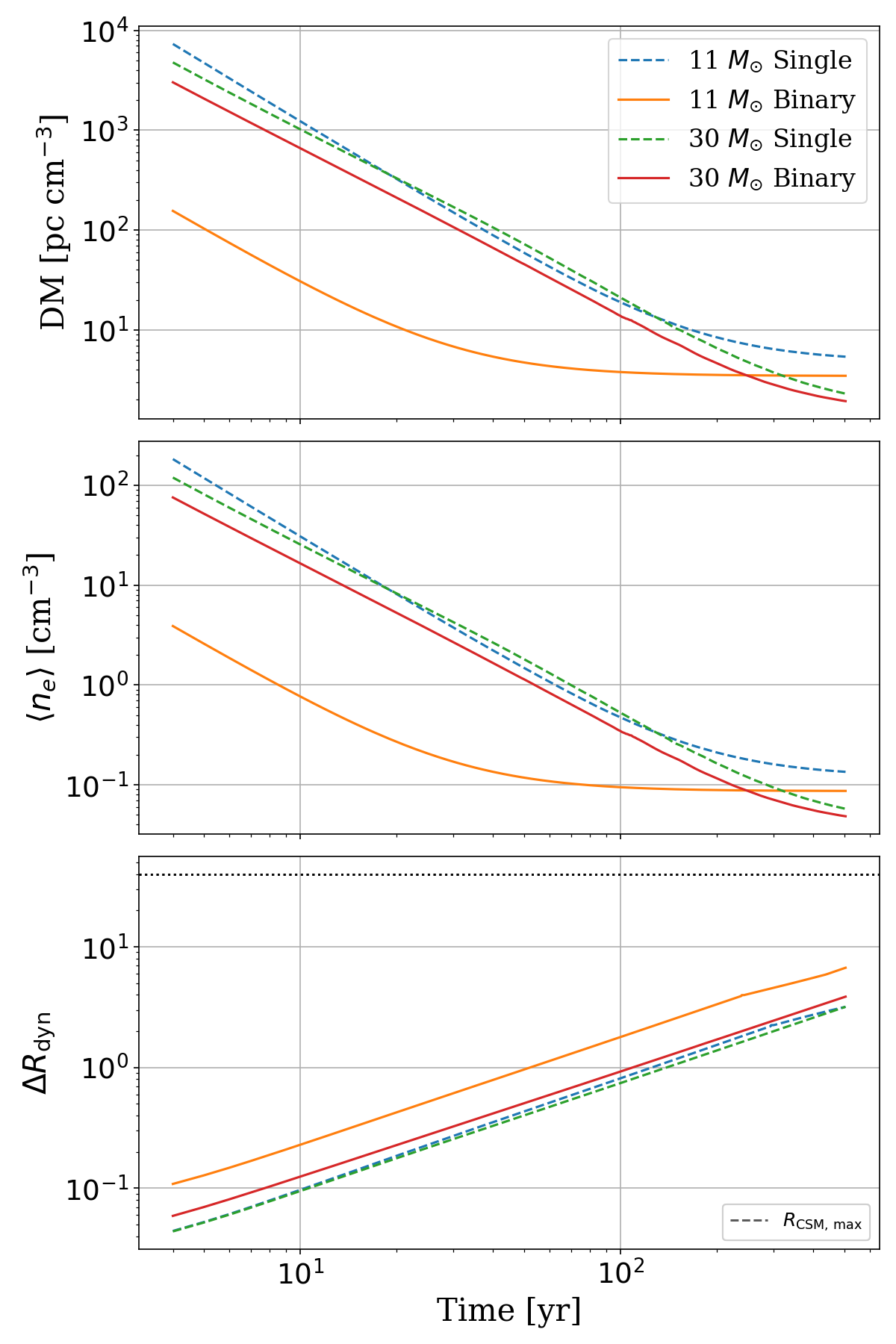}
\caption{Time evolution of the characteristic quantities for the full-region DM calculation, analogous to Fig.~\ref{fig:dm_mass_model_comparison}. The results correspond to the fiducial HH case with $\chi_{e,\mathrm{unej}}=\chi_{e,\mathrm{ISM}}=0.1$. Here we adopt a dynamical path length $\Delta R_{\rm dyn}=\Delta R_{\rm ej}+\Delta R_{\rm sh,CSM}$. In the third panel, $R_{\rm CSM,max}$ is a constant maximum integration radius for the ISM/CSM contribution.}
    \label{fig:dm_ne_deltar_full}
\end{figure}

Figure~\ref{fig:dm_ne_deltar_full} shows the time evolution of the characteristic quantities entering our full-region DM estimate. In this setup, the ionization fraction in the unshocked components is treated as a fixed parameter, and we use $\mu_a$ for the unshocked regions.
The fiducial case adopted in this work is HH, with $\chi_{e,\mathrm{unej}}=\chi_{e,\mathrm{ISM}}=0.1$ (hereafter $\chi_{e,\mathrm{unsh}}=0.1$, unless otherwise stated). We also adopt a low-ionization approximation for the unshocked ejecta in which each atom contributes at most one free electron (effectively singly ionized), so the composition dependence enters primarily through $\mu_a$ (i.e., $\mu_e\simeq\mu_a$).

During the ejecta-dominated phase, the SS models yield larger DMs than the BS models, while the relative ordering between the $11\,M_\odot$ and $30\,M_\odot$ SS cases changes with time. In the later CSM-dominated phase, the ordering is set by the CSM contribution and does not necessarily follow the same SS--BS ranking. At early times ($t\lesssim 20$~yr), the $11\,M_\odot$ SS model produces a larger DM than the $30\,M_\odot$ SS model even though the latter has a higher unshocked-ejecta mass density (see the middle panels of Fig.~\ref{fig:dm_cum_single_binary}a,b). This arises because the electron density scales as
$n_e=\chi_e\,\rho/(\mu_e m_p)$: the $30\,M_\odot$ ejecta has a larger $\mu_e$ (due to its thinner H envelope and higher He/metal fraction; see the bottom panels of Fig.~\ref{fig:dm_cum_single_binary}a,b), which reduces $n_e$ and offsets its density advantage, leading to $n_{e,30}<n_{e,11}$ and hence ${\rm DM}_{30}<{\rm DM}_{11}$ at early times.
As the remnant expands, dynamical dilution becomes dominant; the lower ejecta mass in the $11\,M_\odot$ model implies a higher characteristic expansion velocity and a faster density drop, so the DM curves cross around $t\simeq 20$~yr and ${\rm DM}_{11}<{\rm DM}_{30}$ thereafter. Throughout the evolution, the DM decline remains approximately power-law, with the unshocked ejecta dominating at early times (to $\lesssim 100$~yr) before the CSM contribution grows in importance; for the $11\,M_\odot$ BS case, this transition occurs much earlier, at $\sim 20$~yr.
\begin{figure*}[htbp]
    \centering
    \includegraphics[width=\textwidth]{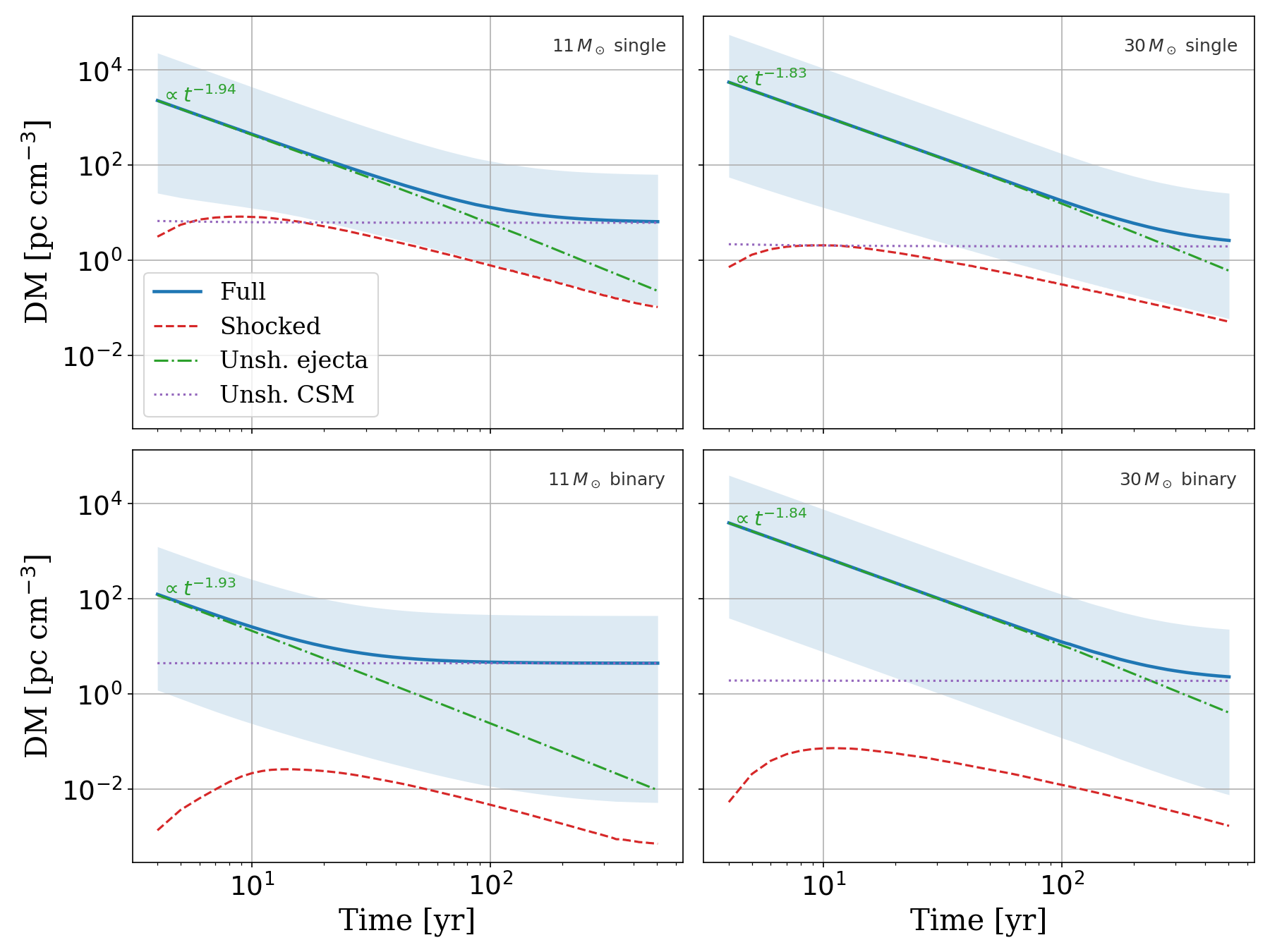}
\caption{Decomposition of the total SNR DM into individual components in the full-ionized-region calculation. The time-scaling shown for the unshocked-ejecta component is obtained from our fit to the unshocked-ejecta evolution.}
    \label{fig:dm_components_full}
\end{figure*}

All models exhibit an initial power-law-like decline in DM and a progressively smoother evolution at late times. These two behaviors correspond, respectively, to the transition from the ED phase to the later stage in which the swept-up CSM increasingly governs the remnant dynamics.

To identify which physical region dominates ${\rm DM}_{\rm SNR}$, and when the dominance transitions in each model, we decompose ${\rm DM}_{\rm SNR}$ into contributions from the unshocked ejecta, the unshocked CSM, and the shocked region (Fig.~\ref{fig:dm_components_full}). At early times, the unshocked ejecta provide the dominant contribution and exhibit an approximately power-law decline, ${\rm DM}\propto t^{-\alpha}$, with best-fit $-\alpha$ slopes of $-1.93$ and $-1.94$ for the $11\,M_\odot$ SS and BS models, respectively, and $-1.83$ and $-1.84$ for the $30\,M_\odot$ SS and BS models. The fact that these slopes are close to $-2$ is physically expected: in this epoch the dominant unshocked ejecta are still in free expansion, so if $\chi_e$ is approximately constant and $\mu_e$ evolves only weakly, then $\langle n_e\rangle\propto\rho\propto R^{-3}\propto t^{-3}$, while the characteristic integration length scales as $\Delta R_{\rm dyn}\propto t$. Therefore ${\rm DM}\sim\langle n_e\rangle\Delta R_{\rm dyn}\propto t^{-2}$. At later times, the unshocked CSM becomes increasingly important and eventually dominates; the characteristic transition times differ across models (e.g., $\sim 30$ yr and $\sim 10$ yr for the $11\,M_\odot$ SS and BS models, respectively, and $\sim 100$ yr for the $30\,M_\odot$ model). By contrast, the shocked region remains a subdominant contribution throughout the simulated time span.
\begin{figure*}[!t]
    \centering
    \includegraphics[width=0.85\textwidth]{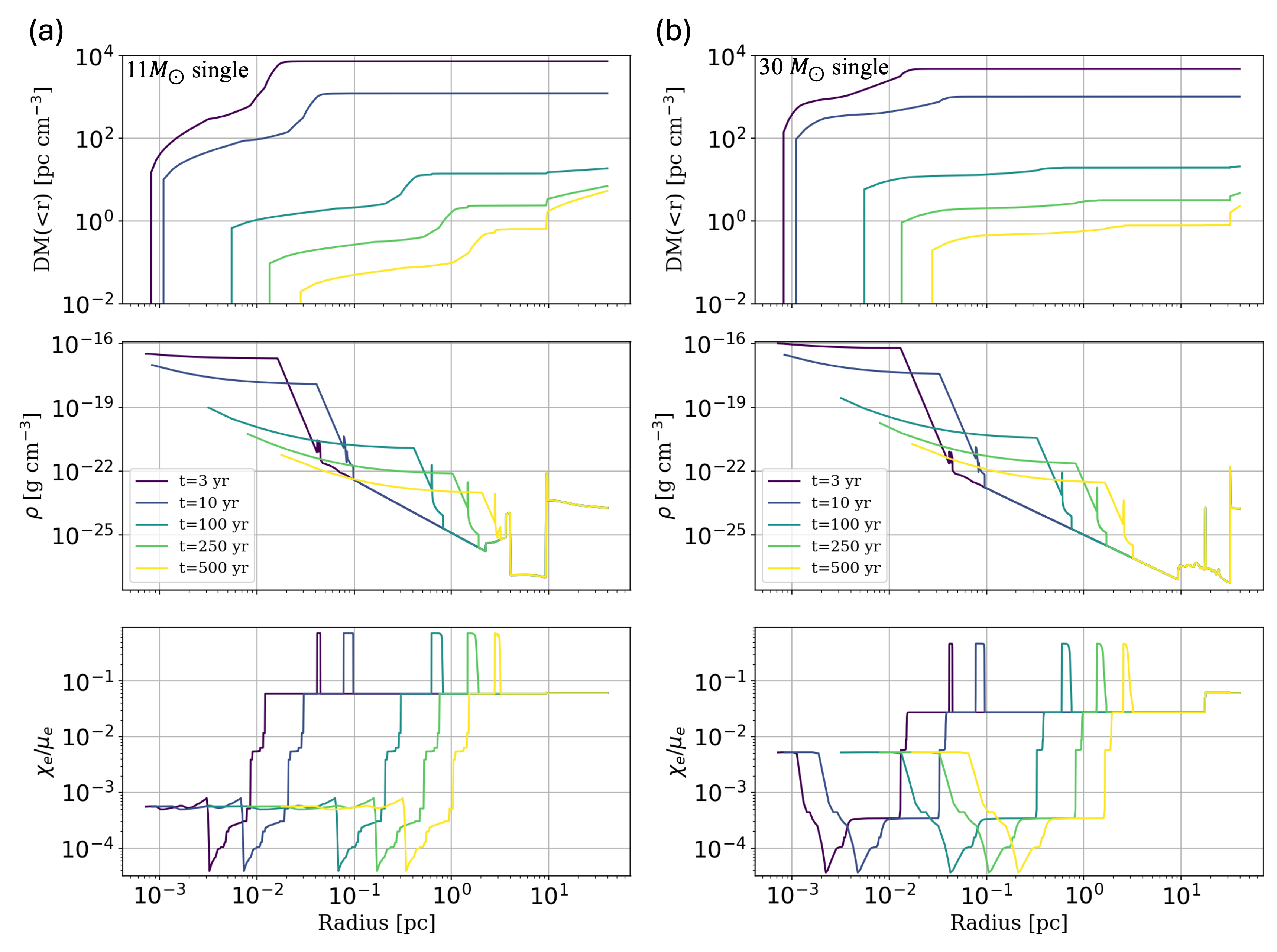}

    \vspace{0.2cm}

    \includegraphics[width=0.85\textwidth]{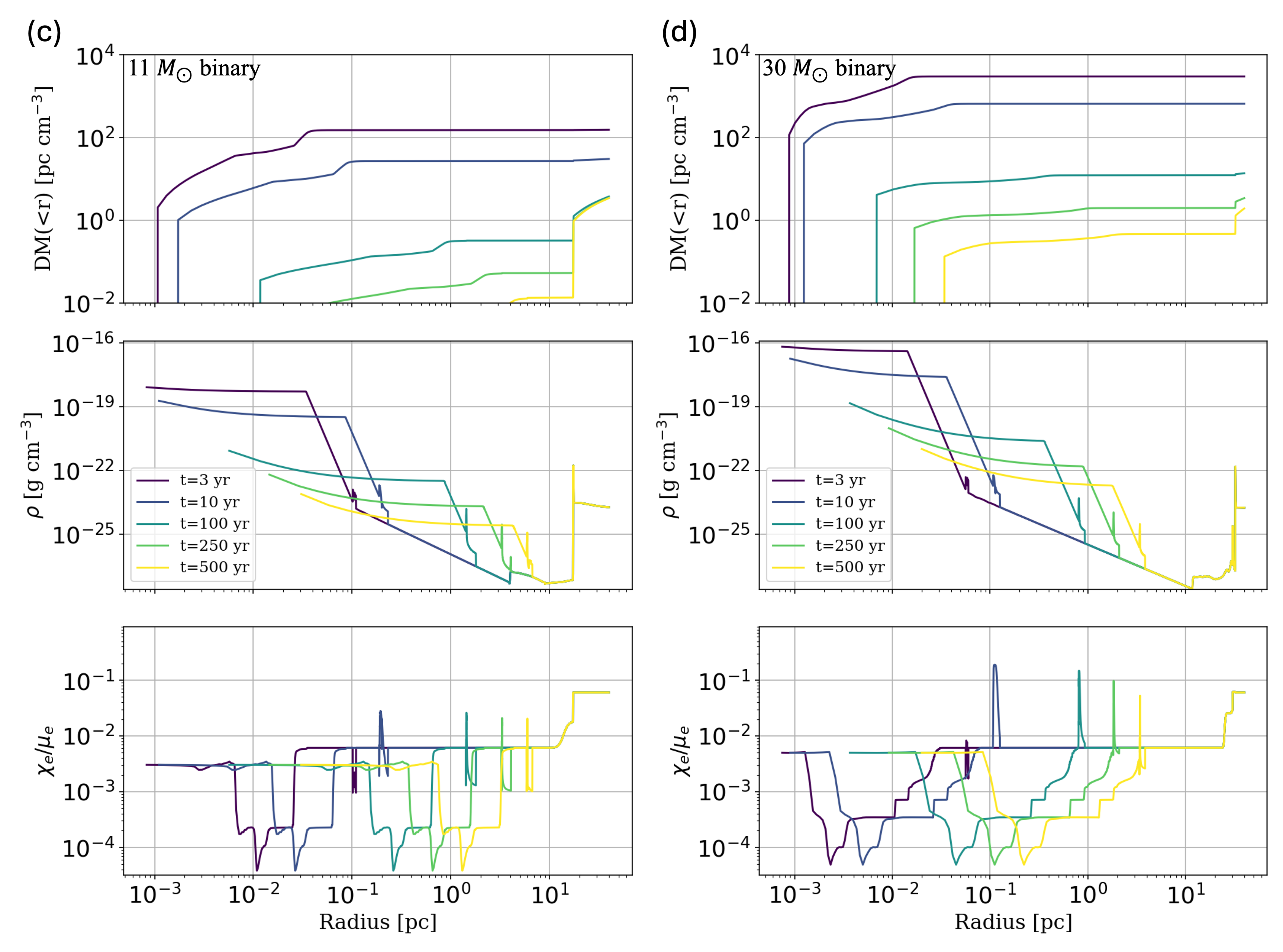}
\caption{Radial cumulative profiles for the full-region calculation, including the cumulative DM as well as the mass density and ionization fraction as functions of radius, evaluated at several characteristic epochs. The top and bottom panels correspond to the single-star (SS) and binary-stripped (BS) models, respectively.}
    \label{fig:dm_cum_single_binary}
\end{figure*}

Figure~\ref{fig:dm_cum_single_binary} provides an independent cross-check of the above interpretation by showing the radial cumulative DM (top panels) together with the corresponding mass-density (middle panels) and ionization-fraction (bottom panels) profiles at several representative epochs. To facilitate comparison with analytic expectations, we interpret the radial structure in a set of characteristic regimes, starting from the density profiles.

\begin{enumerate}[label=(\arabic*),leftmargin=*,itemsep=0.2em]
\item \textit{Inner-core region.}
As an illustrative example, we focus on $t=3\,$yr (the initial condition of our SNR calculations). At this epoch, the ejecta are extremely dense and retain a core--envelope structure close to the adopted ejecta profile (Eq.~\ref{eq:f_w}), i.e., an approximately flat (nearly uniform) inner core surrounded by a steeply declining power-law envelope. As the remnant evolves, the inner-core region gradually develops a modest radial gradient (i.e., a weak decrease with radius). In the inner-core region, the cumulative DM therefore rises rapidly at small radii and is dominated by the unshocked ejecta.

\item \textit{Rapidly declining unshocked-ejecta envelope.}
Outside the nearly uniform core, the unshocked ejecta enter a steep power-law envelope (with the initial decline set by the ejecta index $n$). The density therefore decreases rapidly with radius, and the cumulative DM correspondingly flattens once the line of sight exits the core.

\item \textit{Shocked region.}
After the unshocked-ejecta envelope, the profiles enter the shocked layer. Although the shocked plasma is highly ionized (see $\chi_e$), its DM contribution remains small because the shocked region has a limited path length and a comparatively low mass density, consistent with the decomposition in Fig.~\ref{fig:dm_components_full}.

\item \textit{Wind-dominated region (pre-SN wind$\times$CSM).}
Outside the shocked layer, the density profile transitions to the circumstellar environment shaped by pre-SN mass loss, exhibiting an approximately wind-like scaling $\rho\propto r^{-s}$ over a broad radial range (with $s=2$ in most analytic wind models).

\item \textit{Pre-SN wind--CSM interaction morphologies: SS vs. BS.}
“Beyond the wind-like region, the SS and BS channels develop distinct pre-SN wind--CSM interaction morphologies, including a hot cavity excavated by fast winds and density bumps produced by fast--slow wind interactions.
\begin{itemize}[leftmargin=1.8em,itemsep=0.1em]
\item \textit{SS channel: hot bubble + “double” wall + WR-driven bump.}
In the SS models (Fig.~\ref{fig:dm_cum_single_binary}a,b), the hot bubble is primarily generated by the interaction between the fast OB-type wind and the ambient ISM: the OB wind sweeps up the ambient medium into an outer dense shell (``ISM wall''), while the reverse shock propagating back into the wind heats the shocked wind material and inflates a high-pressure hot bubble. At later stages, the dense and slower RSG wind is confined by the hot-bubble thermal pressure, producing an additional inner density enhancement (``bubble wall''). A WR wind, when present (in the $30\,M_\odot$ case), can interact with the RSG wind and imprint a modest bump in the cavity profile.

\item \textit{BS channel: ISM wall + WR-like bump(with an absent/weak bubble wall).}
In the BS models (Fig.~\ref{fig:dm_cum_single_binary}c,d), the H envelope is stripped quickly, so the progenitor largely skips the RSG-wind stage (especially in the $11\,M_\odot$ case) and the evolution is dominated by a WR-like He-star wind, broadly similar to the WR-wind phase in the $30\,M_\odot$ SS model. In the $30\,M_\odot$ BS case in particular, a brief yet physical RSG-like slow-wind phase can still occur and form an inner ``bubble wall,'' but the subsequent WR-like fast wind rapidly disrupts this structure and drives it outward, leaving it nearly coincident with the ``ISM wall'' and thus hard to distinguish in the radial profiles. The remaining prominent feature is typically a WR-like bump.
\end{itemize}

\item \textit{Outer ambient ISM: late-time growth of the cumulative DM.}
At sufficiently large radii---beyond the wind-bubble structure---the ambient medium approaches an approximately constant density. At early times ($t\sim 3$--$10\,$yr) this outer material contributes little to the cumulative DM, whereas at late times ($t\gtrsim 100\,$yr) the declining ejecta density becomes comparable to that of the ambient medium and the cumulative DM can increase again toward the outer boundary.
\end{enumerate}
\subsubsection{The derivatives of DM}
\label{subsubsec:dDM_dt}
\begin{figure*}[!t]
    \centering
    \includegraphics[width=0.495\textwidth]{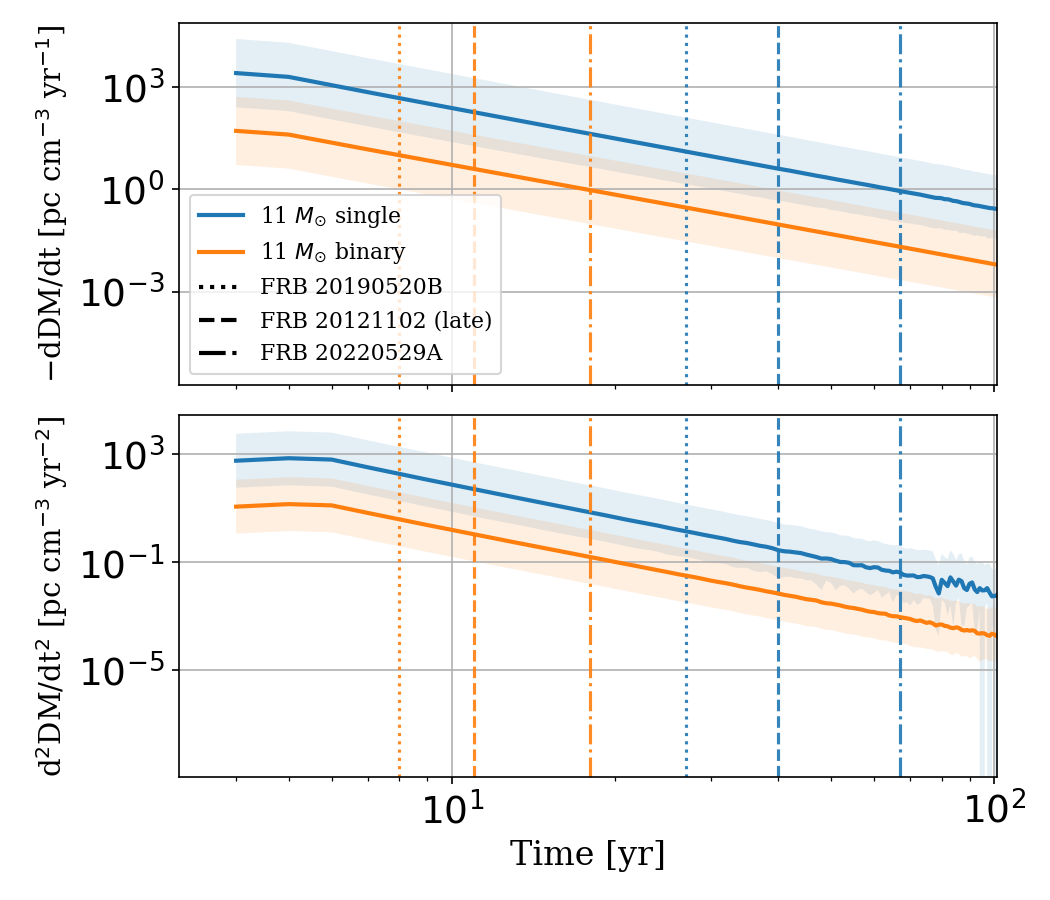}\hfill
    \includegraphics[width=0.495\textwidth]{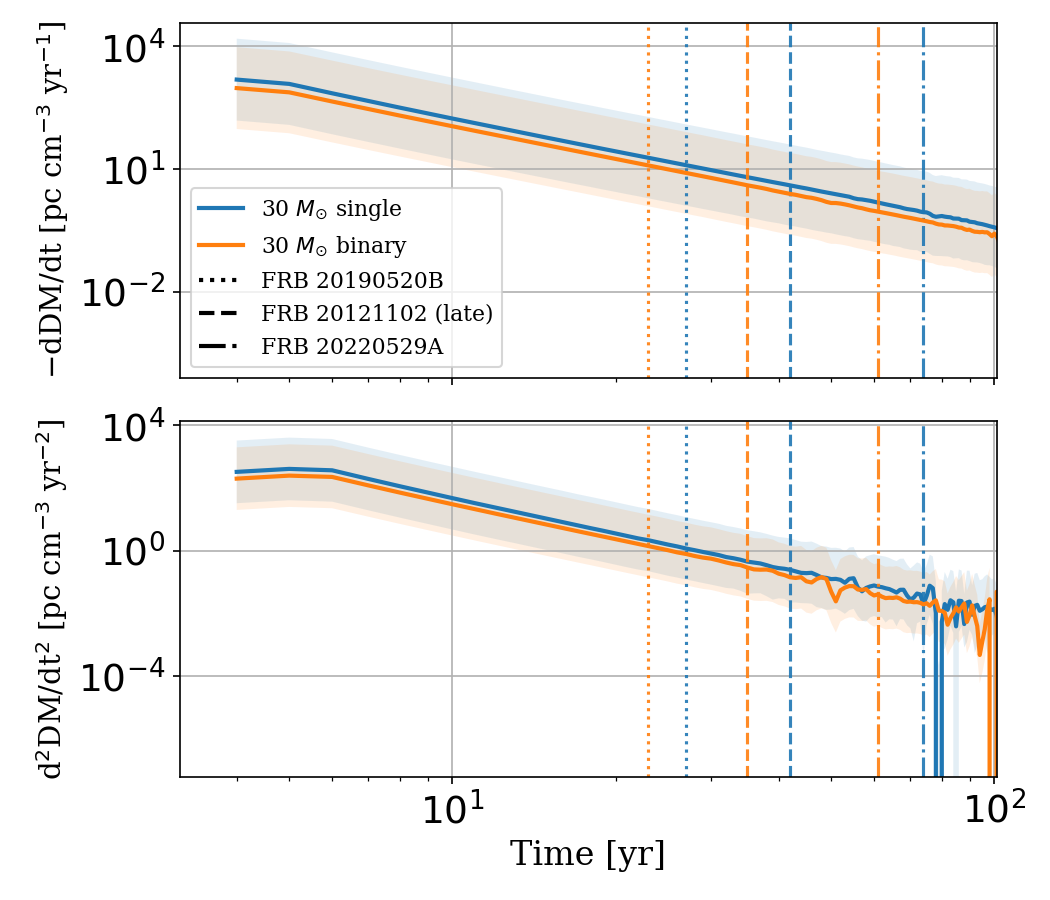}
    \caption{Time evolution of the DM time derivative in the full-region calculation for the $11\,M_\odot$ (left) and $30\,M_\odot$ (right) models. {For the fiducial (HH) configuration, $t_{\rm occur}$ for FRB~20190520B (Table~\ref{tab:toccur_tesc}) is $27$ (SS) and $8$ (BS) yr for $11\,M_\odot$, and $27$ (SS) and $23$ (BS) yr for $30\,M_\odot$; for FRB~20220529A (Table~\ref{tab:toccur_tesc_20220529A}), it is $67$ (SS) and $18$ (BS) yr for $11\,M_\odot$, and $74$ (SS) and $61$ (BS) yr for $30\,M_\odot$; for FRB~20121102 (Table~\ref{tab:toccur_tesc_20121102}), it is $40$ (SS) and $11$ (BS) yr for $11\,M_\odot$, and $42$ (SS) and $35$ (BS) yr for $30\,M_\odot$.}}
    \label{fig:dDM_full_11_30}
\end{figure*}

To characterize the secular evolution of ${\rm DM}_{\rm SNR}(t)$ and identify the observational time window in which an FRB can exhibit a measurable DM drift, we compute the first and second time derivatives of the simulated DM using Eq.~\eqref{eq:dm_derivatives}. Figure~\ref{fig:dDM_full_11_30} shows that both ${\rm dDM}/{\rm d}t$ and ${\rm d}^2{\rm DM}/{\rm d}t^2$ approximately follow a power-law decline in time over the smooth part of the evolution. At $t\gtrsim 100\,$yr, small-amplitude fluctuations become visible, and they are especially prominent in the second derivative. This behavior is expected because finite-difference estimates of higher-order derivatives amplify snapshot-to-snapshot irregularities in ${\rm DM}(t)$; physically, additional non-smooth features can also be introduced when the shock structure encounters density inhomogeneities (shells/bubbles) in the progenitor-shaped CSM.

Motivated by the observed secular decreases in FRB~20190520B ($\mathrm{dDM}/\mathrm{d}t\simeq -12.4\,{\rm pc\,cm^{-3}\,yr^{-1}}$; \citealt{Niu2025}), FRB~20121102 ($\mathrm{dDM}/\mathrm{d}t=-3.93\,{\rm pc\,cm^{-3}\,yr^{-1}}$; \citealt{Wang2025arXiv250715790W}), and {FRB~20220529A ($\mathrm{dDM}/\mathrm{d}t=-0.881\pm0.001\,{\rm pc\,cm^{-3}\,yr^{-1}}$; \citealt{Pandhi2026}),} we identify the corresponding time window after the SN explosion at which an FRB would need to occur. We denote this FRB occurrence time as $t_{\rm occur}$, i.e., the elapsed time between the SN explosion and the observed FRB activity. {Assuming the DM drift is dominated by the local SNR contribution and using the fiducial parameter values adopted in this section, we find $t_{\rm occur}\simeq 27\,$yr (SS) and $t_{\rm occur}\simeq 8\,$yr (BS) for the $11\,M_\odot$ models, and $t_{\rm occur}\simeq 27\,$yr (SS) and $t_{\rm occur}\simeq 23\,$yr (BS) for the $30\,M_\odot$ models for FRB~20190520B. For FRB~20121102, the same fiducial matching gives systematically later epochs, $t_{\rm occur}\simeq 40\,$yr (11 SS), $11\,$yr (11 BS), $42\,$yr (30 SS), and $35\,$yr (30 BS). For the more slowly evolving FRB~20220529A, the fiducial matching shifts to still later epochs: $t_{\rm occur}\simeq 67\,$yr (11 SS), $18\,$yr (11 BS), $74\,$yr (30 SS), and $61\,$yr (30 BS). The corresponding fiducial ${\rm DM}_{\rm SNR}$ values for all three sources are summarized in Table~\ref{tab:dm_fiducial_threefrb}.} Notably, the significantly lower DM in the $11\,M_\odot$ BS case compared with the other progenitors is primarily driven by its much smaller ejecta mass (see Table~\ref{tab:models}), which directly lowers the unshocked-ejecta electron column. These values correspond to the very early SNR phase, close to the transition from optically thick to optically thin conditions for GHz radio emission (see Section~\ref{subsubsec:tau_full} and Table~\ref{tab:toccur_tesc}). The short duration of this ``ambiguous'' phase may naturally contribute to the rarity of FRBs showing a clear SNR-like secular DM signature.

\begin{table*}[!t]
\centering
\caption{{Local SNR DM values at $t_{\rm occur}$ for three FRBs in the fiducial (HH) model.}}
\label{tab:dm_fiducial_threefrb}
\begin{tabular}{lccc}
\toprule
Progenitor / ${\rm DM}_{\rm SNR}$ (pc\,cm$^{-3}$) & FRB~20190520B & FRB~20220529A & FRB~20121102 \\
\midrule
$11\,M_\odot$ SS & 183.6 & 33.6 & 88.5 \\
$11\,M_\odot$ BS & 44.9  & 12.4 & 26.2 \\
$30\,M_\odot$ SS & 202.8 & 36.1 & 97.7 \\
$30\,M_\odot$ BS & 167.5 & 32.4 & 83.0 \\
\bottomrule
\end{tabular}
\end{table*}

{The simulated second-derivative DM curvature evaluated at $t_{\rm occur}$ is summarized in Table~\ref{tab:toccur_tesc} for FRB~20190520B and in Appendix Tables~\ref{tab:toccur_tesc_20121102} and \ref{tab:toccur_tesc_20220529A} for FRB~20121102 and FRB~20220529A, respectively. At these early epochs, the curvature for FRB~20190520B is typically $\mathrm{d}^2{\rm DM}/\mathrm{d}t^2\sim 1$--$4\,{\rm pc\,cm^{-3}\,yr^{-2}}$ across the four models, while for FRB~20121102 it is generally smaller, $\sim0.1$--$1.4\,{\rm pc\,cm^{-3}\,yr^{-2}}$. For FRB~20220529A, the inferred curvature spans an even broader low-amplitude range, from $\sim 7\times10^{-3}$ up to $\sim0.4\,{\rm pc\,cm^{-3}\,yr^{-2}}$, depending on the ionization configuration. Following the order-of-magnitude estimate used by \citet{Niu2025}, we define $\left|\mathrm{d}^2{\rm DM}/\mathrm{d}t^2\right|_{\rm est}\equiv \alpha\,|\mathrm{dDM}/\mathrm{d}t|/t_{\rm SNR}$ and evaluate it using $t_{\rm SNR}\simeq t_{\rm occur}$. For all three sources, the estimator remains broadly consistent with the simulated trends at the factor-of-few level, although a few FRB~20220529A cases show larger deviations because the absolute curvature is very small and can even change sign. This behavior is partly due to numerical uncertainties in the second derivative at late times.}

\begin{figure*}[!t]
    \centering
    \includegraphics[width=0.495\textwidth]{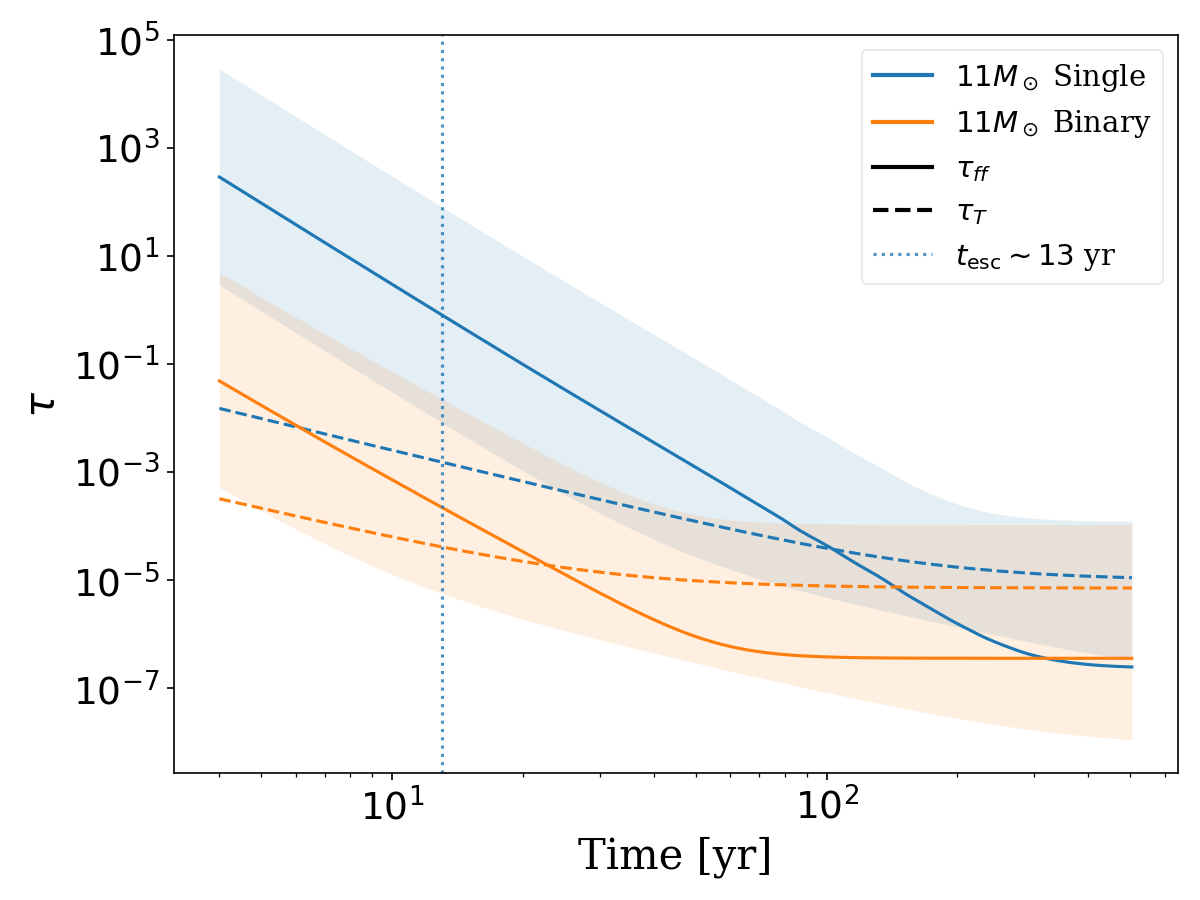}\hfill
    \includegraphics[width=0.495\textwidth]{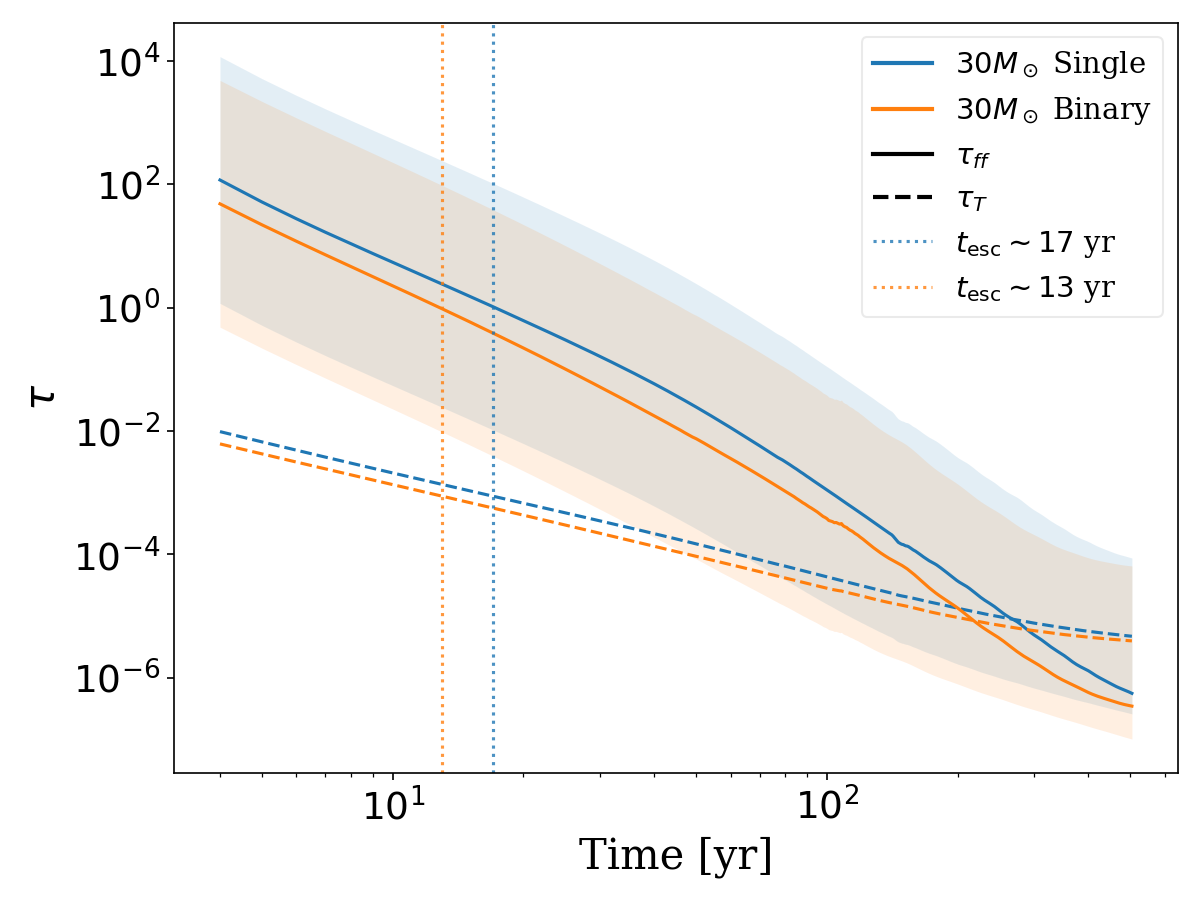}
\caption{Time evolution of the optical depth in the full region for the $11\,M_\odot$ (left) and $30\,M_\odot$ (right) models, including contributions from free--free absorption and Thomson scattering.}
    \label{fig:tau_ff_full_11_30}
\end{figure*}
\subsubsection{Optical depth evolution in the full region}
\label{subsubsec:tau_full}
{To assess whether FRB~20190520B, FRB~20220529A, and FRB~20121102 can plausibly escape from an SNR environment during the inferred occurrence time $t_{\rm occur}$, we compute the time evolution of the optical depth in the full-region calculation and indicate the epoch at which the system becomes optically thin.} Figure~\ref{fig:tau_ff_full_11_30} shows that at early times the total optical depth is dominated by free--free absorption, so we define the GHz escape time $t_{\rm esc}$ by the condition $\tau_{\rm ff}(t_{\rm esc})=1$.

For the models shown in Figure~\ref{fig:tau_ff_full_11_30}, we find $t_{\rm esc}=13\,$yr for the $11\,M_\odot$ SS model and $t_{\rm esc}=17\,$yr for the $30\,M_\odot$ SS model. For the $30\,M_\odot$ BS model, we find $t_{\rm esc}=13\,$yr. In contrast, the $11\,M_\odot$ BS model is essentially transparent from the beginning ($t_{\rm esc}\simeq 0\,$yr; $\tau_{\rm tot}<1$ at all simulated times).

\begin{table*}[!t]
\centering
\caption{Summary of characteristic timescales and DM derivatives evaluated at $t_{\rm occur}$ for different assumed ionization configurations of the unshocked ejecta ($\chi_{e,\rm unej}$) and the unshocked ISM ($\chi_{e,\rm ISM}$).}
\label{tab:toccur_tesc}
\begin{tabular}{llcccc}
\toprule
Ionization config. & Quantity & $11\,M_\odot$ SS & $11\,M_\odot$ BS & $30\,M_\odot$ SS & $30\,M_\odot$ BS\\
\midrule
\multirow{4}{*}{FF: $\chi_{e,\rm unej}=1$, $\chi_{e,\rm ISM}=1$}
& $t_{\rm esc}$ (yr) & 32 & 6 & 62 & 46\\
& $t_{\rm occur}$ (yr) & 60 & 16 & 65 & 55\\
& $\mathrm{d}^2{\rm DM}/\mathrm{d}t^2$ (pc\,cm$^{-3}$\,yr$^{-2}$) & 0.65 & 2.45 & 0.51 & 0.73\\
& $\left|\mathrm{d}^2{\rm DM}/\mathrm{d}t^2\right|_{\rm est}$ (pc\,cm$^{-3}$\,yr$^{-2}$) & 0.40 & 1.50 & 0.35 & 0.41\\
\midrule
\multirow{4}{*}{HF: $\chi_{e,\rm unej}=0.1$, $\chi_{e,\rm ISM}=1$}
& $t_{\rm esc}$ (yr) & 13 & 0 & 17 & 13\\
& $t_{\rm occur}$ (yr) & 27 & 8 & 27 & 23\\
& $\mathrm{d}^2{\rm DM}/\mathrm{d}t^2$ (pc\,cm$^{-3}$\,yr$^{-2}$) & 1.40 & 3.88 & 1.17 & 1.40\\
& $\left|\mathrm{d}^2{\rm DM}/\mathrm{d}t^2\right|_{\rm est}$ (pc\,cm$^{-3}$\,yr$^{-2}$) & 0.89 & 3.01 & 0.84 & 0.99\\
\midrule
\multirow{3}{*}{MF: $\chi_{e,\rm unej}=0.01$, $\chi_{e,\rm ISM}=1$}
& $t_{\rm esc}$ (yr) & 5 & 0 & 4 & 0\\
& $t_{\rm occur}$ (yr) & 13 & -- & 11 & 10\\
& $\mathrm{d}^2{\rm DM}/\mathrm{d}t^2$ (pc\,cm$^{-3}$\,yr$^{-2}$) & 2.60 & -- & 3.28 & 3.04\\
& $\left|\mathrm{d}^2{\rm DM}/\mathrm{d}t^2\right|_{\rm est}$ (pc\,cm$^{-3}$\,yr$^{-2}$) & 1.84 & -- & 2.06 & 2.28\\
\midrule
\multirow{4}{*}{FH: $\chi_{e,\rm unej}=1$, $\chi_{e,\rm ISM}=0.1$}
& $t_{\rm esc}$ (yr) & 32 & 6 & 62 & 46\\
& $t_{\rm occur}$ (yr) & 60 & 16 & 65 & 55\\
& $\mathrm{d}^2{\rm DM}/\mathrm{d}t^2$ (pc\,cm$^{-3}$\,yr$^{-2}$) & 0.065 & 2.45 & 0.51 & 0.73\\
& $\left|\mathrm{d}^2{\rm DM}/\mathrm{d}t^2\right|_{\rm est}$ (pc\,cm$^{-3}$\,yr$^{-2}$) & 0.40 & 1.50 & 0.35 & 0.41\\
\midrule
\multirow{3}{*}{HH: $\chi_{e,\rm unej}=0.1$, $\chi_{e,\rm ISM}=0.1$}
& $t_{\rm esc}$ (yr) & 13 & 0 & 17 & 13\\
& $t_{\rm occur}$ (yr) & 27 & 8 & 27 & 23\\
& $\mathrm{d}^2{\rm DM}/\mathrm{d}t^2$ (pc\,cm$^{-3}$\,yr$^{-2}$) & 1.41 & 3.88 & 1.17 & 1.40\\
& $\left|\mathrm{d}^2{\rm DM}/\mathrm{d}t^2\right|_{\rm est}$ (pc\,cm$^{-3}$\,yr$^{-2}$) & 0.89 & 3.01 & 0.84 & 0.99\\
\midrule
\multirow{3}{*}{MH: $\chi_{e,\rm unej}=0.01$, $\chi_{e,\rm ISM}=0.1$}
& $t_{\rm esc}$ (yr) & 5 & 0 & 4 & 0\\
& $t_{\rm occur}$ (yr) & 13 & -- & 11 & 10\\
& $\mathrm{d}^2{\rm DM}/\mathrm{d}t^2$ (pc\,cm$^{-3}$\,yr$^{-2}$) & 2.59 & -- & 3.28 & 3.03\\
& $\left|\mathrm{d}^2{\rm DM}/\mathrm{d}t^2\right|_{\rm est}$ (pc\,cm$^{-3}$\,yr$^{-2}$) & 1.84 & -- & 2.06 & 2.28\\
\bottomrule
\end{tabular}
\begin{tablenotes}
\item In the ionization configuration labels, H/M/L denote High/Middle/Low ionization levels, and F denotes Fully ionized.
\item $t_{\rm occur}$ is determined by matching the observed DM derivative of FRB~20190520B, i.e., imposing $\mathrm{dDM}/\mathrm{d}t=-12.4\,$pc\,cm$^{-3}$\,yr$^{-1}$. ``--'' indicates that no solution for $t_{\rm occur}$ exists under this condition for the stated ionization assumption.
\item $t_{\rm esc}$ is defined by $\tau_{\rm ff}(t_{\rm esc})=1$; $t_{\rm esc}=0$ indicates the model is transparent at all simulated times.
\end{tablenotes}
\end{table*}
\subsection{RM results}
\label{subsec:rm_evolution}
\begin{figure*}[!t]
    \centering
    \includegraphics[width=0.49\textwidth]{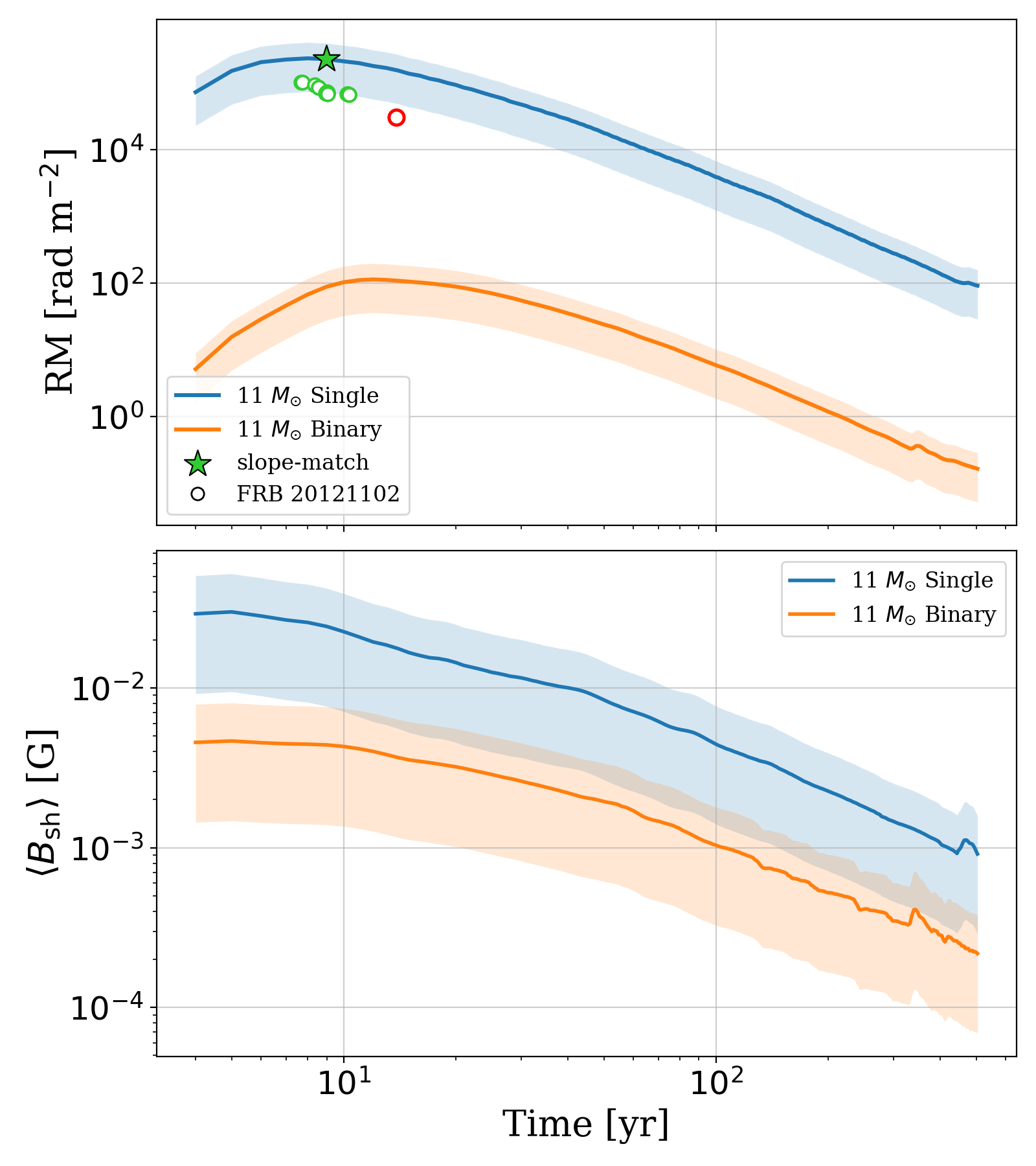}
    \hfill
    \includegraphics[width=0.49\textwidth]{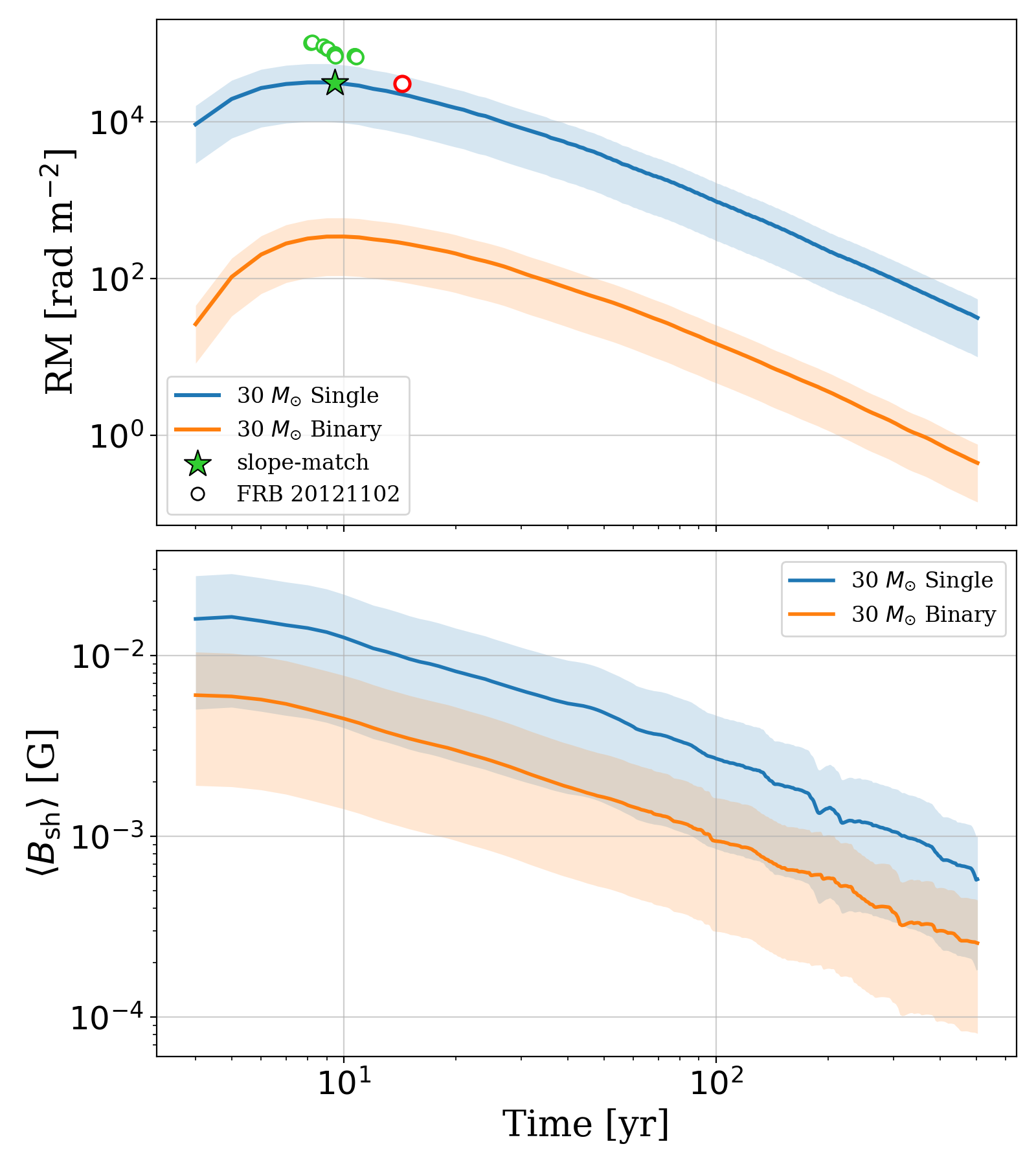}
    \caption{Time evolution of RM and magnetic field for the $11\,M_\odot$ (left) and $30\,M_\odot$ (right) models, with the FRB~20121102 RM data mapped onto the model time axis (see text). In the top panels, blue circles denote RM measurements from \citet{Hilmarsson2021}, while red circles denote the newly added RM measurements from \citet{Wang2025arXiv250715790W}. The top panels show RM evolution, and the bottom panels show magnetic-field evolution, where $\langle B_{\rm sh}\rangle$ is the electron-density-weighted mean field in the shocked region. Shaded bands mark the observed RM range and the allowed $\langle B_{\rm sh}\rangle$ range.}
    \label{fig:rm_vs_time}
\end{figure*}

The RM and magnetic-field evolutions are shown in Fig.~\ref{fig:rm_vs_time}. In this shocked-region-only RM framework, only the $11\,M_\odot$ SS model reproduces the observed RM range and decay trend of FRB~20121102; the $30\,M_\odot$ SS model and both BS models remain below the data and do not satisfy the joint amplitude--slope constraints.

The strong RM contrast between SS and BS models is physically consistent with the shocked-shell dynamics discussed in Section~\ref{subsec:shocked}. In particular, the dominant driver is the large difference in shocked mass $M_{\rm sh}$ (Fig.~\ref{fig:dm_mass_model_comparison}e), which translates into a large density contrast in the shocked region. Although BS models can have a higher shock velocity $v_{\rm sh}$ in a more rarefied environment (Fig.~\ref{fig:dm_mass_model_comparison}d), this does not compensate for their much lower shocked density. As a result, the ram pressure $\rho v_{\rm sh}^2$---and hence the amplified magnetic field in our prescription---remains significantly smaller than in SS models. In addition, the denser SS shocked layer produces a substantially higher post-shock electron density $n_e$ (see Fig.~\ref{fig:dm_mass_model_comparison}b), which further boosts $\int n_e B_\parallel\,\mathrm{d}l$ and amplifies the final RM separation between SS and BS channels.

Therefore, in a deliberately conservative sense, if we include only magnetic-field amplification in the SNR shock region and neglect other nonlinear effects or additional magnetic/ionization enhancement (e.g., from a MWN or photoionization), the current RM data for FRB~20121102 select the $11\,M_\odot$ SS case as the only successful model in our grid, despite studies arguing for a strong MWN environment for this source \citep{Beloborodov2017, Yang2019}.

\subsection{Simulation v.s. analytic models}
\label{subsec:SNR_derivitave_evolution}
\begin{deluxetable*}{ccc}
\tablecaption{Physical parameters adopted in different analytic/semi-analytic DM prescriptions.\label{tab:analytic_params_clean}}
\tablehead{
\colhead{Model} &
\colhead{Parameters in formula} &
\colhead{Adopted values}
}
\startdata
YZ17: ionized ejecta (FE; Eq.~4)
& $M_1,\,E_{51},\,\chi_e,\,\mu_m$
& $\mu_m=1.2$
\\
\tableline
PG18: shocked ionized (Eq.~5--6)
& \makecell[c]{$M_1,\,E_{51}$ shared\\
ISM: $n_0,\,\mu_m,\,\mu_e$\\
wind: $\mu_e,\,K(\dot M,v_w)$}
& \makecell[c]{ISM: $n_0=1~\mathrm{cm^{-3}}$\\
SS: $\mu_m=0.62,\,\mu_e=1.18$\\
BS: $\mu_m=1.34,\,\mu_e=2.0$\\
wind: $\dot M=10^{-5}\,M_\odot\,\mathrm{yr^{-1}}$\\
$v_w=10~\mathrm{km\,s^{-1}}$}
\\
\tableline
Zhao+2021: SSDW wind (Eq.~8--12)
& \makecell[c]{$M_{\rm ej},\,E_{51},\,n,\,s,$
$\mu_a,\,\chi_e,\,\dot M,\,v_w$}
& \makecell[c]{$n=10,\ s=2$\\ $w_{\rm core}=0.1,\ \mu_a=1$\\
$\dot M=10^{-5}\,M_\odot\,\mathrm{yr^{-1}}$\\
$v_w=10~\mathrm{km\,s^{-1}}$}
\\
\enddata
\end{deluxetable*}

\begin{figure*}
    \centering
    \begin{minipage}{0.48\linewidth}
        \centering
        \includegraphics[width=\linewidth]{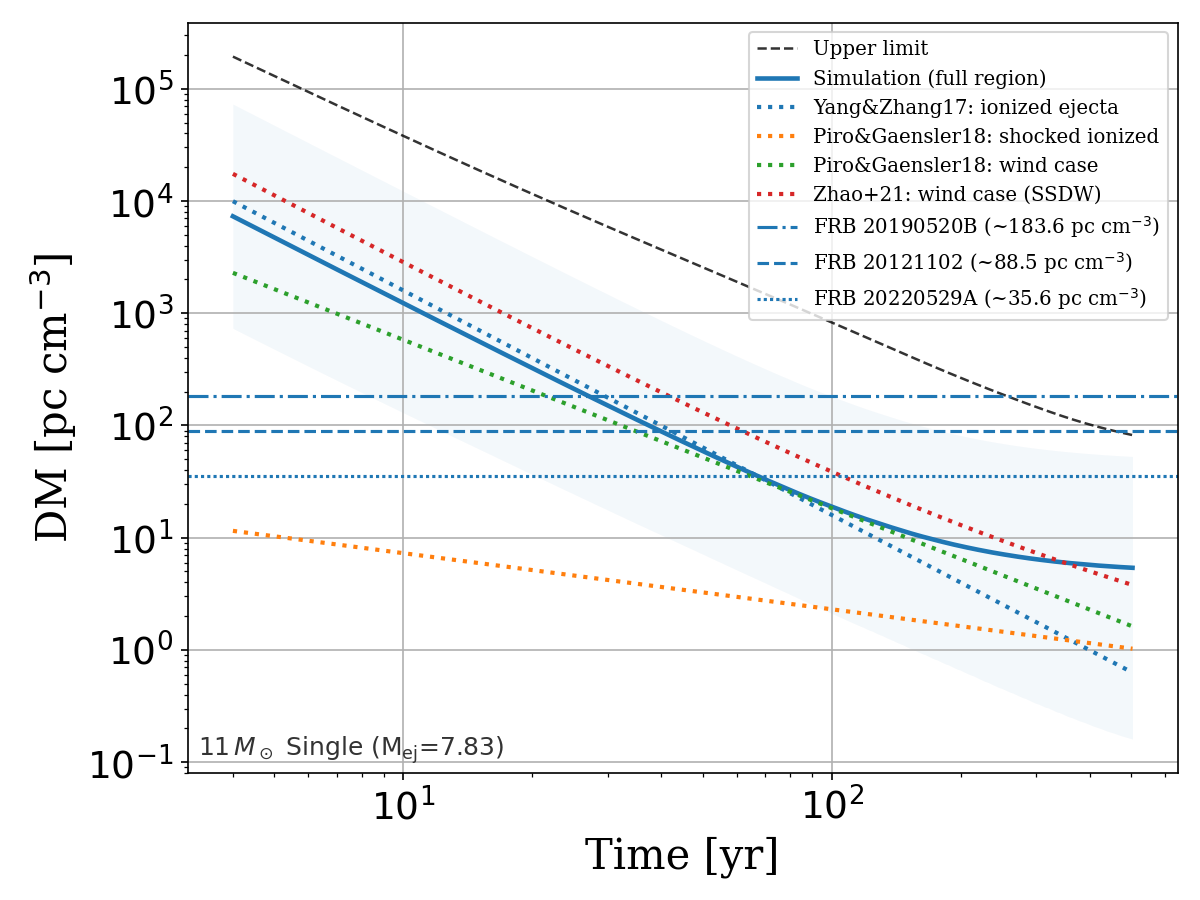}
    \end{minipage}
    \hfill
    \begin{minipage}{0.48\linewidth}
        \centering
        \includegraphics[width=\linewidth]{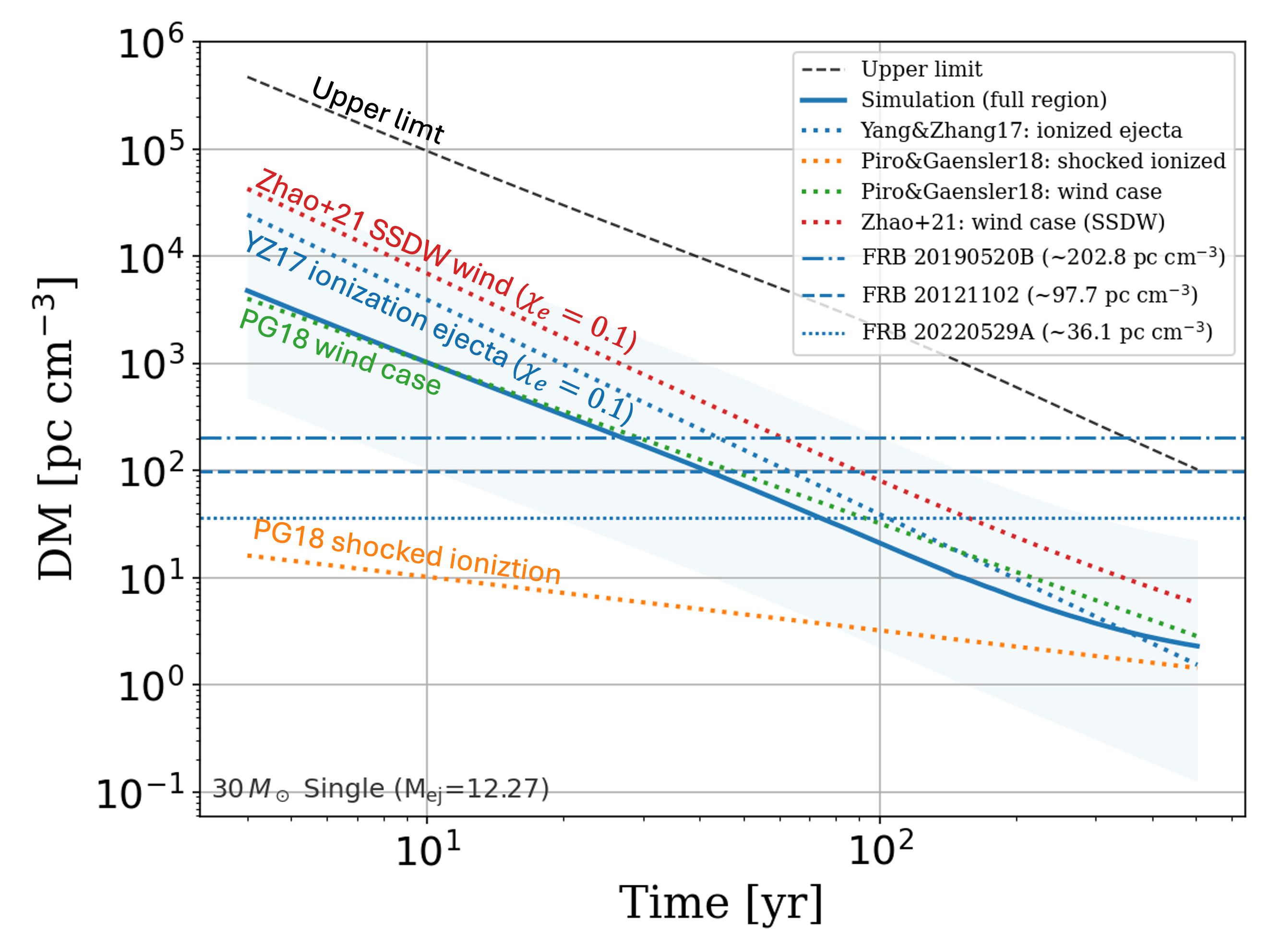}
    \end{minipage}

    \vspace{-0.25cm}

    \begin{minipage}{0.48\linewidth}
        \centering
        \includegraphics[width=\linewidth]{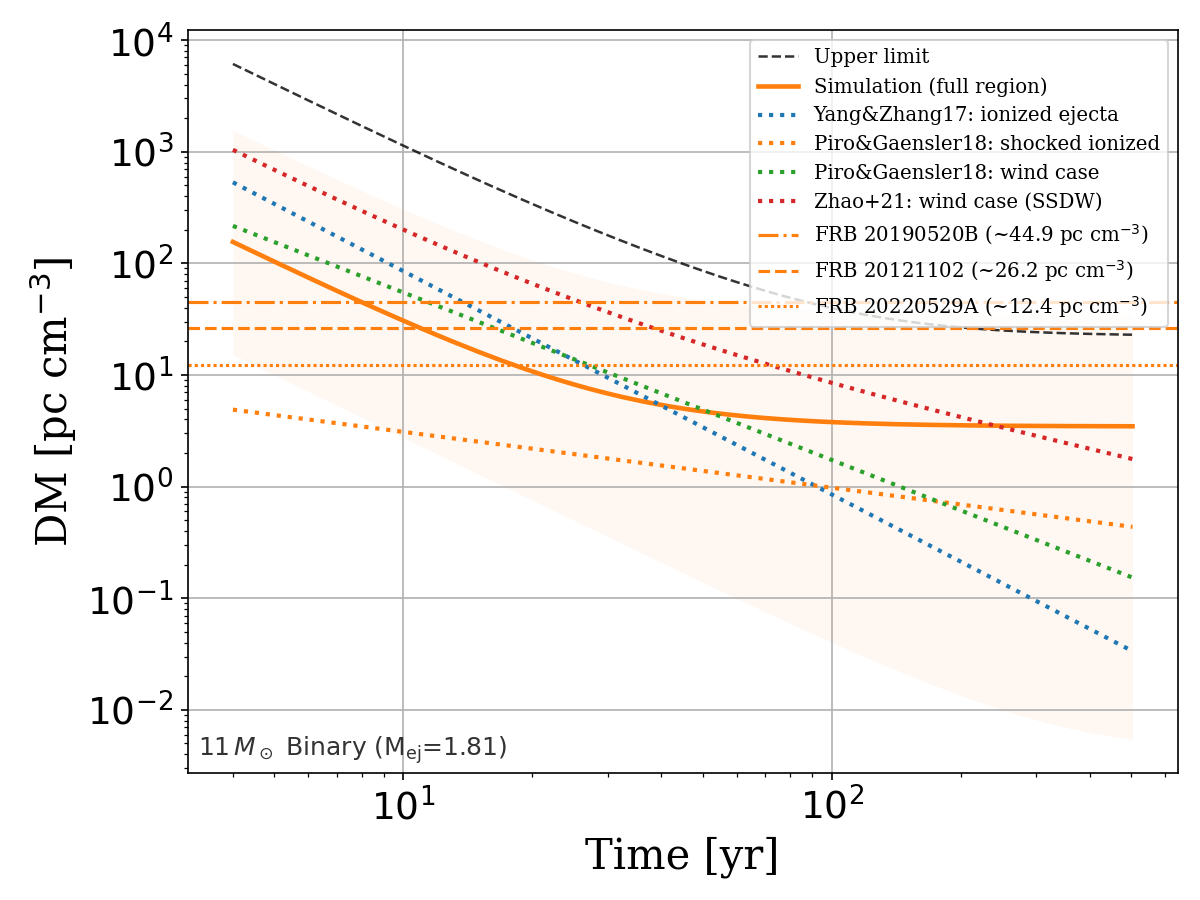}
    \end{minipage}
    \hfill
    \begin{minipage}{0.48\linewidth}
        \centering
        \includegraphics[width=\linewidth]{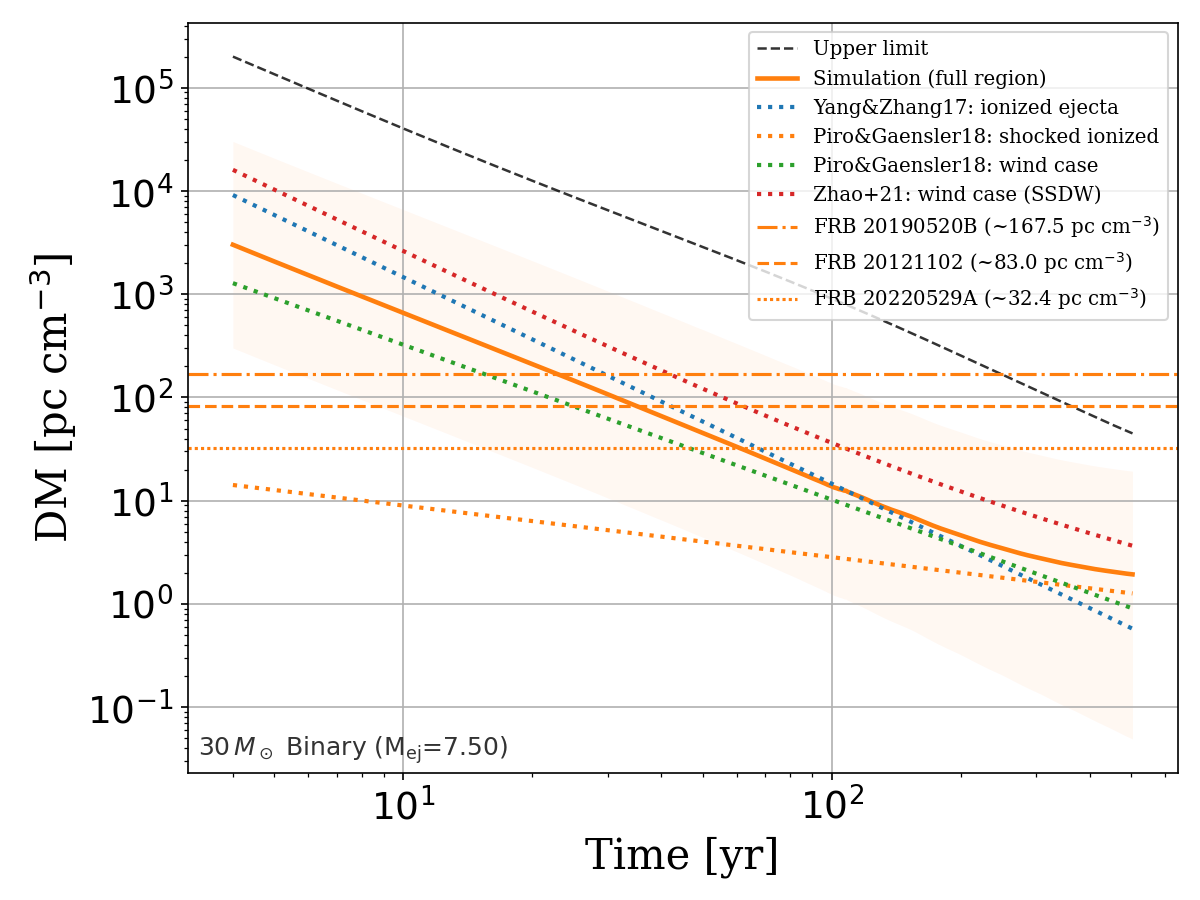}
    \end{minipage}

    \caption{
    Time evolution of the SNR DM comparing numerical simulations with analytic expectations.
    Left panels show $11\,M_\odot$ models and right panels show $30\,M_\odot$ models; the top row corresponds to SS progenitors and the bottom row to BS progenitors.
    The ``upper limit'' curve assumes that the entire SNR material is fully ionized and that the ejecta are composed purely of hydrogen.
    For Zhao+21 and YZ17, we adopt the fiducial assumption $\chi_{e,\mathrm{unsh}}=0.1$
    }
    \label{fig:dm_single_binary_11}
\end{figure*}
Fig.~\ref{fig:dm_single_binary_11} compares the DM evolution obtained from our numerical SNR simulations with several commonly adopted analytic prescriptions. Hereafter, we denote \citet{Yang&Zhang2017} as YZ17 and \citet{Piro&Gaensler2018} as PG18.

We evaluate the analytic prescriptions using the same explosion and wind parameters adopted in our progenitor setup (Table~\ref{tab:models}). For the remaining analytic-model parameters, we follow the values adopted in the original literature where they are explicitly specified (Table~\ref{tab:analytic_params_clean}). Here we only restate the mathematical definitions: $n_e=\rho/(\mu_e m_p)$, $n_{\rm tot}=\rho/(\mu_m m_p)$, and $\mu_a=\sum_i n_i A_i/\sum_i n_i$, where $n_{\rm tot}$ denotes the total particle number density (ions + electrons).

For specific models, \citet{Yang&Zhang2017} adopt $\mu_m=1.2$ (solar-composition, approximately neutral material), while \citet{ZYZhao2021b} state $\mu_a\simeq 1$ for their fiducial H-dominated ejecta. When required, we compute $\mu_a$ directly from the simulated elemental composition. For the PG18 analytic curves, which do not explicitly state $\mu_m$ or $\mu_e$, we adopt fiducial composition assumptions; for a fully ionized H/He mixture, where $X$ and $Y$ are the hydrogen and helium mass fractions ($X+Y=1$), the standard relations are $\mu_e = 2/(1+X)$ and $\mu_m = 1/(2X + 3Y/4)$. We then use: (i) SS case with H+He-dominated ejecta, $X\simeq 0.7$, $Y\simeq 0.3$, giving $\mu_m\simeq 0.62$ and $\mu_e\simeq 1.18$; (ii) BS case with He-dominated ejecta (H envelope lost), giving $\mu_m\simeq 1.34$ and $\mu_e\simeq 2.0$. These adopted values are the ones used in our PG18 comparisons.

We also evaluate the analytic prescriptions using the simulated, composition-dependent $\mu_a$, $\mu_e$, and $\mu_m$. These results are presented in Appendix~\ref{app:analytic_mu_comparison} and Fig.~\ref{fig:dm_single_binary_11_app}. 

In Fig.~\ref{fig:dm_single_binary_11}, the black dashed curve denotes the upper limit, which assumes H-dominated unshocked ejecta and CSM ($\mu_e=\mu_a=1$) that are fully ionized ($\chi_{e}=1$). Among the cases shown, the largest deviation from the numerical results occurs for the shocked-ionized ejecta scenario, whereas most analytic prescriptions fall within the upper and lower envelopes of our DM evolution. Under our fiducial assumptions in Fig.~\ref{fig:dm_single_binary_11}, YZ17 most closely matches the $11\,M_\odot$ SS case, while the PG18 wind case is closer for the $30\,M_\odot$ SS case. These "best-match" identifications are dependent on progenitor-model choices and parameter assumptions and are not intended to indicate a general preference among analytic prescriptions. Notably, when we adopt simulation-based, composition-dependent $\mu_a$, $\mu_e$, and $\mu_m$ (Fig.~\ref{fig:dm_single_binary_11_app}), the $11\,M_\odot$ SS case becomes more consistent with the Zhao+2021 prediction.

Using the $t_{\rm occur}$ values constrained by Fig.~\ref{fig:dDM_full_11_30}, we obtain the corresponding ${\rm DM}_{\rm SNR}$ values of $183.6\,{\rm pc\,cm^{-3}}$ ($11\,M_\odot$ SS), $202.8\,{\rm pc\,cm^{-3}}$ ($30\,M_\odot$ SS), $44.9\,{\rm pc\,cm^{-3}}$ ($11\,M_\odot$ BS), and $167.5\,{\rm pc\,cm^{-3}}$ ($30\,M_\odot$ BS). This indicates that the DM constraint varies significantly not only with model parameters but also with progenitor channel. Nevertheless, all inferred contributions lie in the tens-to-hundreds ${\rm pc\,cm^{-3}}$ range, representing a non-negligible extra component that can materially affect FRB cosmological analyses.

\subsection{Shocked-region electron source and ionization states}
\label{subsec:ionization}
In addition to the total DM$_{\rm sh}$, our simulations track the contribution to the electron density from different elements and ionization stages in the shocked region physically.
\begin{figure}
    \centering
    \includegraphics[width=1\linewidth]{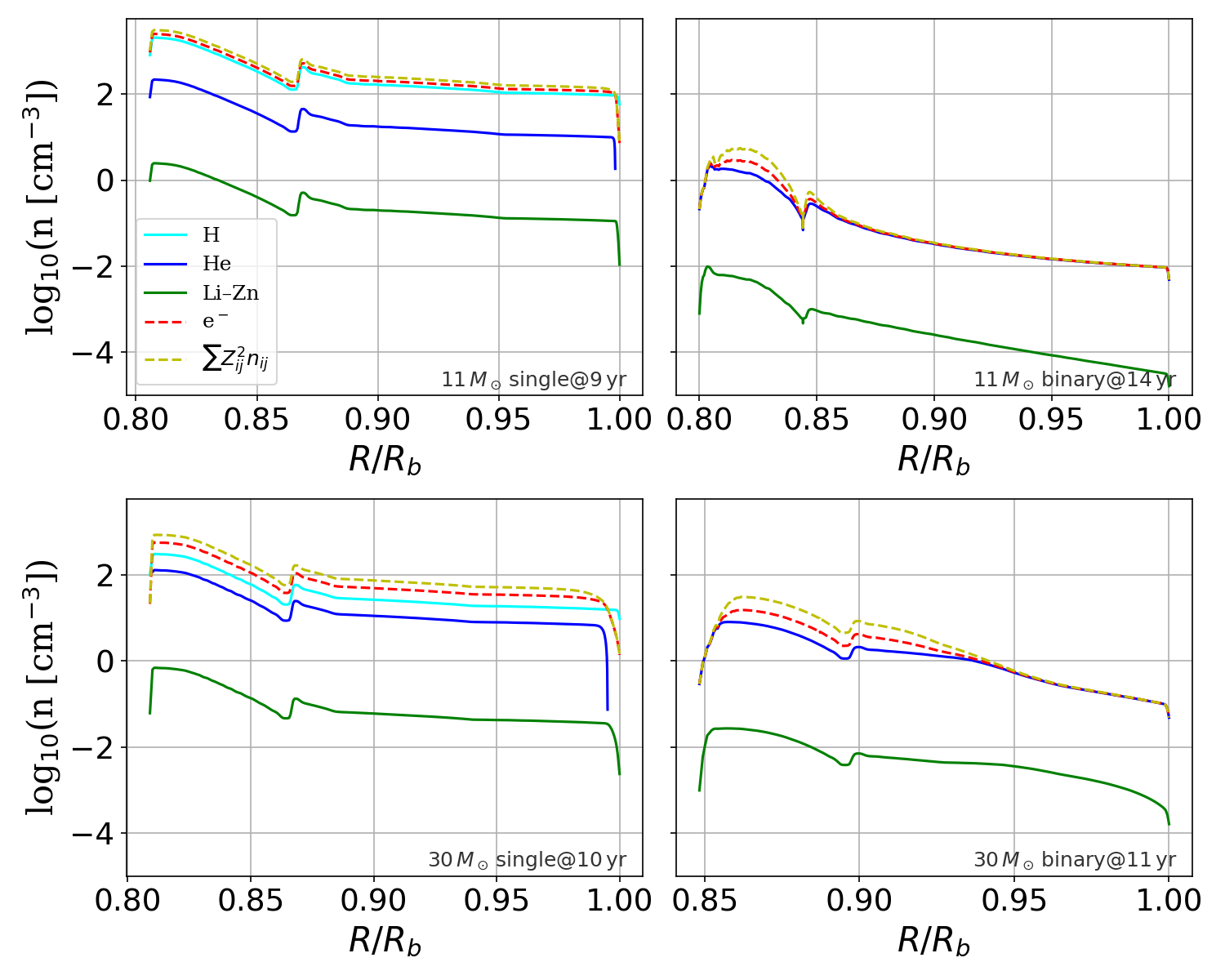}
    \caption{Radial profiles in the \textit{shocked region} at the DM$_{\rm sh}$ peak, showing number densities of electrons (red) and selected elements for the four models considered in this work. The cyan and blue curves correspond to H and He, respectively.}
    \label{fig:elements_density}
\end{figure}
Fig.~\ref{fig:elements_density} shows the radial number-density profiles in the shocked region at the DM$_{\rm sh}$ peak (see Fig.~\ref{fig:dm_mass_model_comparison}a).
A clear channel dependence is visible: in the SS models, the shocked composition is generally H-rich with He as the secondary contributor, whereas in the BS models He dominates and metals remain subdominant. Notably, compared with the $11\,M_\odot$ SS model, the $30\,M_\odot$ SS model has a smaller hydrogen fraction and is therefore more Type~IIb-like than a typical H-rich Type~IIP case.
This difference ultimately reflects the distinct progenitor-envelope structures: SS ejecta are largely composed of the progenitor's H-rich envelope, while in the BS channel binary interaction efficiently removes the H envelope prior to core collapse (e.g., via common-envelope evolution and Roche-lobe overflow), leaving a He-rich progenitor whose ejecta contain comparatively little H.
As a result, in the BS models the post-shock electron budget is primarily supplied by ionized He, whereas in the SS models it is dominated by ionized H.

This composition difference also helps explain why the BS models reach a larger $\langle Z\rangle$ at the peak (bottom panels of Fig.~\ref{fig:optical_depth_evolution}) than the SS models: helium can contribute up to two electrons per ion once ionized, raising the typical ionic charge and hence $\langle Z\rangle$ relative to an H-dominated composition.

\begin{figure*}[!t]
    \centering
    \vspace{-0.3cm}

    \begin{subfigure}{\textwidth}
        \centering
        \includegraphics[width=0.73\linewidth]{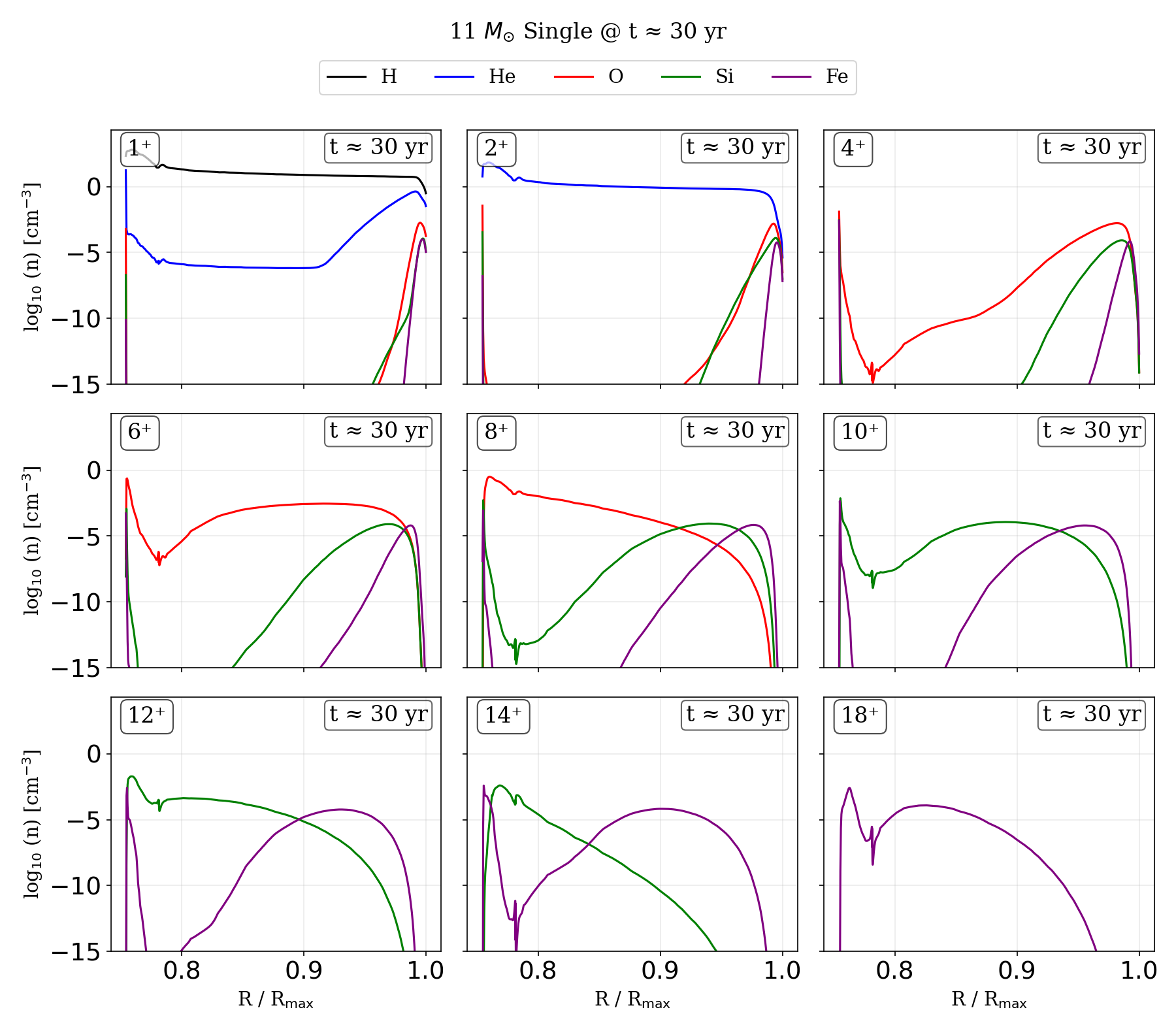}
        \caption{$11\,M_\odot$ SS}
    \end{subfigure}

    \vspace{-0.25cm}

    \begin{subfigure}{\textwidth}
        \centering
        \includegraphics[width=0.73\linewidth]{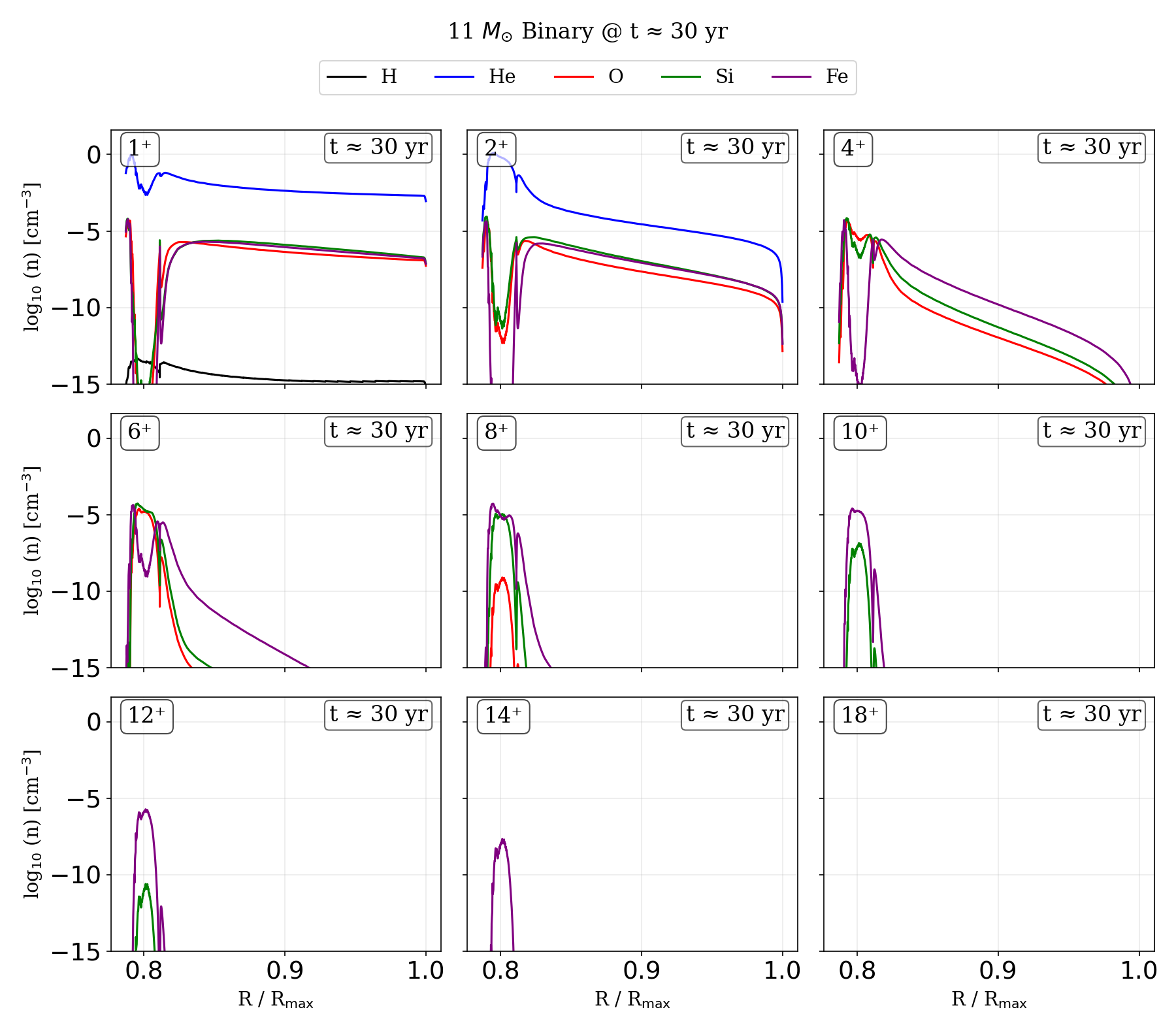}
        \caption{$11\,M_\odot$ BS}
    \end{subfigure}

    \vspace{-0.3cm}
    \caption{Ionization-state profiles in the shocked region at $t=30\,\mathrm{yr}$ for the $11\,M_\odot$ SS and BS models.}
    \label{fig:ionization_panels_11_30t_unsh}
\end{figure*}

This interpretation is further supported by the ionization-state structure of the shocked region in Fig.~\ref{fig:ionization_panels_11_30t_unsh}: in panel (a), corresponding to the $11\,M_\odot$ SS model, the peak number density of H$^{+}$ exceeds that of He$^{2+}$, whereas in panel (b), corresponding to the $11\,M_\odot$ BS model, the shocked region is instead dominated by He$^{+}$ and He$^{2+}$.

\section{Model Limitations and Future Directions}
\label{sec:discussion}
In this section, we outline some of the main implications of our preliminary results and discuss directions for future refinement.
\begin{itemize}
    \item The ionization and thermal evolution history of the unshocked ejecta is treated approximately in the current setup. In particular, we do not yet include a fully self-consistent ionization--recombination balance, nor possible CCO-related heating/ionization channels, in the unshocked region. This limitation introduces a relatively broad uncertainty when constraining source-environment parameters using DM evolution and should be addressed in future work.
    \item In our current 1D SNR simulations, we do not enable cosmic-ray (CR) acceleration and transport. As a result, potentially important CR-related processes---including escaping CR particles and CR-driven magnetic-field amplification (e.g., streaming instabilities and Bell-type instabilities)---are absent. {CR-related processes, such as escaping CRs could in principle contribute additional ionization and enhance the electron density, but these effects are not included in the present model, and their quantitative impact remains uncertain. We note that CR-driven magnetic-field amplification may also affect the RM, particularly in the shocked region.}
    Incorporating these effects in future calculations may provide new physical insight into the magnitude and time evolution of RM.

    \item The present 1D framework assumes spherical symmetry and a stationary central engine. Real SNR evolution is intrinsically 3D and often asymmetric: binary stripping through Roche-lobe overflow is non-spherical, the SN explosion and external CSM can be nonuniform and anisotropic, and a young magnetar may have a substantial natal kick velocity. In addition, the binary-stripped channel considered here represents only one pathway within a much broader binary-interaction parameter space (e.g., different mass ratios, orbital separations, mass-transfer histories, and common-envelope outcomes). These effects can produce significant sightline-to-sightline variance in shell structure and, therefore, in DM and RM. Extending the model to multidimensional calculations and a broader binary-model grid will be essential for quantifying these geometric and progenitor-channel systematics.

    \item The apparent rarity of FRBs with a clear SNR-like secular DM signature can be naturally understood as a short observational window effect, consistent with the population-level transient-association and timescale arguments presented by \citet{Dong2025}.
\end{itemize}

\section{Summary}
\label{sec:conclusions}

In this paper, we develop a framework to quantify the local SNR contribution to FRB DM using 1D HD+NEI simulations for both $11\,M_\odot$ and $30\,M_\odot$ progenitors in single-star (SS) and binary-stripped (BS) channels. Our main conclusions are as follows.

\begin{itemize}
    \item Considering only the shocked region, the DM contribution is limited (${\lesssim}10\,{\rm pc\,cm^{-3}}$) and is especially negligible for BS models. By contrast, even under this conservative setup---using the simulated shocked-plasma ionization state and ram-pressure-based magnetic amplification---the predicted RM can still reach the high values observed in FRB~20121102 for the $11\,M_\odot$ SS model. Within our model grid, this shocked-region RM comparison selects only the $11\,M_\odot$ SS case and disfavors the $30\,M_\odot$ SS and BS progenitor models as primary explanations of the FRB~20121102 RM data.

    \item For DM evolution, the dominant contribution comes from the unshocked ejecta. Its early-time decline is governed by free expansion and follows ${\rm DM}\propto t^{-\alpha}$ with $\alpha\simeq 1.8$--$1.9$ (slightly shallower than 2), while the late-time behavior transitions toward a regime increasingly influenced by the approximately constant-density ambient ISM/CSM contribution.

    \item BS models generally yield smaller DM than SS models at the same $M_{\rm ZAMS}$ because H-envelope stripping reduces ejecta mass and therefore the available electron column in the unshocked ejecta, which dominates ${\rm DM}_{\rm SNR}$ in our simulations. This SS--BS contrast is additionally modulated by reverse-shock development and CSM structure: BS models (and also the massive SS model) can evolve in lower-density wind-blown cavities, delaying strong reverse-shock formation at ages of a few hundred years and further reducing the shocked-ejecta DM contribution.

    \item The total optical depth is generally dominated by free--free absorption. Using the condition $\tau_{\rm ff}=1$, we infer $t_{\rm esc}\lesssim70\,$yr across the explored ionization configurations. For weakly ionized ejecta ($\chi_{e,\rm unej}\sim0.01$), the SNR is nearly transparent to FRB emission from very early times.

    \item {Using the observed DM decay rates of FRB~20190520B, FRB~20121102, and FRB~20220529A to constrain $t_{\rm occur}$, we find $t_{\rm occur}\lesssim100\,$yr (measured from the SN explosion) in all fiducial models. Within the same progenitor model, FRB~20190520B occurs earliest, FRB~20220529A latest, and FRB~20121102 lies in between; in the great majority of cases, all three sources can be transparent to GHz FRB escape.}

    \item The mass dependence is not always monotonic. In our models, DM is dominated by weakly ionized unshocked ejecta, so under the low-ionization assumption (approximately fixed $\chi_e$, effectively at most one free electron per atom), larger mean atomic/electron weights in metal-rich ejecta can reduce $n_e$ and produce non-monotonic trends across progenitors. This behavior can change or even reverse when the reverse shock penetrates more ejecta and/or reverse-shock-related radiation (photoionization/PR) further ionizes heavy elements, increasing the electron yield.
    
    \item {A common conclusion from FRB~20190520B, FRB~20220529A, and FRB~20121102 is that the inferred local SNR contribution is non-negligible, spanning $\sim10$ to a few $10^2$ ${\rm pc\,cm^{-3}}$. At the fiducial-value $t_{\rm occur}$ constrained from the observed DM slopes, the inferred ${\rm DM}_{\rm SNR}$ shows a clear model split for all three sources: for FRB~20190520B, the $11\,M_\odot$ BS model gives only $\sim44.9\,{\rm pc\,cm^{-3}}$, while the other models are $\sim200\,{\rm pc\,cm^{-3}}$; for FRB~20121102, the $11\,M_\odot$ BS model gives $\sim26.2\,{\rm pc\,cm^{-3}}$, while the others are $\sim100\,{\rm pc\,cm^{-3}}$; and for FRB~20220529A, all fiducial models shift downward to $\sim12.4$--$36.1\,{\rm pc\,cm^{-3}}$. The significantly smaller DM in the $11\,M_\odot$ BS case is mainly due to its much smaller ejecta mass. Therefore, the early CCSN/SNR local contribution to ${\rm DM}_{\rm source}$ can materially affect FRB cosmological inferences.} {If non-repeating FRBs are associated with older source populations, our results suggest that the local SNR contribution to DM would be significantly reduced at late times, typically dropping to values of order a few pc cm$^{-3}$ or less. In this regime, the ${\rm DM}_{\rm source}$ term may become subdominant compared to other contributions. However, we emphasize that the age distribution and physical origins of non-repeating FRBs remain uncertain, and the contribution from sources may not be universally negligible.}

    \item For FRB~20121102, the coexistence of strong RM evolution and a two-stage DM-slope evolution (early weak rise and late decline) suggests that a pure single-component interpretation is incomplete. A hybrid SNR+MWN picture is favored, in which the late-time DM decline is consistent with SNR expansion, while additional local plasma from MWN-related processes can contribute to the early-stage slope turnover and magneto-ionic variability.
\end{itemize}

These results underscore the need for physically consistent choices of progenitor and local-environment models when estimating and marginalizing the local DM term in FRB cosmology. Future work will expand the progenitor grid and include additional local components (e.g., MWN and host \ion{H}{2} regions) to further tighten constraints on ${\rm DM}_{\rm source}$.

\section*{acknowledgments}
Z.J.Z. appreciates the helpful assistance and constructive discussions from Kunihito Ioka, Daisuke Toyouchi, Zhenyin Zhao, Shengyu Yan, Yoshiyuki Inoue and Shotaro Yamasaki, and also thanks the Keihan Astrophysics Meeting for facilitating this meaningful collaboration.
This work used computational resources provided by the SQUID at the D3 Center of the University of Osaka, through the HPCI System Research Project (Project IDs: hp240141, hp250119, hp260040). This work is supported by the MEXT/JSPS KAKENHI Grant Numbers JP22K21349, 24H00002, 24H00241, and 25K01032 (K.N.).

\section*{Data and Code Availability}

The SNR simulations presented in this work are performed using a customized code developed for specific scientific applications and are not publicly available at this time.

However, the derived data products supporting the main results of this paper, including the time evolution of DM and RM, as well as the numerical data underlying the figures, can be made available from the corresponding author upon reasonable request. These data products include, for example, the time-dependent DM and RM profiles used in the analysis and plotting of the main results.

The stellar evolution inputs used in this work are based on publicly available MESA models. The corresponding data products are available via Zenodo \href{https://doi.org/10.5281/zenodo.5929871}{DOI: 10.5281/zenodo.5929871} \citep{farmer_2023_5929871}.

The processed datasets and figure-reproduction materials used in this work, including DM, RM, optical-depth measurements, and theoretical benchmark comparison data, are publicly available through Zenodo at \dataset[DOI: 10.5281/zenodo.20486710]{https://doi.org/10.5281/zenodo.20486710}. The source code used to generate these data products is available at \url{https://github.com/zhaojosephzhang/SNR_FRB}.

In addition, the ARCOS (Analysis Repository for Cosmological FRB Studies in Simulations) software framework, developed for FRB cosmology and simulation-based studies using the CROCODILE simulations, is publicly available through Zenodo at \href{https://doi.org/10.5281/zenodo.20487842}{DOI: 10.5281/zenodo.20487842} \citep{10.5281/zenodo.20487842} and GitHub at \url{https://github.com/zhaojosephzhang/ARCOS}.

\newpage
\appendix
\section{Gaunt Factor for Free–Free Absorption}

\label{App:Gaunt}
The free–free absorption coefficient used in Eq.~(6) depends on the 
velocity–averaged Gaunt factor $g_{\rm ff}(\nu, T_{\rm e}, Z)$, which corrects 
the classical bremsstrahlung cross section for quantum and kinematic effects.  
The asymptotic behavior of the Gaunt factor was derived by \citet{NovikovThorne1973} and later summarized in Fig.~5.2 of \citet{RybickiLightman1979}.  
Its dependence can be expressed in terms of two dimensionless parameters:

\begin{equation}
u = \frac{h\nu}{k_{\rm B}T_{\rm e}}, 
\qquad
\eta = \frac{k_{\rm B}T_{\rm e}}{Z^{2} R_{\rm y}},
\tag{A1}
\end{equation}
where $R_{\rm y}=13.6{\rm \,eV}$ is the Rydberg energy.

The Gaunt factor admits three analytic limits:

\paragraph{(1) Large–angle region}
The transition to large–angle Rutherford scattering occurs when the impact 
parameter reaches the regime where
\begin{equation}
1\;\gtrsim\; u \;\gtrsim\; \eta^{1/2}.
\end{equation}
In this regime the Gaunt factor is of order unity,
\begin{equation}
g_{\rm ff}^{\rm (LA)} \simeq 1.
\tag{A2}
\end{equation}

\paragraph{(2) Small–angle classical region ($u\ll 1$ and $\eta < 1$)}
\begin{equation}
g_{\rm ff}^{\rm (cl)}
= \frac{\sqrt{3}}{\pi}
  \ln\!\left[
      \frac{1}{2 \,\xi^{5/2}}
      \left(\frac{k_{\rm B}T_{\rm e}}{Z^{2}R_{\rm y}}\right)^{1/2}
      \left(\frac{k_{\rm B}T_{\rm e}}{h\nu}\right)
  \right],
\tag{A3}\label{eq:gff_cl}
\end{equation}
where $\xi = e^{\gamma} \simeq 1.781$ and $\gamma$ is Euler’s constant.

\paragraph{(3) Small–angle U.P. region ($u\ll 1$ and $\eta > 1$)}
\begin{equation}
g_{\rm ff}^{\rm (UP)}
= \frac{\sqrt{3}}{\pi}
  \ln\!\left[
      \frac{1}{\xi^{2}}
      \left(\frac{k_{\rm B}T_{\rm e}}{h\nu}\right)
  \right].
\tag{A4}
\end{equation}

Because the SNR spans a broad range of temperatures 
($T_{\rm e}\sim 10^{4}$--$10^{7}$\,K) and the FRB frequencies lie in the GHz band, 
different radial layers naturally fall into different asymptotic regimes.  
To ensure numerical smoothness, we evaluate the Gaunt factor using a blended 
interpolation across the boundaries $u=\eta^{1/2}$ and $\eta=1$. Throughout this work, we compute $g_{\rm ff}$ at $\nu=10^{9}\,{\rm Hz}$.

\section{Characteristic Scales and Self-Similar Constants in the Wind Case}
\label{app:wind_self_similar}

In this appendix, we summarize the characteristic scales and dimensionless constants adopted in the self-similar driven wave (SSDW) solution for a supernova remnant (SNR) expanding into a wind-like circumstellar medium (CSM) with density profile $\rho_{\rm w} \propto r^{-2}$ ($s=2$). 

\subsection{Characteristic Radius and Timescale}
\label{APP:ch}
For a wind environment characterized by
$K \equiv \dot{M}/(4\pi v_{\rm w})$, we define the normalized parameter
$K_{13} \equiv K/(10^{13}\,{\rm g\,cm^{-1}})$.
Following the standard SSDW scaling (e.g., \citealt{Tang&Chevalier2017}; \citealt{ZYZhao2021a}), the characteristic radius and timescale are given by
\begin{equation}
R_{\rm ch} \simeq
6.58\times10^{2}\,\mathrm{pc}\;
M_1
\left(\frac{5.1}{K_{13}}\right),
\label{eq:Rch}
\end{equation}
and
\begin{equation}
t_{\rm ch} \simeq
2.86\times10^{5}\,\mathrm{yr}\;
E_{51}^{-1/2}\,
M_1^{3/2}
\left(\frac{5.1}{K_{13}}\right),
\label{eq:tch}
\end{equation}
The numerical factor $5.1$ arises from the fiducial normalization
$\dot M=10^{-5}\,M_\odot\,\mathrm{yr^{-1}}$ and
$v_w=10\,\mathrm{km\,s^{-1}}$, for which
$K\simeq 5.1\times10^{13}\,\mathrm{g\,cm^{-1}}$.
These characteristic scales are used to normalize the self-similar evolution of the forward shock, reverse shock, and contact discontinuity.

\subsection{Contact Discontinuity and the Constant $\zeta_c$}
\label{APP:Contact_constant}
We follow the standard parametrization of the ejecta and ambient-medium density profiles (e.g., TM99), writing
\begin{equation}
\rho(r,t)=
\begin{cases}
\rho_{\rm ej}(r,t), & r\le R_{\rm ej}(t),\\[4pt]
\rho_{\rm a}(r), & r>R_{\rm ej}(t),
\end{cases}
\label{eq:rho_piecewise}
\end{equation}
where the ambient medium is described by a power law
\begin{equation}
\rho_{\rm a}(r)=\eta_s\,r^{-s},
\label{eq:rho_ambient}
\end{equation}
and the freely expanding ejecta take the form
\begin{equation}
\rho_{\rm ej}(r,t)=\frac{M_{\rm ej}}{R_{\rm ej}^3(t)}\,f\!\left(\frac{r}{R_{\rm ej}(t)}\right).
\label{eq:rho_ejecta_general}
\end{equation}
Here $R_{\rm ej}(t)$ is the outer radius of the ejecta and
$w\equiv r/R_{\rm ej}$.
The ejecta structure function $f(w)$ is specified by a flat inner core and
a power-law outer envelope,
\begin{equation}
f(w)=
\begin{cases}
f_0, & 0\le w\le w_{\rm core},\\[4pt]
f_0\left(\dfrac{w_{\rm core}}{w}\right)^{n}, & w_{\rm core}\le w\le 1,
\end{cases}
\qquad (n>5),
\label{eq:f_w}
\end{equation}
where $w_{\rm core}$ denotes the fractional core radius and $n$ is the outer
power-law index of the ejecta.
The normalization constant $f_0$ is fixed by the mass conservation condition
$\int_0^1 4\pi w^2 f(w)\,dw = 1$, yielding
\begin{equation}
f_0=\frac{3}{4\pi w_{\rm core}^n}
\left[
\frac{1-n/3}{1-(n/3)\,w_{\rm core}^{\,3-n}}
\right].
\label{eq:f0_norm}
\end{equation}

The radius of the contact discontinuity can be written as \citep{Tang&Chevalier2017}
\begin{equation}
R_c^*(t^*)=\left[\left(\lambda_c\, t^*\right)^{-\tilde{\alpha}}
+\left(c\, t^{*\beta}\right)^{-\tilde{\alpha}}
\right]^{-1/\tilde{\alpha}},
\label{eq:Rc_star_TC17}
\end{equation}
where $R_c^*\equiv R_c/R_{\rm ch}$ and $t^*\equiv t/t_{\rm ch}$. Here $\tilde{\alpha}$ is the interpolation exponent in the fitting formula of \citet{Tang&Chevalier2017} and is unrelated to the DM time-scaling index $\alpha$ (defined in the main text via ${\rm DM}\propto t^{-\alpha}$).
This expression interpolates between the early-time free-expansion behavior $R_c^*\simeq \lambda_c t^*$ and the self-similar ejecta-dominated regime
$R_c^*\simeq c\,t^{*\beta}$, with $\beta=(n-3)/(n-s)$.
In the SSDW limit ($t\ll t_{\rm ch}$), $R_c$ reduce to
\begin{equation}
R_c(t)\simeq \zeta_c\,R_{\rm ch}
\left(\frac{t}{t_{\rm ch}}\right)^{\frac{n-3}{n-s}},
\label{eq:Rc_self_similar_limit}
\end{equation}
where $s=2$ for the wind case.
The dimensionless normalization constant $\zeta_c$ encapsulates the detailed
structure of the self-similar solution and is given by
\begin{equation}
\zeta_c = \left(
A\,f_0\,w_{\rm core}^{\,n}\,\lambda_c^{\,n-3}
\right)^{\frac{1}{n-s}}.
\label{eq:zeta_c_def}
\end{equation}
where the constant $A$ is a dimensionless coefficient that depends on $(n,s)$, with values tabulated in \citet{Chevalier1982}. In this work, we adopt $A=0.067$, corresponding to $n=10$ and $s=2$. The parameter $\lambda_c$ is a dimensionless constant related to the density profile of the ejecta and can be expressed as
\begin{equation}
\lambda_c^2 =
2\,w_{\rm core}^{-2}
\left(\frac{5-n}{3-n}\right)
\left(
\frac{w_{\rm core}^{\,n-3}-n/3}{w_{\rm core}^{\,n-5}-n/5}
\right).
\label{eq:lambda_c_def}
\end{equation}

\section{Analytic-Model Comparison Using Simulated Mean Molecular Weights}
\label{app:analytic_mu_comparison}

In these appendix figures, the analytic prescriptions are evaluated using the composition-dependent quantities extracted from the simulations. For the unshocked ejecta and unshocked CSM we adopt $\mu_a$, under the assumption that each atom in the unshocked ejecta contributes only one free electron (hence $\mu_e=\mu_a$), while for the shocked region we use $\mu_e$ and $\mu_m$ computed from the self-consistent ionization evolution. In the PG18 wind case, the time-dependent $\mu_e(t)$ in the shocked region introduces an early-time variation that is qualitatively consistent with the NEI features seen in our shocked-region simulations. By contrast, the PG18 shocked-ionized case shows a nearly negligible NEI imprint because the DM evolution depends primarily on the ratio $\mu_m/\mu_e$ (see Eq.~\ref{eq:dm_shock}); this ratio varies less than $\mu_e$ or $\mu_m$ individually, thereby suppressing the apparent ionization-driven modulation in ${\rm DM}(t)$.

\begin{figure*}
    \centering
    \begin{minipage}{0.48\linewidth}
        \centering
        \includegraphics[width=\linewidth]{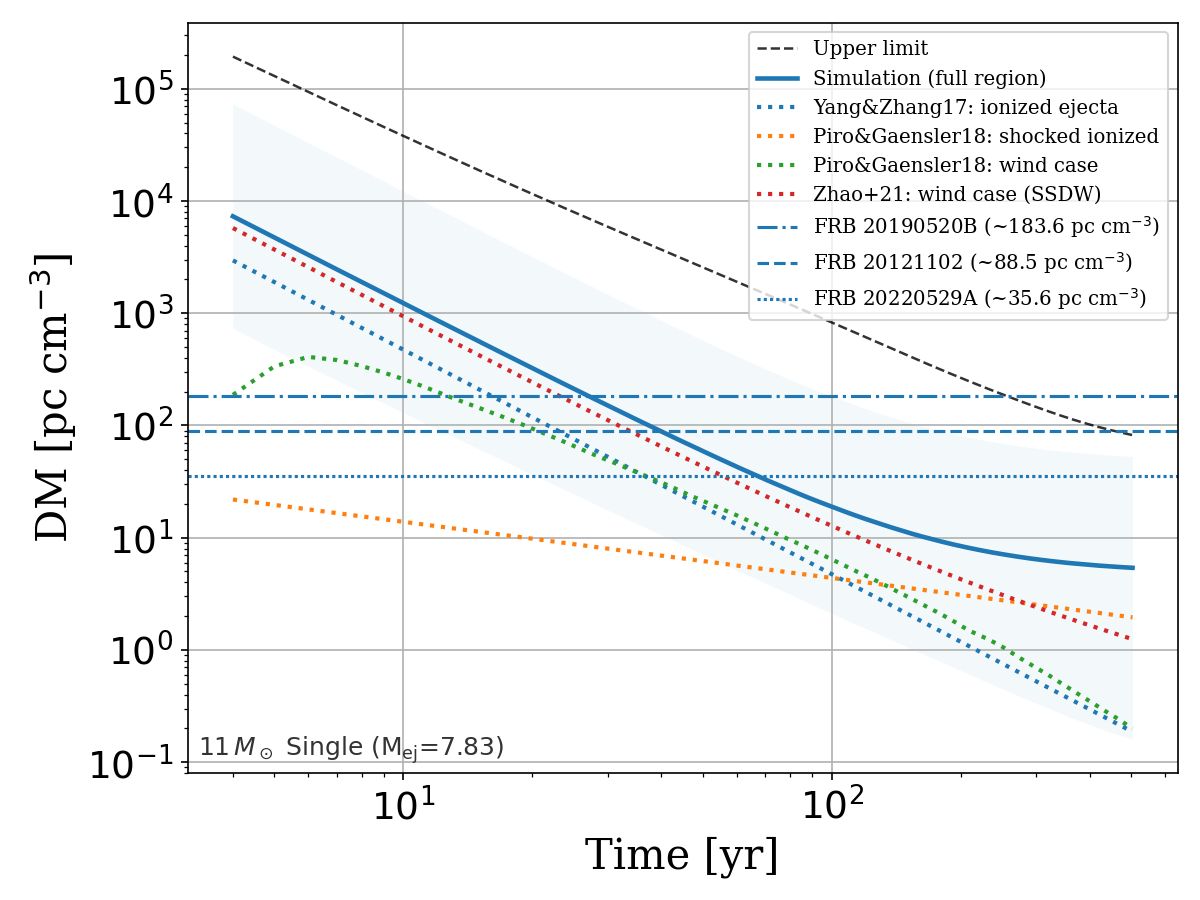}
    \end{minipage}
    \hfill
    \begin{minipage}{0.48\linewidth}
        \centering
        \includegraphics[width=\linewidth]{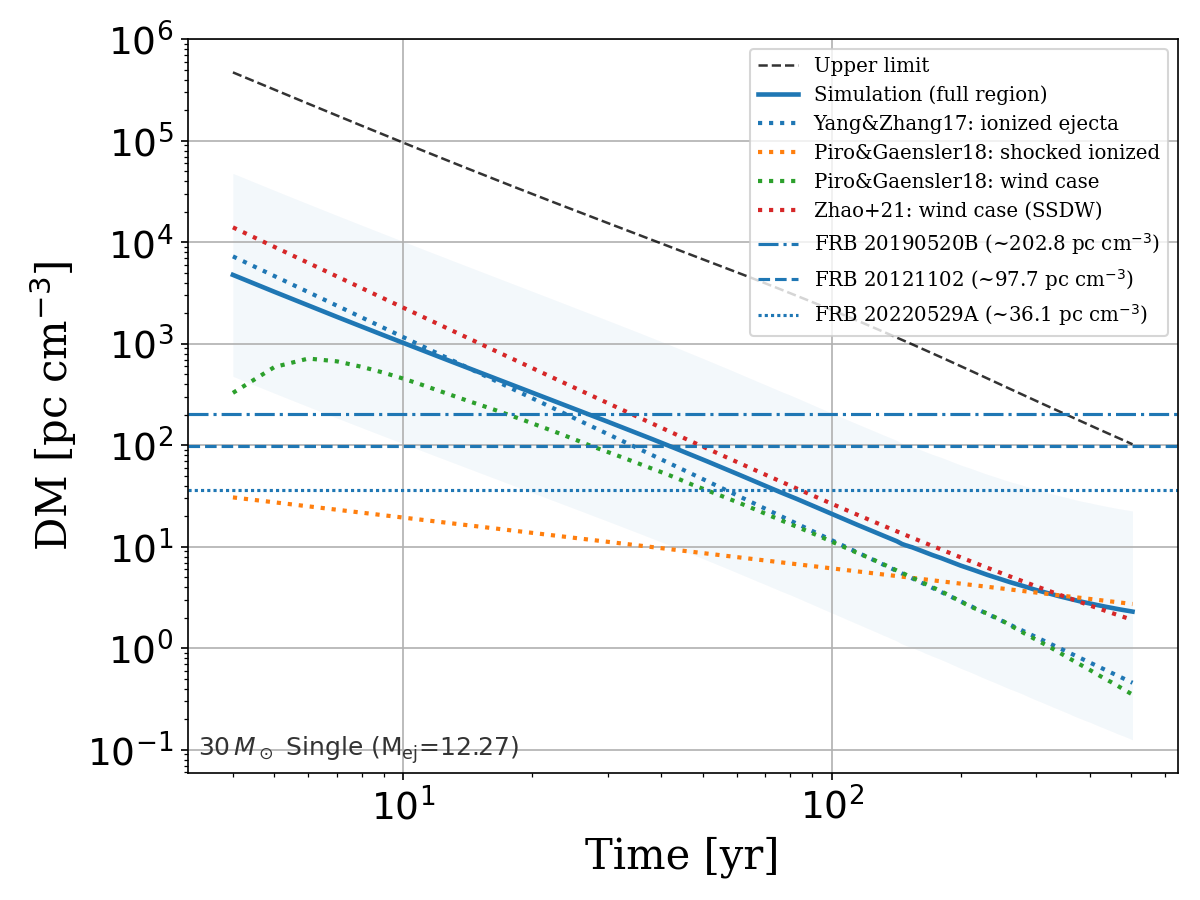}
    \end{minipage}

    \vspace{-0.25cm}

    \begin{minipage}{0.48\linewidth}
        \centering
        \includegraphics[width=\linewidth]{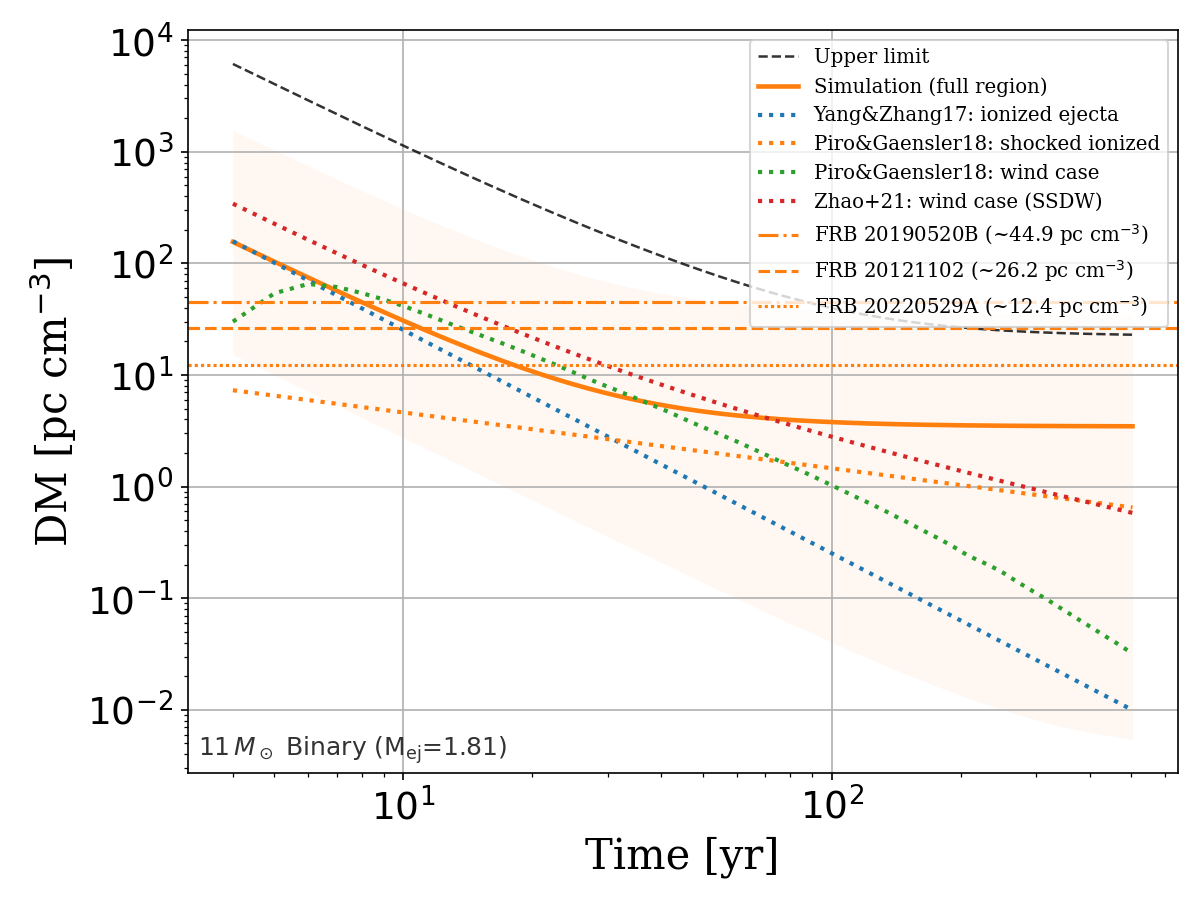}
    \end{minipage}
    \hfill
    \begin{minipage}{0.48\linewidth}
        \centering
        \includegraphics[width=\linewidth]{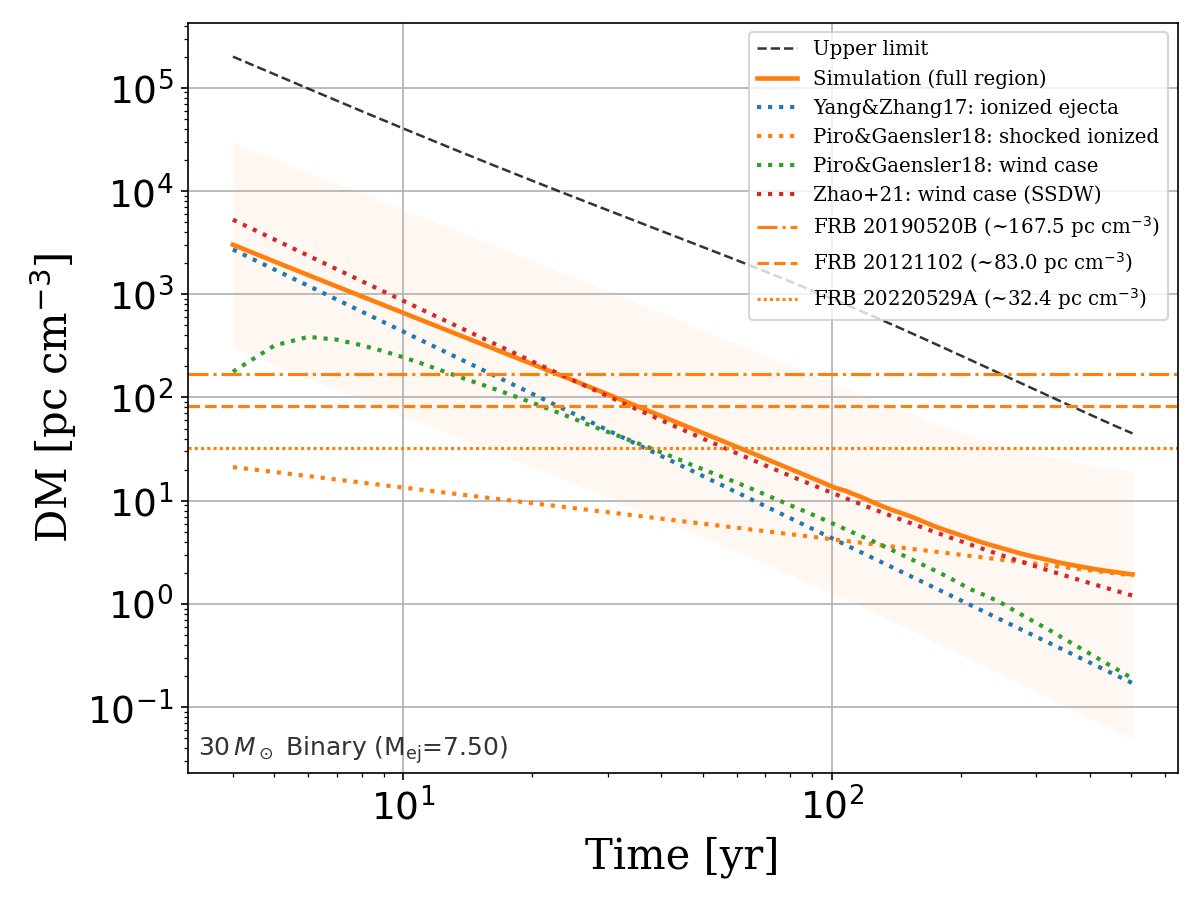}
    \end{minipage}

    \caption{
    Same format as Fig.~\ref{fig:dm_single_binary_11}, but the analytic prescriptions adopt the simulated mean molecular weights ($\mu_a,\,\mu_e,\,\mu_m$) for each model, rather than the fiducial values used in Section~\ref{subsec:SNR_derivitave_evolution}.
    }
    \label{fig:dm_single_binary_11_app}
\end{figure*}

\section{Occurrence-Time and Curvature Table for FRB~20121102}
\label{app:table_20121102}

In this appendix table, $t_{\rm occur}$ and the DM-derivative quantities are evaluated at the epoch where $\mathrm{dDM}/\mathrm{d}t=-3.93\,{\rm pc\,cm^{-3}\,yr^{-1}}$, corresponding to the second-stage DM evolution of FRB~20121102.

\begin{table*}[!t]
\centering
\caption{Same format as Table~\ref{tab:toccur_tesc}, but for FRB~20121102. Here $t_{\rm occur}$ and DM-derivative quantities are defined at $\mathrm{dDM}/\mathrm{d}t=-3.93\,{\rm pc\,cm^{-3}\,yr^{-1}}$ (the second-stage DM evolution of FRB~20121102).}
\label{tab:toccur_tesc_20121102}
\begin{tabular}{llcccc}
\toprule
Ionization config. & Quantity & $11\,M_\odot$ SS & $11\,M_\odot$ BS & $30\,M_\odot$ SS & $30\,M_\odot$ BS\\
\midrule
\multirow{4}{*}{FF: $\chi_{e,\rm unej}=1$, $\chi_{e,\rm ISM}=1$}
& $t_{\rm esc}$ (yr) & 32 & 6 & 62 & 46\\
& $t_{\rm occur}$ (yr) & 87 & 24 & 98 & 84\\
& $\mathrm{d}^2{\rm DM}/\mathrm{d}t^2$ (pc\,cm$^{-3}$\,yr$^{-2}$) & 0.21 & 0.50 & 0.15 & 0.10\\
& $\left|\mathrm{d}^2{\rm DM}/\mathrm{d}t^2\right|_{\rm est}$ (pc\,cm$^{-3}$\,yr$^{-2}$) & 0.09 & 0.33 & 0.07 & 0.09\\
\midrule
\multirow{4}{*}{HF: $\chi_{e,\rm unej}=0.1$, $\chi_{e,\rm ISM}=1$}
& $t_{\rm esc}$ (yr) & 13 & 0 & 17 & 13\\
& $t_{\rm occur}$ (yr) & 40 & 11 & 42 & 35\\
& $\mathrm{d}^2{\rm DM}/\mathrm{d}t^2$ (pc\,cm$^{-3}$\,yr$^{-2}$) & 0.28 & 1.07 & 0.24 & 0.29\\
& $\left|\mathrm{d}^2{\rm DM}/\mathrm{d}t^2\right|_{\rm est}$ (pc\,cm$^{-3}$\,yr$^{-2}$) & 0.19 & 0.69 & 0.17 & 0.21\\
\midrule
\multirow{3}{*}{MF: $\chi_{e,\rm unej}=0.01$, $\chi_{e,\rm ISM}=1$}
& $t_{\rm esc}$ (yr) & 5 & 0 & 4 & 0\\
& $t_{\rm occur}$ (yr) & 19 & 5 & 17 & 15\\
& $\mathrm{d}^2{\rm DM}/\mathrm{d}t^2$ (pc\,cm$^{-3}$\,yr$^{-2}$) & 0.60 & 1.43 & 0.65 & 0.67\\
& $\left|\mathrm{d}^2{\rm DM}/\mathrm{d}t^2\right|_{\rm est}$ (pc\,cm$^{-3}$\,yr$^{-2}$) & 0.40 & 1.56 & 0.45 & 0.46\\
\midrule
\multirow{4}{*}{FH: $\chi_{e,\rm unej}=1$, $\chi_{e,\rm ISM}=0.1$}
& $t_{\rm esc}$ (yr) & 32 & 6 & 62 & 46\\
& $t_{\rm occur}$ (yr) & 87 & 24 & 98 & 84\\
& $\mathrm{d}^2{\rm DM}/\mathrm{d}t^2$ (pc\,cm$^{-3}$\,yr$^{-2}$) & 0.21 & 0.50 & 0.15 & 0.10\\
& $\left|\mathrm{d}^2{\rm DM}/\mathrm{d}t^2\right|_{\rm est}$ (pc\,cm$^{-3}$\,yr$^{-2}$) & 0.09 & 0.33 & 0.07 & 0.09\\
\midrule
\multirow{3}{*}{HH: $\chi_{e,\rm unej}=0.1$, $\chi_{e,\rm ISM}=0.1$}
& $t_{\rm esc}$ (yr) & 13 & 0 & 17 & 13\\
& $t_{\rm occur}$ (yr) & 40 & 11 & 42 & 35\\
& $\mathrm{d}^2{\rm DM}/\mathrm{d}t^2$ (pc\,cm$^{-3}$\,yr$^{-2}$) & 0.22 & 1.08 & 0.24 & 0.29\\
& $\left|\mathrm{d}^2{\rm DM}/\mathrm{d}t^2\right|_{\rm est}$ (pc\,cm$^{-3}$\,yr$^{-2}$) & 0.20 & 0.69 & 0.17 & 0.21\\
\midrule
\multirow{3}{*}{MH: $\chi_{e,\rm unej}=0.01$, $\chi_{e,\rm ISM}=0.1$}
& $t_{\rm esc}$ (yr) & 5 & 0 & 4 & 0\\
& $t_{\rm occur}$ (yr) & 19 & 5 & 17 & 15\\
& $\mathrm{d}^2{\rm DM}/\mathrm{d}t^2$ (pc\,cm$^{-3}$\,yr$^{-2}$) & 0.60 & 1.43 & 0.65 & 0.67\\
& $\left|\mathrm{d}^2{\rm DM}/\mathrm{d}t^2\right|_{\rm est}$ (pc\,cm$^{-3}$\,yr$^{-2}$) & 0.40 & 1.56 & 0.45 & 0.46\\
\bottomrule
\end{tabular}
\end{table*}

\section{{Occurrence-Time and Curvature Table for FRB~20220529A}}
\label{app:table_20220529A}

{In this appendix table, $t_{\rm occur}$ and the DM-derivative quantities are evaluated at the epoch where $\mathrm{dDM}/\mathrm{d}t=-0.88\,{\rm pc\,cm^{-3}\,yr^{-1}}$, corresponding to the secular DM evolution of FRB~20220529A.}

\begin{table*}[!t]
\centering
\caption{{Same format as Table~\ref{tab:toccur_tesc}, but for FRB~20220529A. Here $t_{\rm occur}$ and DM-derivative quantities are defined at $\mathrm{dDM}/\mathrm{d}t=-0.88\,{\rm pc\,cm^{-3}\,yr^{-1}}$.}}
\label{tab:toccur_tesc_20220529A}
\begin{tabular}{llcccc}
\toprule
Ionization config. & Quantity & $11\,M_\odot$ SS & $11\,M_\odot$ BS & $30\,M_\odot$ SS & $30\,M_\odot$ BS\\
\midrule
\multirow{4}{*}{FF: $\chi_{e,\rm unej}=1$, $\chi_{e,\rm ISM}=1$}
& $t_{\rm esc}$ (yr) & 32 & 6 & 62 & 46\\
& $t_{\rm occur}$ (yr) & 146 & 41 & 150 & 138\\
& $\mathrm{d}^2{\rm DM}/\mathrm{d}t^2$ (pc\,cm$^{-3}$\,yr$^{-2}$) & 0.02 & 0.06 & -0.05 & 0.01\\
& $\left|\mathrm{d}^2{\rm DM}/\mathrm{d}t^2\right|_{\rm est}$ (pc\,cm$^{-3}$\,yr$^{-2}$) & 0.01 & 0.04 & 0.05 & 0.01\\
\midrule
\multirow{4}{*}{HF: $\chi_{e,\rm unej}=0.1$, $\chi_{e,\rm ISM}=1$}
& $t_{\rm esc}$ (yr) & 13 & 0 & 17 & 13\\
& $t_{\rm occur}$ (yr) & 67 & 18 & 74 & 61\\
& $\mathrm{d}^2{\rm DM}/\mathrm{d}t^2$ (pc\,cm$^{-3}$\,yr$^{-2}$) & 0.04 & 0.15 & 0.02 & 0.04\\
& $\left|\mathrm{d}^2{\rm DM}/\mathrm{d}t^2\right|_{\rm est}$ (pc\,cm$^{-3}$\,yr$^{-2}$) & 0.03 & 0.10 & 0.02 & 0.03\\
\midrule
\multirow{4}{*}{MF: $\chi_{e,\rm unej}=0.01$, $\chi_{e,\rm ISM}=1$}
& $t_{\rm esc}$ (yr) & 5 & 0 & 4 & 0\\
& $t_{\rm occur}$ (yr) & 32 & 8 & 31 & 26\\
& $\mathrm{d}^2{\rm DM}/\mathrm{d}t^2$ (pc\,cm$^{-3}$\,yr$^{-2}$) & 0.09 & 0.39 & 0.07 & 0.09\\
& $\left|\mathrm{d}^2{\rm DM}/\mathrm{d}t^2\right|_{\rm est}$ (pc\,cm$^{-3}$\,yr$^{-2}$) & 0.05 & 0.21 & 0.05 & 0.06\\
\midrule
\multirow{4}{*}{FH: $\chi_{e,\rm unej}=1$, $\chi_{e,\rm ISM}=0.1$}
& $t_{\rm esc}$ (yr) & 32 & 6 & 62 & 46\\
& $t_{\rm occur}$ (yr) & 146 & 41 & 150 & 138\\
& $\mathrm{d}^2{\rm DM}/\mathrm{d}t^2$ (pc\,cm$^{-3}$\,yr$^{-2}$) & 0.02 & 0.06 & -0.05 & 0.05\\
& $\left|\mathrm{d}^2{\rm DM}/\mathrm{d}t^2\right|_{\rm est}$ (pc\,cm$^{-3}$\,yr$^{-2}$) & 0.01 & 0.04 & 0.01 & 0.01\\
\midrule
\multirow{4}{*}{HH: $\chi_{e,\rm unej}=0.1$, $\chi_{e,\rm ISM}=0.1$}
& $t_{\rm esc}$ (yr) & 13 & 0 & 17 & 13\\
& $t_{\rm occur}$ (yr) & 67 & 18 & 74 & 61\\
& $\mathrm{d}^2{\rm DM}/\mathrm{d}t^2$ (pc\,cm$^{-3}$\,yr$^{-2}$) & 0.04 & 0.15 & 0.02 & 0.04\\
& $\left|\mathrm{d}^2{\rm DM}/\mathrm{d}t^2\right|_{\rm est}$ (pc\,cm$^{-3}$\,yr$^{-2}$) & 0.03 & 0.10 & 0.02 & 0.03\\
\midrule
\multirow{4}{*}{MH: $\chi_{e,\rm unej}=0.01$, $\chi_{e,\rm ISM}=0.1$}
& $t_{\rm esc}$ (yr) & 5 & 0 & 4 & 0\\
& $t_{\rm occur}$ (yr) & 32 & 8 & 31 & 26\\
& $\mathrm{d}^2{\rm DM}/\mathrm{d}t^2$ (pc\,cm$^{-3}$\,yr$^{-2}$) & 0.08 & 0.39 & 0.007 & 0.09\\
& $\left|\mathrm{d}^2{\rm DM}/\mathrm{d}t^2\right|_{\rm est}$ (pc\,cm$^{-3}$\,yr$^{-2}$) & 0.05 & 0.21 & 0.05 & 0.06\\
\bottomrule
\end{tabular}
\end{table*}

\clearpage	
\bibliography{ms.bib}{}

@ARTICLE{KHRYKIN2024,
       author = {{Khrykin}, Ilya S. and {Ata}, Metin and {Lee}, Khee-Gan and {Simha}, Sunil and {Huang}, Yuxin and {Prochaska}, J. Xavier and {Tejos}, Nicolas and {Bannister}, Keith W. and {Cooke}, Jeff and {Day}, Cherie K. and {Deller}, Adam and {Glowacki}, Marcin and {Gordon}, Alexa C. and {James}, Clancy W. and {Marnoch}, Lachlan and {Shannon}, Ryan. M. and {Zhang}, Jielai and {Bernales-Cortes}, Lucas},
        title = "{FLIMFLAM DR1: The First Constraints on the Cosmic Baryon Distribution from Eight Fast Radio Burst Sight Lines}",
      journal = {\apj},
         year = 2024,
        month = oct,
       volume = {973},
       number = {2},
          eid = {151},
        pages = {151},
          doi = {10.3847/1538-4357/ad6567},
archivePrefix = {arXiv},
       eprint = {2402.00505},
 primaryClass = {astro-ph.GA},
       adsurl = {https://ui.adsabs.harvard.edu/abs/2024ApJ...973..151K}
}

@ARTICLE{Lee2022,
       author = {{Lee}, Khee-Gan and {Ata}, Metin and {Khrykin}, Ilya S. and {Huang}, Yuxin and {Prochaska}, J. Xavier and {Cooke}, Jeff and {Zhang}, Jielai and {Batten}, Adam},
        title = "{Constraining the Cosmic Baryon Distribution with Fast Radio Burst Foreground Mapping}",
      journal = {\apj},
         year = 2022,
        month = mar,
       volume = {928},
       number = {1},
          eid = {9},
        pages = {9},
          doi = {10.3847/1538-4357/ac4f62},
archivePrefix = {arXiv},
       eprint = {2109.00386},
 primaryClass = {astro-ph.CO},
       adsurl = {https://ui.adsabs.harvard.edu/abs/2022ApJ...928....9L}
}

@ARTICLE{Lee2023,
       author = {{Lee}, Khee-Gan and {Khrykin}, Ilya S. and {Simha}, Sunil and {Ata}, Metin and {Huang}, Yuxin and {Prochaska}, J. Xavier and {Tejos}, Nicolas and {Cooke}, Jeff and {Nagamine}, Kentaro and {Zhang}, Jielai},
        title = "{The FRB 20190520B Sight Line Intersects Foreground Galaxy Clusters}",
      journal = {\apjl},
         year = 2023,
        month = sep,
       volume = {954},
       number = {1},
          eid = {L7},
        pages = {L7},
          doi = {10.3847/2041-8213/acefb5},
archivePrefix = {arXiv},
       eprint = {2306.05403},
 primaryClass = {astro-ph.GA},
       adsurl = {https://ui.adsabs.harvard.edu/abs/2023ApJ...954L...7L}
}

@ARTICLE{Lorimer2007,
       author = {{Lorimer}, D.~R. and {Bailes}, M. and {McLaughlin}, M.~A. and {Narkevic}, D.~J. and {Crawford}, F.},
        title = "{A Bright Millisecond Radio Burst of Extragalactic Origin}",
      journal = {Science},
         year = 2007,
        month = nov,
       volume = {318},
       number = {5851},
        pages = {777},
          doi = {10.1126/science.1147532},
archivePrefix = {arXiv},
       eprint = {0709.4301},
 primaryClass = {astro-ph},
       adsurl = {https://ui.adsabs.harvard.edu/abs/2007Sci...318..777L}
}

@ARTICLE{Macquart2020Nature,
       author = {{Macquart}, J. -P. and {Prochaska}, J.~X. and {McQuinn}, M. and {Bannister}, K.~W. and {Bhandari}, S. and {Day}, C.~K. and {Deller}, A.~T. and {Ekers}, R.~D. and {James}, C.~W. and {Marnoch}, L. and {Os{\l}owski}, S. and {Phillips}, C. and {Ryder}, S.~D. and {Scott}, D.~R. and {Shannon}, R.~M. and {Tejos}, N.},
        title = "{A census of baryons in the Universe from localized fast radio bursts}",
      journal = {\nat},
         year = 2020,
        month = may,
       volume = {581},
       number = {7809},
        pages = {391-395},
          doi = {10.1038/s41586-020-2300-2},
archivePrefix = {arXiv},
       eprint = {2005.13161},
 primaryClass = {astro-ph.CO},
       adsurl = {https://ui.adsabs.harvard.edu/abs/2020Natur.581..391M}
}

@ARTICLE{Tendulkar2017,
       author = {{Tendulkar}, S.~P. and {Bassa}, C.~G. and {Cordes}, J.~M. and {Bower}, G.~C. and {Law}, C.~J. and {Chatterjee}, S. and {Adams}, E.~A.~K. and {Bogdanov}, S. and {Burke-Spolaor}, S. and {Butler}, B.~J. and {Demorest}, P. and {Hessels}, J.~W.~T. and {Kaspi}, V.~M. and {Lazio}, T.~J.~W. and {Maddox}, N. and {Marcote}, B. and {McLaughlin}, M.~A. and {Paragi}, Z. and {Ransom}, S.~M. and {Scholz}, P. and {Seymour}, A. and {Spitler}, L.~G. and {van Langevelde}, H.~J. and {Wharton}, R.~S.},
        title = "{The Host Galaxy and Redshift of the Repeating Fast Radio Burst FRB 121102}",
      journal = {\apjl},
         year = 2017,
        month = jan,
       volume = {834},
       number = {2},
          eid = {L7},
        pages = {L7},
          doi = {10.3847/2041-8213/834/2/L7},
archivePrefix = {arXiv},
       eprint = {1701.01100},
 primaryClass = {astro-ph.HE},
       adsurl = {https://ui.adsabs.harvard.edu/abs/2017ApJ...834L...7T}
}

@ARTICLE{Niu2022,
       author = {{Niu}, C. -H. and {Aggarwal}, K. and {Li}, D. and {Zhang}, X. and {Chatterjee}, S. and {Tsai}, C. -W. and {Yu}, W. and {Law}, C.~J. and {Burke-Spolaor}, S. and {Cordes}, J.~M. and {Zhang}, Y. -K. and {Ocker}, S.~K. and {Yao}, J. -M. and {Wang}, P. and {Feng}, Y. and {Niino}, Y. and {Bochenek}, C. and {Cruces}, M. and {Connor}, L. and {Jiang}, J. -A. and {Dai}, S. and {Luo}, R. and {Li}, G. -D. and {Miao}, C. -C. and {Niu}, J. -R. and {Anna-Thomas}, R. and {Sydnor}, J. and {Stern}, D. and {Wang}, W. -Y. and {Yuan}, M. and {Yue}, Y. -L. and {Zhou}, D. -J. and {Yan}, Z. and {Zhu}, W. -W. and {Zhang}, B.},
        title = "{A repeating fast radio burst associated with a persistent radio source}",
      journal = {\nat},
         year = 2022,
        month = jun,
       volume = {606},
       number = {7916},
        pages = {873-877},
          doi = {10.1038/s41586-022-04755-5},
archivePrefix = {arXiv},
       eprint = {2110.07418},
 primaryClass = {astro-ph.HE},
       adsurl = {https://ui.adsabs.harvard.edu/abs/2022Natur.606..873N}
}

@ARTICLE{Medlock2024,
       author = {{Medlock}, Isabel and {Nagai}, Daisuke and {Singh}, Priyanka and {Oppenheimer}, Benjamin and {Angl{\'e}s-Alc{\'a}zar}, Daniel and {Villaescusa-Navarro}, Francisco},
        title = "{Probing the Circumgalactic Medium with Fast Radio Bursts: Insights from CAMELS}",
      journal = {\apj},
         year = 2024,
        month = may,
       volume = {967},
       number = {1},
          eid = {32},
        pages = {32},
          doi = {10.3847/1538-4357/ad3070},
archivePrefix = {arXiv},
       eprint = {2403.02313},
 primaryClass = {astro-ph.GA},
       adsurl = {https://ui.adsabs.harvard.edu/abs/2024ApJ...967...32M}
}

@ARTICLE{Medlock2021,
       author = {{Medlock}, Isabel and {Cen}, Renyue},
        title = "{Dispersion measure distributions of fast radio bursts due to the intergalactic medium}",
      journal = {\mnras},
         year = 2021,
        month = apr,
       volume = {502},
       number = {3},
        pages = {3664-3669},
          doi = {10.1093/mnras/stab288},
       adsurl = {https://ui.adsabs.harvard.edu/abs/2021MNRAS.502.3664M}
}

@ARTICLE{McQuinn2014,
       author = {{McQuinn}, Matthew},
        title = "{Locating the ``Missing'' Baryons with Extragalactic Dispersion Measure Estimates}",
      journal = {\apjl},
         year = 2014,
        month = jan,
       volume = {780},
       number = {2},
          eid = {L33},
        pages = {L33},
          doi = {10.1088/2041-8205/780/2/L33},
archivePrefix = {arXiv},
       eprint = {1309.4451},
 primaryClass = {astro-ph.CO},
       adsurl = {https://ui.adsabs.harvard.edu/abs/2014ApJ...780L..33M}
}

@ARTICLE{Connor2024,
       author = {{Connor}, Liam and {Ravi}, Vikram and {Sharma}, Kritti and {Ocker}, Stella Koch and {Faber}, Jakob and {Hallinan}, Gregg and {Harnach}, Charlie and {Hellbourg}, Greg and {Hobbs}, Rick and {Hodge}, David and {Hodges}, Mark and {Kosogorov}, Nikita and {Lamb}, James and {Law}, Casey and {Rasmussen}, Paul and {Sherman}, Myles and {Somalwar}, Jean and {Weinreb}, Sander and {Woody}, David},
        title = "{A gas rich cosmic web revealed by partitioning the missing baryons}",
      journal = {arXiv e-prints},
         year = 2024,
        month = sep,
          eid = {arXiv:2409.16952},
        pages = {arXiv:2409.16952},
          doi = {10.48550/arXiv.2409.16952},
archivePrefix = {arXiv},
       eprint = {2409.16952},
 primaryClass = {astro-ph.CO},
       adsurl = {https://ui.adsabs.harvard.edu/abs/2024arXiv240916952C}
}

@article{Li2019...876...146,
doi = {10.3847/1538-4357/ab18fe},
url = {https://dx.doi.org/10.3847/1538-4357/ab18fe},
year = {2019},
month = {may},
publisher = {The American Astronomical Society},
volume = {876},
number = {2},
pages = {146},
author = {Li, Zhengxiang and Gao, He and Wei, Jun-Jie and Yang, Yuan-Pei and Zhang, Bing and Zhu, Zong-Hong},
title = {Cosmology-independent Estimate of the Fraction of Baryon Mass in the IGM from Fast Radio Burst Observations},
journal = {The Astrophysical Journal}
}

@ARTICLE{DZ2014,
       author = {{Deng}, Wei and {Zhang}, Bing},
        title = "{Cosmological Implications of Fast Radio Burst/Gamma-Ray Burst Associations}",
      journal = {\apjl},
         year = 2014,
        month = mar,
       volume = {783},
       number = {2},
          eid = {L35},
        pages = {L35},
          doi = {10.1088/2041-8205/783/2/L35},
archivePrefix = {arXiv},
       eprint = {1401.0059},
 primaryClass = {astro-ph.HE},
       adsurl = {https://ui.adsabs.harvard.edu/abs/2014ApJ...783L..35D}
}

@ARTICLE{Zheng2014,
       author = {{Zheng}, Z. and {Ofek}, E.~O. and {Kulkarni}, S.~R. and {Neill}, J.~D. and {Juric}, M.},
        title = "{Probing the Intergalactic Medium with Fast Radio Bursts}",
      journal = {\apj},
         year = 2014,
        month = dec,
       volume = {797},
       number = {1},
          eid = {71},
        pages = {71},
          doi = {10.1088/0004-637X/797/1/71},
archivePrefix = {arXiv},
       eprint = {1409.3244},
 primaryClass = {astro-ph.HE},
       adsurl = {https://ui.adsabs.harvard.edu/abs/2014ApJ...797...71Z}
}

@ARTICLE{Zhu2021,
       author = {{Zhu}, Weishan and {Feng}, Long-Long},
        title = "{The Dispersion Measure and Scattering of Fast Radio Bursts: Contributions from the Intergalactic Medium, Foreground Halos, and Hosts}",
      journal = {\apj},
         year = 2021,
        month = jan,
       volume = {906},
       number = {2},
          eid = {95},
        pages = {95},
          doi = {10.3847/1538-4357/abcb90},
archivePrefix = {arXiv},
       eprint = {2011.08519},
 primaryClass = {astro-ph.HE},
       adsurl = {https://ui.adsabs.harvard.edu/abs/2021ApJ...906...95Z}
}

@ARTICLE{Yang&Zhang2017,
       author = {{Yang}, Yuan-Pei and {Zhang}, Bing},
        title = "{Dispersion Measure Variation of Repeating Fast Radio Burst Sources}",
      journal = {\apj},
         year = 2017,
        month = sep,
       volume = {847},
       number = {1},
          eid = {22},
        pages = {22},
          doi = {10.3847/1538-4357/aa8721},
archivePrefix = {arXiv},
       eprint = {1707.02923},
 primaryClass = {astro-ph.HE},
       adsurl = {https://ui.adsabs.harvard.edu/abs/2017ApJ...847...22Y}
}

@article{ZZ2025,
doi = {10.3847/1538-4357/ae00c2},
url = {https://doi.org/10.3847/1538-4357/ae00c2},
year = {2025},
month = {oct},
publisher = {The American Astronomical Society},
volume = {993},
number = {2},
pages = {162},
author = {Zhang, Zhao Joseph and Nagamine, Kentaro and Oku, Yuri and Lee, Khee-Gan and Fukushima, Keita and Tomaru, Kazuki and Zhang, Bing and Medlock, Isabel and Nagai, Daisuke},
title = {Probing the Cosmic Baryon Distribution and the Impact of Active Galactic Nuclei Feedback with Fast Radio Bursts in CROCODILE Simulation},
journal = {The Astrophysical Journal}
}

@ARTICLE{Konietzka2025arXiv,
       author = {{Konietzka}, Ralf M. and {Connor}, Liam and {Semenov}, Vadim A. and {Beane}, Angus and {Springel}, Volker and {Hernquist}, Lars},
        title = "{Ray-tracing Fast Radio Bursts Through IllustrisTNG: Cosmological Dispersion Measures from Redshift 0 to 5.5}",
      journal = {arXiv e-prints},
         year = 2025,
        month = jul,
          eid = {arXiv:2507.07090},
        pages = {arXiv:2507.07090},
          doi = {10.48550/arXiv.2507.07090},
archivePrefix = {arXiv},
       eprint = {2507.07090},
 primaryClass = {astro-ph.CO},
       adsurl = {https://ui.adsabs.harvard.edu/abs/2025arXiv250707090K}
}

@article{Piro2016,
doi = {10.3847/2041-8205/824/2/L32},
url = {https://doi.org/10.3847/2041-8205/824/2/L32},
year = {2016},
month = {jun},
publisher = {The American Astronomical Society},
volume = {824},
number = {2},
pages = {L32},
author = {Piro, Anthony L.},
title = {THE IMPACT OF A SUPERNOVA REMNANT ON FAST RADIO BURSTS},
journal = {The Astrophysical Journal Letters}
}

@ARTICLE{Piro&Gaensler2018,
       author = {{Piro}, Anthony L. and {Gaensler}, B.~M.},
        title = "{The Dispersion and Rotation Measure of Supernova Remnants and Magnetized Stellar Winds: Application to Fast Radio Bursts}",
      journal = {\apj},
         year = 2018,
        month = jul,
       volume = {861},
       number = {2},
          eid = {150},
        pages = {150},
          doi = {10.3847/1538-4357/aac9bc},
archivePrefix = {arXiv},
       eprint = {1804.01104},
 primaryClass = {astro-ph.HE},
       adsurl = {https://ui.adsabs.harvard.edu/abs/2018ApJ...861..150P}
}

@ARTICLE{Margalit2018,
       author = {{Margalit}, Ben and {Metzger}, Brian D.},
        title = "{A Concordance Picture of FRB 121102 as a Flaring Magnetar Embedded in a Magnetized Ion-Electron Wind Nebula}",
      journal = {\apjl},
         year = 2018,
        month = nov,
       volume = {868},
       number = {1},
          eid = {L4},
        pages = {L4},
          doi = {10.3847/2041-8213/aaedad},
archivePrefix = {arXiv},
       eprint = {1808.09969},
 primaryClass = {astro-ph.HE},
       adsurl = {https://ui.adsabs.harvard.edu/abs/2018ApJ...868L...4M}
}

@article{Mahlmann2022,
doi = {10.3847/2041-8213/ac7156},
url = {https://doi.org/10.3847/2041-8213/ac7156},
year = {2022},
month = {jun},
publisher = {The American Astronomical Society},
volume = {932},
number = {2},
pages = {L20},
author = {Mahlmann, J. F. and Philippov, A. A. and Levinson, A. and Spitkovsky, A. and Hakobyan, H.},
title = {Electromagnetic Fireworks: Fast Radio Bursts from Rapid Reconnection in the Compressed Magnetar Wind},
journal = {The Astrophysical Journal Letters}
}

@ARTICLE{Yang2016,
       author = {{Yang}, Yuan-Pei and {Zhang}, Bing and {Dai}, Zi-Gao},
        title = "{Synchrotron Heating by a Fast Radio Burst in a Self-absorbed Synchrotron Nebula and Its Observational Signature}",
      journal = {\apjl},
         year = 2016,
        month = mar,
       volume = {819},
       number = {1},
          eid = {L12},
        pages = {L12},
          doi = {10.3847/2041-8205/819/1/L12},
archivePrefix = {arXiv},
       eprint = {1602.05013},
 primaryClass = {astro-ph.HE},
       adsurl = {https://ui.adsabs.harvard.edu/abs/2016ApJ...819L..12Y}
}

@ARTICLE{Yang2017ApJ...839L..25Y,
       author = {{Yang}, Yuan-Pei and {Luo}, Rui and {Li}, Zhuo and {Zhang}, Bing},
        title = "{Large Host-galaxy Dispersion Measure of Fast Radio Bursts}",
      journal = {\apjl},
         year = 2017,
        month = apr,
       volume = {839},
       number = {2},
          eid = {L25},
        pages = {L25},
          doi = {10.3847/2041-8213/aa6c2e},
archivePrefix = {arXiv},
       eprint = {1701.06465},
 primaryClass = {astro-ph.HE},
       adsurl = {https://ui.adsabs.harvard.edu/abs/2017ApJ...839L..25Y}
}

@ARTICLE{Michilli2018Natur,
       author = {{Michilli}, D. and {Seymour}, A. and {Hessels}, J.~W.~T. and {Spitler}, L.~G. and {Gajjar}, V. and {Archibald}, A.~M. and {Bower}, G.~C. and {Chatterjee}, S. and {Cordes}, J.~M. and {Gourdji}, K. and {Heald}, G.~H. and {Kaspi}, V.~M. and {Law}, C.~J. and {Sobey}, C. and {Adams}, E.~A.~K. and {Bassa}, C.~G. and {Bogdanov}, S. and {Brinkman}, C. and {Demorest}, P. and {Fernandez}, F. and {Hellbourg}, G. and {Lazio}, T.~J.~W. and {Lynch}, R.~S. and {Maddox}, N. and {Marcote}, B. and {McLaughlin}, M.~A. and {Paragi}, Z. and {Ransom}, S.~M. and {Scholz}, P. and {Siemion}, A.~P.~V. and {Tendulkar}, S.~P. and {van Rooy}, P. and {Wharton}, R.~S. and {Whitlow}, D.},
        title = "{An extreme magneto-ionic environment associated with the fast radio burst source FRB 121102}",
      journal = {\nat},
         year = 2018,
        month = jan,
       volume = {553},
       number = {7687},
        pages = {182-185},
          doi = {10.1038/nature25149},
archivePrefix = {arXiv},
       eprint = {1801.03965},
 primaryClass = {astro-ph.HE},
       adsurl = {https://ui.adsabs.harvard.edu/abs/2018Natur.553..182M}
}

@ARTICLE{Bassa2017,
       author = {{Bassa}, C.~G. and {Tendulkar}, S.~P. and {Adams}, E.~A.~K. and {Maddox}, N. and {Bogdanov}, S. and {Bower}, G.~C. and {Burke-Spolaor}, S. and {Butler}, B.~J. and {Chatterjee}, S. and {Cordes}, J.~M. and {Hessels}, J.~W.~T. and {Kaspi}, V.~M. and {Law}, C.~J. and {Marcote}, B. and {Paragi}, Z. and {Ransom}, S.~M. and {Scholz}, P. and {Spitler}, L.~G. and {van Langevelde}, H.~J.},
        title = "{FRB 121102 Is Coincident with a Star-forming Region in Its Host Galaxy}",
      journal = {\apjl},
         year = 2017,
        month = jul,
       volume = {843},
       number = {1},
          eid = {L8},
        pages = {L8},
          doi = {10.3847/2041-8213/aa7a0c},
archivePrefix = {arXiv},
       eprint = {1705.07698},
 primaryClass = {astro-ph.HE},
       adsurl = {https://ui.adsabs.harvard.edu/abs/2017ApJ...843L...8B}
}

@ARTICLE{Metzger2019,
       author = {{Metzger}, Brian D. and {Margalit}, Ben and {Sironi}, Lorenzo},
        title = "{Fast radio bursts as synchrotron maser emission from decelerating relativistic blast waves}",
      journal = {\mnras},
         year = 2019,
        month = may,
       volume = {485},
       number = {3},
        pages = {4091-4106},
          doi = {10.1093/mnras/stz700},
archivePrefix = {arXiv},
       eprint = {1902.01866},
 primaryClass = {astro-ph.HE},
       adsurl = {https://ui.adsabs.harvard.edu/abs/2019MNRAS.485.4091M}
}

@article{Niu2025,
title = {A Persistently Active Fast Radio Burst source Embedded in an Expanding Supernova Remnant},
journal = {Science Bulletin},
year = {2025},
issn = {2095-9273},
doi = {https://doi.org/10.1016/j.scib.2025.11.023},
url = {https://www.sciencedirect.com/science/article/pii/S209592732501151X},
author = {Chen-Hui Niu and Di Li and Yuan-Pei Yang and Yuhao Zhu and Yongkun Zhang and Jia-Heng Zhang and Zexin Du and Jumei Yao and Xiaoping Zheng and Pei Wang and Yi Feng and Bing Zhang and Weiwei Zhu and Wenfei Yu and Ji-An Jiang and Shi Dai and Chao-Wei Tsai and A Ming Chen and Yijun Hou and Jiarui Niu and Weiyang Wang and Chenchen Miao and Xinming Li and Junshuo Zhang}
}

@ARTICLE{Murase2016,
       author = {{Murase}, Kohta and {Kashiyama}, Kazumi and {M{\'e}sz{\'a}ros}, Peter},
        title = "{A burst in a wind bubble and the impact on baryonic ejecta: high-energy gamma-ray flashes and afterglows from fast radio bursts and pulsar-driven supernova remnants}",
      journal = {\mnras},
         year = 2016,
        month = sep,
       volume = {461},
       number = {2},
        pages = {1498-1511},
          doi = {10.1093/mnras/stw1328},
archivePrefix = {arXiv},
       eprint = {1603.08875},
 primaryClass = {astro-ph.HE},
       adsurl = {https://ui.adsabs.harvard.edu/abs/2016MNRAS.461.1498M}
}

@ARTICLE{Kashiyama2017,
       author = {{Kashiyama}, Kazumi and {Murase}, Kohta},
        title = "{Testing the Young Neutron Star Scenario with Persistent Radio Emission Associated with FRB 121102}",
      journal = {\apjl},
         year = 2017,
        month = apr,
       volume = {839},
       number = {1},
          eid = {L3},
        pages = {L3},
          doi = {10.3847/2041-8213/aa68e1},
archivePrefix = {arXiv},
       eprint = {1701.04815},
 primaryClass = {astro-ph.HE},
       adsurl = {https://ui.adsabs.harvard.edu/abs/2017ApJ...839L...3K}
}

@ARTICLE{Masui2015Natur,
       author = {{Masui}, Kiyoshi and {Lin}, Hsiu-Hsien and {Sievers}, Jonathan and {Anderson}, Christopher J. and {Chang}, Tzu-Ching and {Chen}, Xuelei and {Ganguly}, Apratim and {Jarvis}, Miranda and {Kuo}, Cheng-Yu and {Li}, Yi-Chao and {Liao}, Yu-Wei and {McLaughlin}, Maura and {Pen}, Ue-Li and {Peterson}, Jeffrey B. and {Roman}, Alexander and {Timbie}, Peter T. and {Voytek}, Tabitha and {Yadav}, Jaswant K.},
        title = "{Dense magnetized plasma associated with a fast radio burst}",
      journal = {\nat},
         year = 2015,
        month = dec,
       volume = {528},
       number = {7583},
        pages = {523-525},
          doi = {10.1038/nature15769},
archivePrefix = {arXiv},
       eprint = {1512.00529},
 primaryClass = {astro-ph.HE},
       adsurl = {https://ui.adsabs.harvard.edu/abs/2015Natur.528..523M}
}

@article{Mckinven2023,
doi = {10.3847/1538-4357/acc65f},
url = {https://doi.org/10.3847/1538-4357/acc65f},
year = {2023},
month = {jun},
publisher = {The American Astronomical Society},
volume = {950},
number = {1},
pages = {12},
author = {Mckinven, R. and Gaensler, B. M. and Michilli, D. and Masui, K. and Kaspi, V. M. and Bhardwaj, M. and Cassanelli, T. and Chawla, P. and Dong, F. (Adam) and Fonseca, E. and Leung, C. and Li, D. Z. and Ng, C. and Patel, C. and Petroff, E. and Pearlman, A. B. and Pleunis, Z. and Rafiei-Ravandi, M. and Rahman, M. and Sand, K. R. and Shin, K. and Scholz, P. and Stairs, I. H. and Smith, K. and Su, J. and Tendulkar, S.},
title = {A Large-scale Magneto-ionic Fluctuation in the Local Environment of Periodic Fast Radio Burst Source FRB 20180916B},
journal = {The Astrophysical Journal}
}

@ARTICLE{Wang2022NatCo,
       author = {{Wang}, F.~Y. and {Zhang}, G.~Q. and {Dai}, Z.~G. and {Cheng}, K.~S.},
        title = "{Repeating fast radio burst 20201124A originates from a magnetar/Be star binary}",
      journal = {Nature Communications},
         year = 2022,
        month = sep,
       volume = {13},
          eid = {4382},
        pages = {4382},
          doi = {10.1038/s41467-022-31923-y},
archivePrefix = {arXiv},
       eprint = {2204.08124},
 primaryClass = {astro-ph.HE},
       adsurl = {https://ui.adsabs.harvard.edu/abs/2022NatCo..13.4382W}
}

@article{Hilmarsson2021,
doi = {10.3847/2041-8213/abdec0},
url = {https://doi.org/10.3847/2041-8213/abdec0},
year = {2021},
month = {feb},
publisher = {The American Astronomical Society},
volume = {908},
number = {1},
pages = {L10},
author = {Hilmarsson, G. H. and Michilli, D. and Spitler, L. G. and Wharton, R. S. and Demorest, P. and Desvignes, G. and Gourdji, K. and Hackstein, S. and Hessels, J. W. T. and Nimmo, K. and Seymour, A. D. and Kramer, M. and Mckinven, R.},
title = {Rotation Measure Evolution of the Repeating Fast Radio Burst Source FRB 121102},
journal = {The Astrophysical Journal Letters}
}

@ARTICLE{McKee&Truelove1995PhR,
       author = {{McKee}, C.~F. and {Truelove}, J.~K.},
        title = "{Explosions in the interstellar medium.}",
      journal = {\physrep},
         year = 1995,
        month = may,
       volume = {256},
       number = {1},
        pages = {157-172},
          doi = {10.1016/0370-1573(94)00106-D},
       adsurl = {https://ui.adsabs.harvard.edu/abs/1995PhR...256..157M}
}

@article{Cordes2017,
doi = {10.3847/1538-4357/aa74da},
url = {https://doi.org/10.3847/1538-4357/aa74da},
year = {2017},
month = {jun},
publisher = {The American Astronomical Society},
volume = {842},
number = {1},
pages = {35},
author = {Cordes, J. M. and Wasserman, I. and Hessels, J. W. T. and Lazio, T. J. W. and Chatterjee, S. and Wharton, R. S.},
title = {Lensing of Fast Radio Bursts by Plasma Structures in Host Galaxies},
journal = {The Astrophysical Journal}
}

@ARTICLE{Main2018Natur,
       author = {{Main}, Robert and {Yang}, I.-Sheng and {Chan}, Victor and {Li}, Dongzi and {Lin}, Fang Xi and {Mahajan}, Nikhil and {Pen}, Ue-Li and {Vanderlinde}, Keith and {van Kerkwijk}, Marten H.},
        title = "{Pulsar emission amplified and resolved by plasma lensing in an eclipsing binary}",
      journal = {\nat},
         year = 2018,
        month = may,
       volume = {557},
       number = {7706},
        pages = {522-525},
          doi = {10.1038/s41586-018-0133-z},
archivePrefix = {arXiv},
       eprint = {1805.09348},
 primaryClass = {astro-ph.HE},
       adsurl = {https://ui.adsabs.harvard.edu/abs/2018Natur.557..522M}
}

@ARTICLE{Lee+2012,
       author = {{Lee}, Shiu-Hang and {Ellison}, Donald C. and {Nagataki}, Shigehiro},
        title = "{A Generalized Model of Nonlinear Diffusive Shock Acceleration Coupled to an Evolving Supernova Remnant}",
      journal = {\apj},
         year = 2012,
        month = may,
       volume = {750},
       number = {2},
          eid = {156},
        pages = {156},
          doi = {10.1088/0004-637X/750/2/156},
archivePrefix = {arXiv},
       eprint = {1203.3614},
 primaryClass = {astro-ph.HE},
       adsurl = {https://ui.adsabs.harvard.edu/abs/2012ApJ...750..156L}
}

@ARTICLE{Jacovich+2021,
       author = {{Jacovich}, Taylor and {Patnaude}, Daniel and {Slane}, Patrick and {Badenes}, Carles and {Lee}, Shiu-Hang and {Nagataki}, Shigehiro and {Milisavljevic}, Dan},
        title = "{A Grid of Core-collapse Supernova Remnant Models. I. The Effect of Wind-driven Mass Loss}",
      journal = {\apj},
         year = 2021,
        month = jun,
       volume = {914},
       number = {1},
          eid = {41},
        pages = {41},
          doi = {10.3847/1538-4357/abf935},
archivePrefix = {arXiv},
       eprint = {2103.07980},
 primaryClass = {astro-ph.HE},
       adsurl = {https://ui.adsabs.harvard.edu/abs/2021ApJ...914...41J}
}

@ARTICLE{Farmer+2023,
       author = {{Farmer}, R. and {Laplace}, E. and {Ma}, Jing-ze and {de Mink}, S.~E. and {Justham}, S.},
        title = "{Nucleosynthesis of Binary-stripped Stars}",
      journal = {\apj},
         year = 2023,
        month = may,
       volume = {948},
       number = {2},
          eid = {111},
        pages = {111},
          doi = {10.3847/1538-4357/acc315},
archivePrefix = {arXiv},
       eprint = {2303.04520},
 primaryClass = {astro-ph.SR},
       adsurl = {https://ui.adsabs.harvard.edu/abs/2023ApJ...948..111F}
}

@ARTICLE{Laplace+2021,
       author = {{Laplace}, E. and {Justham}, S. and {Renzo}, M. and {G{\"o}tberg}, Y. and {Farmer}, R. and {Vartanyan}, D. and {de Mink}, S.~E.},
        title = "{Different to the core: The pre-supernova structures of massive single and binary-stripped stars}",
      journal = {\aap},
         year = 2021,
        month = dec,
       volume = {656},
          eid = {A58},
        pages = {A58},
          doi = {10.1051/0004-6361/202140506},
archivePrefix = {arXiv},
       eprint = {2102.05036},
 primaryClass = {astro-ph.SR},
       adsurl = {https://ui.adsabs.harvard.edu/abs/2021A&A...656A..58L}
}

@ARTICLE{Sana2012Sci,
       author = {{Sana}, H. and {de Mink}, S.~E. and {de Koter}, A. and {Langer}, N. and {Evans}, C.~J. and {Gieles}, M. and {Gosset}, E. and {Izzard}, R.~G. and {Le Bouquin}, J.-B. and {Schneider}, F.~R.~N.},
        title = "{Binary Interaction Dominates the Evolution of Massive Stars}",
      journal = {Science},
         year = 2012,
        month = jul,
       volume = {337},
       number = {6093},
        pages = {444},
          doi = {10.1126/science.1223344},
archivePrefix = {arXiv},
       eprint = {1207.6397},
 primaryClass = {astro-ph.SR},
       adsurl = {https://ui.adsabs.harvard.edu/abs/2012Sci...337..444S}
}

@ARTICLE{Lee+2014,
       author = {{Lee}, Shiu-Hang and {Patnaude}, Daniel J. and {Ellison}, Donald C. and {Nagataki}, Shigehiro and {Slane}, Patrick O.},
        title = "{Reverse and Forward Shock X-Ray Emission in an Evolutionary Model of Supernova Remnants Undergoing Efficient Diffusive Shock Acceleration}",
      journal = {\apj},
         year = 2014,
        month = aug,
       volume = {791},
       number = {2},
          eid = {97},
        pages = {97},
          doi = {10.1088/0004-637X/791/2/97},
archivePrefix = {arXiv},
       eprint = {1407.0095},
 primaryClass = {astro-ph.HE},
       adsurl = {https://ui.adsabs.harvard.edu/abs/2014ApJ...791...97L}
}

@ARTICLE{Lee+2015,
       author = {{Lee}, Shiu-Hang and {Patnaude}, Daniel J. and {Raymond}, John C. and {Nagataki}, Shigehiro and {Slane}, Patrick O. and {Ellison}, Donald C.},
        title = "{Modeling Bright {\ensuremath{\gamma}}-Ray and Radio Emission at Fast Cloud Shocks}",
      journal = {\apj},
         year = 2015,
        month = jun,
       volume = {806},
       number = {1},
          eid = {71},
        pages = {71},
          doi = {10.1088/0004-637X/806/1/71},
archivePrefix = {arXiv},
       eprint = {1504.05313},
 primaryClass = {astro-ph.HE},
       adsurl = {https://ui.adsabs.harvard.edu/abs/2015ApJ...806...71L}
}

@BOOK{RybickiLightman1979,
       author = {{Rybicki}, George B. and {Lightman}, Alan P.},
        title = "{Radiative processes in astrophysics}",
         year = 1979,
       adsurl = {https://ui.adsabs.harvard.edu/abs/1979rpa..book.....R}
}

@INPROCEEDINGS{NovikovThorne1973,
       author = {{Novikov}, I.~D. and {Thorne}, K.~S.},
        title = "{Astrophysics of black holes.}",
    booktitle = {Black Holes (Les Astres Occlus)},
         year = 1973,
       editor = {{Dewitt}, C. and {Dewitt}, B.~S.},
        month = jan,
        pages = {343-450},
       adsurl = {https://ui.adsabs.harvard.edu/abs/1973blho.conf..343N}
}

@ARTICLE{Lee+2013,
       author = {{Lee}, Shiu-Hang and {Slane}, Patrick O. and {Ellison}, Donald C. and {Nagataki}, Shigehiro and {Patnaude}, Daniel J.},
        title = "{A CR-hydro-NEI Model of Multi-wavelength Emission from the Vela Jr. Supernova Remnant (SNR RX J0852.0-4622)}",
      journal = {\apj},
         year = 2013,
        month = apr,
       volume = {767},
       number = {1},
          eid = {20},
        pages = {20},
          doi = {10.1088/0004-637X/767/1/20},
archivePrefix = {arXiv},
       eprint = {1302.4645},
 primaryClass = {astro-ph.HE},
       adsurl = {https://ui.adsabs.harvard.edu/abs/2013ApJ...767...20L}
}

@ARTICLE{ZYZhao2021a,
       author = {{Zhao}, Z.~Y. and {Zhang}, G.~Q. and {Wang}, Y.~Y. and {Tu}, Zuo-Lin and {Wang}, F.~Y.},
        title = "{Dispersion and Rotation Measures from the Ejecta of Compact Binary Mergers: Clue to the Progenitors of Fast Radio Bursts}",
      journal = {\apj},
         year = 2021,
        month = feb,
       volume = {907},
       number = {2},
          eid = {111},
        pages = {111},
          doi = {10.3847/1538-4357/abd321},
archivePrefix = {arXiv},
       eprint = {2010.10702},
 primaryClass = {astro-ph.HE},
       adsurl = {https://ui.adsabs.harvard.edu/abs/2021ApJ...907..111Z}
}

@ARTICLE{ZYZhao2021b,
       author = {{Zhao}, Z.~Y. and {Wang}, F.~Y.},
        title = "{FRB 190520B Embedded in a Magnetar Wind Nebula and Supernova Remnant: A Luminous Persistent Radio Source, Decreasing Dispersion Measure, and Large Rotation Measure}",
      journal = {\apjl},
         year = 2021,
        month = dec,
       volume = {923},
       number = {1},
          eid = {L17},
        pages = {L17},
          doi = {10.3847/2041-8213/ac3f2f},
archivePrefix = {arXiv},
       eprint = {2112.00935},
 primaryClass = {astro-ph.HE},
       adsurl = {https://ui.adsabs.harvard.edu/abs/2021ApJ...923L..17Z}
}

@ARTICLE{Tang&Chevalier2017,
       author = {{Tang}, Xiaping and {Chevalier}, Roger A.},
        title = "{Shock evolution in non-radiative supernova remnants}",
      journal = {\mnras},
         year = 2017,
        month = mar,
       volume = {465},
       number = {4},
        pages = {3793-3802},
          doi = {10.1093/mnras/stw2978},
archivePrefix = {arXiv},
       eprint = {1607.06391},
 primaryClass = {astro-ph.HE},
       adsurl = {https://ui.adsabs.harvard.edu/abs/2017MNRAS.465.3793T}
}

@ARTICLE{Chevalier1982,
       author = {{Chevalier}, R.~A.},
        title = "{Self-similar solutions for the interaction of stellar ejecta with an external medium.}",
      journal = {\apj},
         year = 1982,
        month = jul,
       volume = {258},
        pages = {790-797},
          doi = {10.1086/160126},
       adsurl = {https://ui.adsabs.harvard.edu/abs/1982ApJ...258..790C}
}

@ARTICLE{McKee&Truelove1999,
       author = {{Truelove}, J. Kelly and {McKee}, Christopher F.},
        title = "{Evolution of Nonradiative Supernova Remnants}",
      journal = {\apjs},
         year = 1999,
        month = feb,
       volume = {120},
       number = {2},
        pages = {299-326},
          doi = {10.1086/313176},
       adsurl = {https://ui.adsabs.harvard.edu/abs/1999ApJS..120..299T}
}

@ARTICLE{Patnaude2010,
       author = {{Patnaude}, Daniel J. and {Slane}, Patrick and {Raymond}, John C. and {Ellison}, Donald C.},
        title = "{The Role of Diffusive Shock Acceleration on Nonequilibrium Ionization in Supernova Remnant Shocks. II. Emitted Spectra}",
      journal = {\apj},
         year = 2010,
        month = dec,
       volume = {725},
       number = {2},
        pages = {1476-1484},
          doi = {10.1088/0004-637X/725/2/1476},
archivePrefix = {arXiv},
       eprint = {1010.3208},
 primaryClass = {astro-ph.HE},
       adsurl = {https://ui.adsabs.harvard.edu/abs/2010ApJ...725.1476P}
}

@INPROCEEDINGS{Diesing2019,
       author = {{Diesing}, R.},
        title = "{On the Spectrum of Electrons Accelerated in Supernova Remnants}",
    booktitle = {36th International Cosmic Ray Conference (ICRC2019)},
         year = 2019,
       series = {International Cosmic Ray Conference},
       volume = {36},
        month = jul,
          eid = {59},
        pages = {59},
       adsurl = {https://ui.adsabs.harvard.edu/abs/2019ICRC...36...59D}
}

@ARTICLE{Diesing2024,
       author = {{Diesing}, Rebecca and {Guo}, Minghao and {Kim}, Chang-Goo and {Stone}, James and {Caprioli}, Damiano},
        title = "{Nonthermal Signatures of Radiative Supernova Remnants}",
      journal = {\apj},
         year = 2024,
        month = oct,
       volume = {974},
       number = {2},
          eid = {201},
        pages = {201},
          doi = {10.3847/1538-4357/ad74f0},
archivePrefix = {arXiv},
       eprint = {2404.15396},
 primaryClass = {astro-ph.HE},
       adsurl = {https://ui.adsabs.harvard.edu/abs/2024ApJ...974..201D}
}

@ARTICLE{Diesing2025,
       author = {{Diesing}, Rebecca and {Gupta}, Siddhartha},
        title = "{Nonthermal Signatures of Radiative Supernova Remnants. II. The Impact of Cosmic Rays and Magnetic Fields}",
      journal = {\apj},
         year = 2025,
        month = feb,
       volume = {980},
       number = {2},
          eid = {167},
        pages = {167},
          doi = {10.3847/1538-4357/ada93d},
archivePrefix = {arXiv},
       eprint = {2411.18679},
 primaryClass = {astro-ph.HE},
       adsurl = {https://ui.adsabs.harvard.edu/abs/2025ApJ...980..167D}
}

@ARTICLE{Blondin2001,
       author = {{Blondin}, John M. and {Chevalier}, Roger A. and {Frierson}, Dargan M.},
        title = "{Pulsar Wind Nebulae in Evolved Supernova Remnants}",
      journal = {\apj},
         year = 2001,
        month = dec,
       volume = {563},
       number = {2},
        pages = {806-815},
          doi = {10.1086/324042},
archivePrefix = {arXiv},
       eprint = {astro-ph/0107076},
 primaryClass = {astro-ph},
       adsurl = {https://ui.adsabs.harvard.edu/abs/2001ApJ...563..806B}
}

@article{Blondin&Ellison2001,
doi = {10.1086/322499},
url = {https://doi.org/10.1086/322499},
year = {2001},
month = {oct},
publisher = {},
volume = {560},
number = {1},
pages = {244},
author = {Blondin, John M. and Ellison, Donald C.},
title = {Rayleigh-Taylor Instabilities in Young Supernova Remnants Undergoing Efficient Particle Acceleration},
journal = {The Astrophysical Journal}
}

@article{Ellison2007,
doi = {10.1086/517518},
url = {https://doi.org/10.1086/517518},
year = {2007},
month = {jun},
publisher = {},
volume = {661},
number = {2},
pages = {879},
author = {Ellison, Donald C. and Patnaude, Daniel J. and Slane, Patrick and Blasi, Pasquale and Gabici, Stefano},
title = {Particle Acceleration in Supernova Remnants and the Production of Thermal and Nonthermal Radiation},
journal = {The Astrophysical Journal}
}

@ARTICLE{Ferrand2010,
       author = {{Ferrand}, G. and {Decourchelle}, A. and {Ballet}, J. and {Teyssier}, R. and {Fraschetti}, F.},
        title = "{3D simulations of supernova remnants evolution including non-linear particle acceleration}",
      journal = {\aap},
         year = 2010,
        month = jan,
       volume = {509},
          eid = {L10},
        pages = {L10},
          doi = {10.1051/0004-6361/200913666},
archivePrefix = {arXiv},
       eprint = {0912.4886},
 primaryClass = {astro-ph.HE},
       adsurl = {https://ui.adsabs.harvard.edu/abs/2010A&A...509L..10F}
}

@ARTICLE{Ferrand2012,
       author = {{Ferrand}, Gilles and {Decourchelle}, Anne and {Safi-Harb}, Samar},
        title = "{Three-dimensional Simulations of the Thermal X-Ray Emission from Young Supernova Remnants Including Efficient Particle Acceleration}",
      journal = {\apj},
         year = 2012,
        month = nov,
       volume = {760},
       number = {1},
          eid = {34},
        pages = {34},
          doi = {10.1088/0004-637X/760/1/34},
archivePrefix = {arXiv},
       eprint = {1210.0085},
 primaryClass = {astro-ph.HE},
       adsurl = {https://ui.adsabs.harvard.edu/abs/2012ApJ...760...34F}
}

@article{Orlando2012,
doi = {10.1088/0004-637X/749/2/156},
url = {https://doi.org/10.1088/0004-637X/749/2/156},
year = {2012},
month = {apr},
publisher = {The American Astronomical Society},
volume = {749},
number = {2},
pages = {156},
author = {Orlando, S. and Bocchino, F. and Miceli, M. and Petruk, O. and Pumo, M. L.},
title = {ROLE OF EJECTA CLUMPING AND BACK-REACTION OF ACCELERATED COSMIC RAYS IN THE EVOLUTION OF TYPE Ia SUPERNOVA REMNANTS},
journal = {The Astrophysical Journal}
}

@ARTICLE{Ferrand2019,
       author = {{Ferrand}, Gilles and {Warren}, Donald C. and {Ono}, Masaomi and {Nagataki}, Shigehiro and {R{\"o}pke}, Friedrich K. and {Seitenzahl}, Ivo R.},
        title = "{From Supernova to Supernova Remnant: The Three-dimensional Imprint of a Thermonuclear Explosion}",
      journal = {\apj},
         year = 2019,
        month = jun,
       volume = {877},
       number = {2},
          eid = {136},
        pages = {136},
          doi = {10.3847/1538-4357/ab1a3d},
archivePrefix = {arXiv},
       eprint = {1904.08062},
 primaryClass = {astro-ph.HE},
       adsurl = {https://ui.adsabs.harvard.edu/abs/2019ApJ...877..136F}
}

@ARTICLE{Patnaude2017,
       author = {{Patnaude}, Daniel J. and {Lee}, Shiu-Hang and {Slane}, Patrick O. and {Badenes}, Carles and {Nagataki}, Shigehiro and {Ellison}, Donald C. and {Milisavljevic}, Dan},
        title = "{The Impact of Progenitor Mass Loss on the Dynamical and Spectral Evolution of Supernova Remnants}",
      journal = {\apj},
         year = 2017,
        month = nov,
       volume = {849},
       number = {2},
          eid = {109},
        pages = {109},
          doi = {10.3847/1538-4357/aa9189},
archivePrefix = {arXiv},
       eprint = {1708.04984},
 primaryClass = {astro-ph.HE},
       adsurl = {https://ui.adsabs.harvard.edu/abs/2017ApJ...849..109P}
}

@ARTICLE{Paczyski1967,
       author = {{Paczy{\'n}ski}, B.},
        title = "{Gravitational Waves and the Evolution of Close Binaries}",
      journal = {\actaa},
         year = 1967,
        month = jan,
       volume = {17},
        pages = {287},
       adsurl = {https://ui.adsabs.harvard.edu/abs/1967AcA....17..287P}
}

@ARTICLE{Heuvel1969,
       author = {{van den Heuvel}, E.~P.~J.},
        title = "{The Expected Fraction of Evolved Close Binaries among Main-Sequence Stars of Spectral Type Earlier than A5}",
      journal = {\aj},
         year = 1969,
        month = nov,
       volume = {74},
        pages = {1095},
          doi = {10.1086/110909},
       adsurl = {https://ui.adsabs.harvard.edu/abs/1969AJ.....74.1095V}
}

@ARTICLE{Court2024,
       author = {{Court}, Travis and {Badenes}, Carles and {Lee}, Shiu-Hang and {Patnaude}, Daniel and {Garc{\'\i}a-Segura}, Guillermo and {Bravo}, Eduardo},
        title = "{Do Type Ia Supernovae Explode inside Planetary Nebulae?}",
      journal = {\apj},
         year = 2024,
        month = feb,
       volume = {962},
       number = {1},
          eid = {63},
        pages = {63},
          doi = {10.3847/1538-4357/ad165f},
archivePrefix = {arXiv},
       eprint = {2309.00572},
 primaryClass = {astro-ph.HE},
       adsurl = {https://ui.adsabs.harvard.edu/abs/2024ApJ...962...63C}
}

@article{Amaku2010,
title = {Decay chain differential equations: Solution through matrix algebra},
journal = {Computer Physics Communications},
volume = {181},
number = {1},
pages = {21-23},
year = {2010},
issn = {0010-4655},
doi = {https://doi.org/10.1016/j.cpc.2009.08.011},
url = {https://www.sciencedirect.com/science/article/pii/S0010465509002628},
author = {M. Amaku and P.R. Pascholati and V.R. Vanin}
}

@misc{ICRP2008NuclearCalculations,
    title = {{Nuclear Decay Data for Dosimetric Calculations}},
    year = {2008},
    booktitle = {ICRP Publication 107},
    author = {{ICRP}},
    number = {(3)},
    pages = {7--96},
    volume = {38(3)},
    doi = {https://doi.org/10.1016/j.icrp.2008.10.004.}
}

@INCOLLECTION{Chevalier2017,
       author = {{Chevalier}, Roger A. and {Fransson}, Claes},
        title = "{Thermal and Non-thermal Emission from Circumstellar Interaction}",
    booktitle = {Handbook of Supernovae},
         year = 2017,
       editor = {{Alsabti}, Athem W. and {Murdin}, Paul},
        pages = {875},
          doi = {10.1007/978-3-319-21846-5_34},
       adsurl = {https://ui.adsabs.harvard.edu/abs/2017hsn..book..875C}
}

@ARTICLE{Laming2020,
       author = {{Laming}, J. Martin and {Temim}, Tea},
        title = "{Element Abundances in the Unshocked Ejecta of Cassiopeia A}",
      journal = {\apj},
         year = 2020,
        month = dec,
       volume = {904},
       number = {2},
          eid = {115},
        pages = {115},
          doi = {10.3847/1538-4357/abc1e5},
archivePrefix = {arXiv},
       eprint = {2010.07718},
 primaryClass = {astro-ph.HE},
       adsurl = {https://ui.adsabs.harvard.edu/abs/2020ApJ...904..115L}
}

@article{Dai2017,
doi = {10.3847/2041-8213/aa6745},
url = {https://doi.org/10.3847/2041-8213/aa6745},
year = {2017},
month = {mar},
publisher = {The American Astronomical Society},
volume = {838},
number = {1},
pages = {L7},
author = {Dai, Z. G. and Wang, J. S. and Yu, Y. W.},
title = {Radio Emission from Pulsar Wind Nebulae without Surrounding Supernova Ejecta: Application to FRB 121102},
journal = {The Astrophysical Journal Letters}
}

@article{Yang2019,
doi = {10.3847/1538-4357/ab48dd},
url = {https://doi.org/10.3847/1538-4357/ab48dd},
year = {2019},
month = {nov},
publisher = {The American Astronomical Society},
volume = {885},
number = {2},
pages = {149},
author = {Yang, Yu-Han and Dai, Zi-Gao},
title = {Emission from a Pulsar Wind Nebula: Application to the Persistent Radio Counterpart of FRB 121102},
journal = {The Astrophysical Journal}
}

@article{Beloborodov2017,
doi = {10.3847/2041-8213/aa78f3},
url = {https://doi.org/10.3847/2041-8213/aa78f3},
year = {2017},
month = {jul},
publisher = {The American Astronomical Society},
volume = {843},
number = {2},
pages = {L26},
author = {Beloborodov, Andrei M.},
title = {A Flaring Magnetar in FRB 121102?},
journal = {The Astrophysical Journal Letters}
}

@article{Dong2025,
doi = {10.3847/1538-4357/adfb74},
url = {https://doi.org/10.3847/1538-4357/adfb74},
year = {2025},
month = {sep},
publisher = {The American Astronomical Society},
volume = {991},
number = {2},
pages = {199},
author = {Dong, Y. and Kilpatrick, C. D. and Fong, W. and Curtin, A. P. and Opoku, S. and Andersen, B. C. and Cook, A. M. and Eftekhari, T. and Fonseca, E. and Gaensler, B. M. and Joseph, R. C. and Kaczmarek, J. F. and Kahinga, L. A. and Kaspi, V. and Lanman, A. E. and Lazda, M. and Leung, C. and Masui, K. W. and Michilli, D. and Nimmo, K. and Pandhi, A. and Pearlman, A. B. and Sammons, M. and Scholz, P. and Shah, V. and Shin, K. and Smith, K.},
title = {Searching for Historical Extragalactic Optical Transients Associated with Fast Radio Bursts},
journal = {The Astrophysical Journal}
}

@article{Paxton2011ModulesMESA,
    title = {{Modules for Experiments in Stellar Astrophysics (MESA)}},
    year = {2011},
    journal = {Astrophysical Journal, Supplement Series},
    author = {Paxton, Bill and Bildsten, Lars and Dotter, Aaron and Herwig, Falk and Lesaffre, Pierre and Timmes, Frank},
    number = {1},
    month = {1},
    volume = {192},
    doi = {10.1088/0067-0049/192/1/3},
    issn = {00670049},
    arxivId = {1009.1622}
}

@article{Paxton2015ModulesExplosions,
    title = {{Modules for Experiments in Stellar Astrophysics (MESA): Binaries, pulsations, and explosions}},
    year = {2015},
    journal = {Astrophysical Journal, Supplement Series},
    author = {Paxton, Bill and Marchant, Pablo and Schwab, Josiah and Bauer, Evan B. and Bildsten, Lars and Cantiello, Matteo and Dessart, Luc and Farmer, R. and Hu, H. and Langer, N. and Townsend, R. H.D. and Townsley, Dean M. and Timmes, F. X.},
    number = {1},
    month = {9},
    volume = {220},
    publisher = {Institute of Physics Publishing},
    doi = {10.1088/0067-0049/220/1/15},
    issn = {00670049},
    arxivId = {1506.03146}
}

@article{Paxton2018ModulesExplosions,
    title = {{Modules for Experiments in Stellar Astrophysics (MESA): Convective Boundaries, Element Diffusion, and Massive Star Explosions}},
    year = {2018},
    author = {Paxton, Bill and Schwab, Josiah and Bauer, Evan B. and Bildsten, Lars and Blinnikov, Sergei and Duffell, Paul and Farmer, R. and Goldberg, Jared A. and Marchant, Pablo and Sorokina, Elena and Thoul, Anne and Townsend, Richard H. D. and Timmes, F. X.},
    month = {1},
    url = {http://arxiv.org/abs/1710.08424 http://dx.doi.org/10.3847/1538-4365/aaa5a8},
    doi = {10.3847/1538-4365/aaa5a8},
    arxivId = {1710.08424}
}

@article{Paxton2013ModulesStars,
    title = {{Modules for experiments in stellar astrophysics (MESA): Planets, oscillations, rotation, and massive stars}},
    year = {2013},
    journal = {Astrophysical Journal, Supplement Series},
    author = {Paxton, Bill and Cantiello, Matteo and Arras, Phil and Bildsten, Lars and Brown, Edward F. and Dotter, Aaron and Mankovich, Christopher and Montgomery, M. H. and Stello, Dennis and Timmes, F. X. and Townsend, Richard},
    number = {1},
    month = {9},
    volume = {208},
    doi = {10.1088/0067-0049/208/1/4},
    issn = {00670049},
    arxivId = {1301.0319}
}

@article{Paxton2019ModulesConservation,
    title = {{Modules for Experiments in Stellar Astrophysics (MESA): Pulsating Variable Stars, Rotation, Convective Boundaries, and Energy Conservation}},
    year = {2019},
    journal = {The Astrophysical Journal Supplement Series},
    author = {Paxton, Bill and Smolec, R. and Schwab, Josiah and Gautschy, A. and Bildsten, Lars and Cantiello, Matteo and Dotter, Aaron and Farmer, R. and Goldberg, Jared A. and Jermyn, Adam S. and Kanbur, S. M. and Marchant, Pablo and Thoul, Anne and Townsend, Richard H. D. and Wolf, William M. and Zhang, Michael and Timmes, F. X.},
    number = {1},
    month = {7},
    pages = {10},
    volume = {243},
    publisher = {American Astronomical Society},
    doi = {10.3847/1538-4365/ab2241},
    issn = {0067-0049},
    arxivId = {1903.01426}
}

@article{Jermyn2023ModulesInfrastructure,
    title = {{Modules for Experiments in Stellar Astrophysics (MESA): Time-dependent Convection, Energy Conservation, Automatic Differentiation, and Infrastructure}},
    year = {2023},
    journal = {The Astrophysical Journal Supplement Series},
    author = {Jermyn, Adam S. and Bauer, Evan B. and Schwab, Josiah and Farmer, R. and Ball, Warrick H. and Bellinger, Earl P. and Dotter, Aaron and Joyce, Meridith and Marchant, Pablo and Mombarg, Joey S. G. and Wolf, William M. and Sunny Wong, Tin Long and Cinquegrana, Giulia C. and Farrell, Eoin and Smolec, R. and Thoul, Anne and Cantiello, Matteo and Herwig, Falk and Toloza, Odette and Bildsten, Lars and Townsend, Richard H. D. and Timmes, F. X.},
    number = {1},
    month = {3},
    pages = {15},
    volume = {265},
    publisher = {American Astronomical Society},
    doi = {10.3847/1538-4365/acae8d},
    issn = {0067-0049},
    arxivId = {2208.03651}
}

@ARTICLE{Takahashi2021MNRAS,
       author = {{Takahashi}, Ryuichi and {Ioka}, Kunihito and {Mori}, Asuka and {Funahashi}, Koki},
        title = "{Statistical modelling of the cosmological dispersion measure}",
      journal = {\mnras},
         year = 2021,
        month = apr,
       volume = {502},
       number = {2},
        pages = {2615-2629},
          doi = {10.1093/mnras/stab170},
archivePrefix = {arXiv},
       eprint = {2010.01560},
 primaryClass = {astro-ph.CO},
       adsurl = {https://ui.adsabs.harvard.edu/abs/2021MNRAS.502.2615T}
}

@ARTICLE{Wang2025arXiv250715790W,
       author = {{Wang}, P. and {Zhang}, J.~S. and {Yang}, Y.~P. and {Zhou}, D.~K. and {Zhang}, Y.~K. and {Feng}, Y. and {Zhao}, Z.~Y. and {Fang}, J.~H. and {Li}, D. and {Zhu}, W.~W. and {Zhang}, B. and {Wang}, F.~Y. and {Huang}, Y.~F. and {Luo}, R. and {Han}, J.~L. and {Lee}, K.~J. and {Tsai}, C.~W. and {Dai}, Z.~G. and {Gao}, H. and {Zheng}, X.~P. and {Cao}, J.~H. and {Chen}, X.~L. and {Gugercinoglu}, E. and {Jiang}, J.~C. and {Jing}, W.~C. and {Li}, Y. and {Li}, J. and {Lu}, W.~J. and {Luo}, J.~W. and {Lyu}, F. and {Miao}, C.~C. and {Niu}, C.~H. and {Niu}, J.~R. and {Qu}, Y. and {Wang}, W.~Y. and {Wang}, Y.~D. and {Wang}, Y.~B. and {Wang}, C.~J. and {Wu}, Q. and {Wu}, Y.~S. and {Weng}, S.~M. and {Xiao}, D. and {Xu}, H. and {Yao}, J.~M. and {Zhang}, C.~F. and {Zhao}, R.~S. and {Liu}, Q.~Z. and {Zhang}, J. and {Zhou}, D.~J. and {Zhang}, L. and {Zhu}, Y.~H.},
        title = "{Decadal evolution of a repeating fast radio burst source}",
      journal = {arXiv e-prints},
         year = 2025,
        month = jul,
          eid = {arXiv:2507.15790},
        pages = {arXiv:2507.15790},
          doi = {10.48550/arXiv.2507.15790},
archivePrefix = {arXiv},
       eprint = {2507.15790},
 primaryClass = {astro-ph.HE},
       adsurl = {https://ui.adsabs.harvard.edu/abs/2025arXiv250715790W}
}

@ARTICLE{Zhan2023RvMP,
       author = {{Zhang}, Bing},
        title = "{The physics of fast radio bursts}",
      journal = {Reviews of Modern Physics},
         year = 2023,
        month = jul,
       volume = {95},
       number = {3},
          eid = {035005},
        pages = {035005},
          doi = {10.1103/RevModPhys.95.035005},
archivePrefix = {arXiv},
       eprint = {2212.03972},
 primaryClass = {astro-ph.HE},
       adsurl = {https://ui.adsabs.harvard.edu/abs/2023RvMP...95c5005Z}
}

@ARTICLE{Simha2020,
       author = {{Simha}, Sunil and {Burchett}, Joseph N. and {Prochaska}, J. Xavier and {Chittidi}, Jay S. and {Elek}, Oskar and {Tejos}, Nicolas and {Jorgenson}, Regina and {Bannister}, Keith W. and {Bhandari}, Shivani and {Day}, Cherie K. and {Deller}, Adam T. and {Forbes}, Angus G. and {Macquart}, Jean-Pierre and {Ryder}, Stuart D. and {Shannon}, Ryan M.},
        title = "{Disentangling the Cosmic Web toward FRB 190608}",
      journal = {\apj},
         year = 2020,
        month = oct,
       volume = {901},
       number = {2},
          eid = {134},
        pages = {134},
          doi = {10.3847/1538-4357/abafc3},
archivePrefix = {arXiv},
       eprint = {2005.13157},
 primaryClass = {astro-ph.GA},
       adsurl = {https://ui.adsabs.harvard.edu/abs/2020ApJ...901..134S}
}

@ARTICLE{Wang2025arXiv251207140W,
       author = {{Wang}, F.~Y. and {Lan}, H.~T. and {Zhao}, Z.~Y. and {Wu}, Q. and {Feng}, Y. and {Yi}, S.~X. and {Dai}, Z.~G. and {Cheng}, K.~S.},
        title = "{Evidence of young magnetars in massive binary embedded in a supernova remnant as sources of active fast radio bursts}",
      journal = {arXiv e-prints},
         year = 2025,
        month = dec,
          eid = {arXiv:2512.07140},
        pages = {arXiv:2512.07140},
archivePrefix = {arXiv},
       eprint = {2512.07140},
 primaryClass = {astro-ph.HE},
       adsurl = {https://ui.adsabs.harvard.edu/abs/2025arXiv251207140W}
}

@ARTICLE{Kawashima2026arXiv,
       author = {{Kawashima}, Gaku and {Lee}, Shiu-Hang and {Maeda}, Keiichi and {Patnaude}, Daniel},
        title = "{Single vs. Binary Origin: The Diversity of Stripped-Envelope Supernova Remnants}",
      journal = {arXiv e-prints},
         year = 2026,
        month = mar,
          eid = {arXiv:2603.28234},
        pages = {arXiv:2603.28234},
          doi = {10.48550/arXiv.2603.28234},
archivePrefix = {arXiv},
       eprint = {2603.28234},
 primaryClass = {astro-ph.HE},
       adsurl = {https://ui.adsabs.harvard.edu/abs/2026arXiv260328234K}
}

@misc{farmer_2023_5929871,
  author       = {Farmer, Robert and
                  Laplace, Eva and
                  Ma, Jing-Ze and
                  de Mink, S. E. and
                  Justham, S},
  title        = {Nucleosynthesis of binary-stripped stars.},
  month        = mar,
  year         = 2023,
  publisher    = {Zenodo},
  doi          = {10.5281/zenodo.5929871},
  url          = {https://doi.org/10.5281/zenodo.5929871},
}

@ARTICLE{Pandhi2026,
       author = {{Pandhi}, Ayush and {Nimmo}, Kenzie and {Andrew}, Shion and {Brar}, Charanjot and {Chatterjee}, Shami and {Cook}, Amanda M. and {Curtin}, Alice and {Gaensler}, B.~M. and {Gawronski}, Marcin and {Hessels}, Jason and {Kaspi}, Victoria M. and {Khan}, Afrokk and {Kirsten}, Franz and {Lazda}, Mattias and {Leung}, Calvin and {Main}, Robert and {Masui}, Kiyoshi W. and {Mckinven}, Ryan and {Michilli}, Daniele and {Ng}, Mason and {Ould-Boukattine}, Omar and {Pearlman}, Aaron B. and {Pleunis}, Ziggy and {Pollak}, Alexander W. and {Pradeep E.~T.}, Sachin and {Puchalska}, Weronika and {Sammons}, Mawson W. and {Scholz}, Paul and {Shah}, Vishwangi and {Shin}, Kaitlyn and {Siegel}, Seth R. and {Smith}, Kendrick},
        title = "{A Steadily Declining Dispersion Measure for the Repeating Fast Radio Burst FRB 20220529A: Evidence for a Fast Radio Burst Engine Embedded in an Expanding Supernova Remnant}",
      journal = {\apjl},
         year = 2026,
        month = apr,
       volume = {1000},
       number = {2},
          eid = {L53},
        pages = {L53},
          doi = {10.3847/2041-8213/ae52f8},
archivePrefix = {arXiv},
       eprint = {2602.22309},
 primaryClass = {astro-ph.HE},
       adsurl = {https://ui.adsabs.harvard.edu/abs/2026ApJ..1000L..53P}
}

@ARTICLE{Margalit2018MNRAS,
       author = {{Margalit}, Ben and {Metzger}, Brian D. and {Berger}, Edo and {Nicholl}, Matt and {Eftekhari}, Tarraneh and {Margutti}, Raffaella},
        title = "{Unveiling the engines of fast radio bursts, superluminous supernovae, and gamma-ray bursts}",
      journal = {\mnras},
         year = 2018,
        month = dec,
       volume = {481},
       number = {2},
        pages = {2407-2426},
          doi = {10.1093/mnras/sty2417},
archivePrefix = {arXiv},
       eprint = {1806.05690},
 primaryClass = {astro-ph.HE},
       adsurl = {https://ui.adsabs.harvard.edu/abs/2018MNRAS.481.2407M}
}

@misc{10.5281/zenodo.20487842,
  author       = {{Zhang}, Zhao Joseph},
  title        = {ARCOS: Analysis Repository for Cosmological FRB Studies in Simulations},
  month        = jun,
  year         = 2026,
  publisher    = {Zenodo},
  doi          = {10.5281/zenodo.20487842},
  url          = {https://doi.org/10.5281/zenodo.20487842}
}
\bibliographystyle{aasjournal}

\end{CJK}
\end{document}